\documentclass[times,twocolumn,final]{elsarticle}

\usepackage{lineno,hyperref}
\usepackage{amsmath,amsfonts,amssymb}
\usepackage{graphicx}
\usepackage{multirow}
\usepackage{subcaption}
\usepackage{pdflscape}
\usepackage{setspace}
\usepackage{tocloft}
\usepackage[export]{adjustbox}

\biboptions{authoryear}

\begin{document}

\begin{frontmatter}

\title{MedShift: identifying shift data for medical dataset curation }


\author[emorycs]{Xiaoyuan Guo}
\ead{xiaoyuan.guo@emory.edu}
\author[emorymed,emoryrad]{Judy Wawira Gichoya}
\ead{judywawira@emory.edu}
\author[emorymed,emoryrad]{Hari Trivedi}
\ead{hari.trivedi@emory.edu}
\author[iupui]{Saptarshi Purkayastha}
\ead{saptpurk@iupui.edu}

\author[mayo,asu]{Imon Banerjee\corref{mycorrespondingauthor}}
\cortext[mycorrespondingauthor]{Corresponding author}
\ead{banerjee.imon@mayo.edu}

\address[emorycs]{Department of Computer Science, Emory University, GA, USA}
\address[emorymed]{School of Medicine, Emory University, GA, USA}
\address[emoryrad]{Department of Radiology and Imaging Sciences, Emory University, GA, USA}
\address[iupui]{Indiana University-Purdue University Indianapolis, School of Informatics and Computing, IN, USA}
\address[mayo]{Department of Radiology, Mayo clinic, Phoenix, AZ, USA}
\address[asu]{School of Computing and Augmented Intelligence, Arizona State University, AZ, USA}

\begin{abstract}
Automated dataset curation in the medical domain has long been demanding as AI technologies are often hungry for annotated data. To curate a high-quality dataset, identifying data variance between the internal and external sources is a fundamental and crucial step as the data distributions from different sources can vary significantly and thus affect the performance of the AI models. However, methods to detect shift or variance in data have not been significantly researched. Challenges to this are the lack of effective approaches to learn dense representation of a dataset by capturing its semantics and difficulties of sharing private data across medical institutions. To overcome the problems, we propose a unified pipeline called \textit{MedShift} to detect the top-level shift samples and thus facilitate the medical curation. Given an internal dataset \textit{A} as the base source, we first train anomaly detectors for each class of dataset \textit{A} to learn internal distributions in an unsupervised way. Second, without exchanging data across sources, we run the trained anomaly detectors on an external dataset \textit{B} for each class. The data samples with high anomaly scores are identified as shift data. To quantify the \textit{shiftness} of the external dataset, we cluster \textit{B}'s data into groups class-wise based on the obtained scores. We then train a multi-class classifier on A and measure the \textit{shiftness} with the classifier's performance variance on B by gradually dropping the group with the largest anomaly score for each class. Additionally, we adapt a dataset quality metric to help inspect the distribution differences for multiple medical sources. We verify the efficacy of \textit{MedShift} with musculoskeletal radiographs (MURA) and chest X-rays datasets from more than one external source. Experiments show our proposed shift data detection pipeline can be beneficial for medical centers to curate high-quality datasets more efficiently. 
An interface introduction video to visualize our results is available at \url{https://youtu.be/V3BF0P1sxQE}.
\end{abstract}

\begin{keyword}
Dataset curation \sep Medical shift data \sep Anomaly detection \sep OOD detection, MURA X-ray \sep Chest X-ray
\end{keyword}

\end{frontmatter}


\section{Introduction}
\label{sect:intro}  
Supervised deep learning has been promising in solving various medical image-related tasks, and often requires well-annotated datasets for training, which highly drives the generation of medical datasets by research institutions and hospitals. Many of them have established or plan to establish research data curation services. When building large data collections for usage in training and validation of machine learning, merely collecting a lot of data is not enough~\cite{van2019quality,yamoah2019data}. It is essential that the quality of the data is sufficient for the intended application in order to obtain valid results~\cite{van2019quality}. The medical datasets from different institutions can be heterogeneous and with distribution shifts. Models trained on an internal dataset \textit{A} from a specific institute may show degraded performance on an external dataset \textit{B} from other sources due to the possible noisy data, distribution shift and poor-quality data, which are called \textit{shift data} in this paper. Dataset/Distribution shift is a common problem in predictive modelling and present in most practical applications, for reasons ranging from the bias in introduced by experimental design the irreproducibility of the testing conditions at training time~\cite{quinonero2009dataset,wang2021embracing}, of which imbalanced data, domain shift, source component shift, may be the most common forms~\cite{storkey2009training}. The shift data introduces out-of-distribution (OOD) in the dataset, and should account for the performance dropping of well-trained models. Thus, identifying the shift data is crucial for cleaning the datasets and helpful in enhancing the model's generalization with future training. Unfortunately, it still lacks an effective way to identify the difference for a bunch of datasets from the same medical domain. The main challenge lies in the inaccessibility to external medical datasets. Privacy concerns around sharing personally identifiable information are a major barrier to data sharing in medical research~\cite{schutte2021overcoming}. To address these privacy concerns, there has been an impressive number of large-scale research collaborations to pool and curate de-identified medical data for open-source research purposes~\cite{clark2013cancer}. Nevertheless, most medical data is still isolated and locally stored in hospitals and laboratories due to the worries associated with sharing patient data~\cite{van2014systematic}.  Therefore, an efficient way of external dataset curation/cleaning without sharing data is desired. 

To overcome the obstacle, we propose \textit{MedShift}, a pipeline for identifying shift data, which takes advantage of the accessible models trained on the internal dataset to gain the in-distribution knowledge. As observed by Ref.~\cite{rabanser2018failing}, domain-discriminating approaches tend to be helpful for characterizing shifts qualitatively and determining if the are harmful. Therefore, we utilize unsupervised anomaly detectors to learn the ``normality" of in-domain features. Suppose the internal dataset has multiple classes, the feature representation of each class is learnt by an OOD detector. Without sharing the internal dataset with others, the shift data is theoretically under-represented and should be detected by the accessible anomaly detectors as outliers from the external datasets. Since the supervised deep learning suffers from the performance dropping when facing the distribution/dataset shifting, especially when training data and test data are from two sources, two intuitions for example, the \textit{shiftness} of the identified data can be reflected via the performance variance of a well-trained model. Instead of checking the shift sample one by one, \textit{MedShift} quantifies the \textit{shiftness} for each class in small groups. Based on the assigned anomaly scores, each class of the external datasets is clustered into multiple groups. Data samples with similar qualities will be grouped together. A multi-class classifier is then trained on the internal dataset and evaluated on the external datasets. Each group of each class in external datasets is gradually dropped in the decreasing order of anomaly scores. Meanwhile, the classification performance on the updated external data is recorded. The corresponding variation in performance, hence, reflects the significance of the distribution shift based on the fact that subtle changes in data distribution may affect the performance of well-trained classifiers. Additionally, we adopt a dataset quality metric (OTDD~\cite{alvarez2020geometric}) for helping facilitate the comparison of differences among a series of datasets coming from the same medical domain. We summarize our contributions as follows:
\begin{enumerate}
\item We propose an automatic pipeline of identifying shift data for medical data curation applications and evaluating the significance of shift data without sharing data between the internal and external organizations;
\item We employ two unsupervised anomaly detectors to learn the internal distribution and identify samples showing the significant \textit{shiftness} for external datasets, and compared their performance; 
\item We quantify the effects of the shift data by training a multi-class classifier that learns internal domain knowledge and evaluating the classification performance for each sub-group of each class in external domains after dropping the shift data;
\item We adapt a data quality metric to quantify the dissimilarity between the internal and external datasets;
\item We experiment on two pairs of representative medical datasets and show effective qualitative and quantitative results, which prove the usefulness of the suggested pipeline for future medical dataset curation.
\end{enumerate}

\section{Methodology}
In Section~\ref{problem_statement} and ~\ref{notation}, we formulate the dataset shift identification problem and introduce the necessary notations. Then, we propose and illustrate the pipeline of shift identification in Section~\ref{shift_identification}; we further dive deep in the \textit{shiftness} evaluation in Section~\ref{shift_quantification}. To complement, we introduce the details of our anomaly detection architecture used for \textit{MedShift} pipeline in Section~\ref{anomaly_detection}. Additionally we introduce the dataset quality measurement in Section~\ref{dataset_quality}.    

\subsection{Problem Statement}~\label{problem_statement}
In view of the fact that the digital healthcare research is hugely limited by the data sharing and privacy issues because of the regulation imposed by Health Insurance Portability and Accountability Act (HIPPA), \textit{MedShift} aims to overcome the barrier by exploiting the advantage of sharing data quality evaluation models across the organizations and inspects the \textit{shiftness} of external datasets based on the learnt internal domain.

\subsection{Formulation and Notation}~\label{notation}
Given two datasets $D_{A}$ and $D_{B}$ of the same medical domain with the same classes (say $c_{1}, c_{2}, ..., c_{n}$, $n$ is the total number of classes) from two intuitions $A$ and $B$ (e.g., a chest X-ray dataset from Emory University $D_{A}$ and a chest X-ray dataset from Stanford University $D_{B}$), let $D_{A}$ be the internal dataset and $D_{B}$ be the external dataset. Dataset distribution shift is termed the situation where $P_{D_{A}}(Y|X)=P_{D_{B}}(Y|X)$ but $P_{D_{A}}(X) \neq P_{D_{B}}(X)$, where $Y$ and $X$ represent the class labels and input data respectively.

Suppose we are given i.i.d. internal data ${\{{X_{c_{i}}^{A}}\}_{i=1}^{n}}$ with $n$ classes, and input samples $\{{x_{j}^{A{c_{i}}}\}_{j=1}^{N^{A}_{c_{i}}}}\subset{{X_{c_{i}}^{A}}}$ (${N^{A}_{c_{i}}}$ is the sample number of dataset \textit{A}'s class $c_{i}$) from the internal input distribution, and i.i.d. external data ${\{{X_{c_{i}}^{B}}\}_{i=1}^{n}}$ and input samples $\{{x_{j}^{B{c_{i}}}\}_{j=1}^{N^{B}_{c_{i}}}}\subset{{X_{c_{i}}^{B}}}$ (${N^{B}_{c_{i}}}$ is the sample number of dataset \textit{B}'s class $c_{i}$) from external distribution, the detection of class-wise distribution shift for dataset $D_{B}$ based on $D_{A}$ is to identify the anomalous samples ${{\bar X_{c_{i}}^{B}}} \subseteq {{X_{c_{i}}^{B}}}$. Take $D_{A}$ class data as in-distribution (ID) data and train machine learning models (e.g. classification models), the models can learn the distribution of $D_{A}$'s classes and make predictions $P(y^{A}_{c_{i}}|x^{A}_{c_{i}})$ for some targets $y^{A}_{c_{i}}$ given data samples $x^{A}_{c_{i}}$ for class $c_{i}$. Theoretically, given the target model trained on the ID data ${{X_{c_{i}}^{A}}}$, the predictions over set ${{X_{c_{i}}^{B}}}-{{\bar X_{c_{i}}^{B}}}$ should produce more relevant results than on the whole set ${{X_{c_{i}}^{B}}}$. 

\subsection{Shift Identification}~\label{shift_identification}
\begin{figure*}[tp]
\begin{center}
  \includegraphics[width=\linewidth]{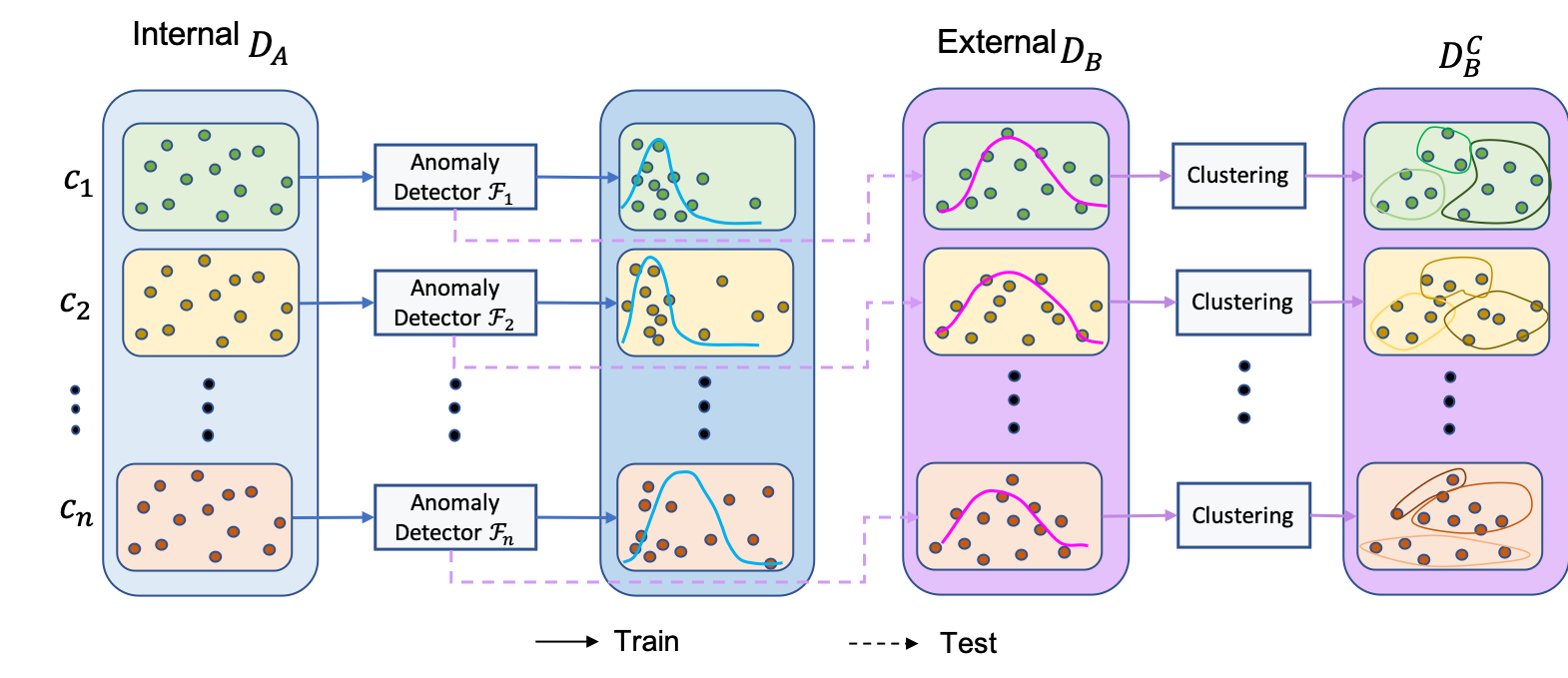}
\end{center}
  \caption{Shift data identification pipeline}
\label{mura-pipeline1}
\end{figure*}

In this section, we introduce the methodology for identification of image data distribution shift to discriminate the poor-quality, noisy and under-represented samples from the external data in an automatic way. The pipeline is built on top of the anomaly detection architecture to leverage the anomaly score as illustrated in the framework in Fig.~\ref{mura-pipeline1}, which involves two separate phases - internal training and test phase. An interesting challenge of shift identification is that the anomaly detectors should be able to identify unknown anomalous patterns of an external dataset without including any anomalous data samples in training since in the real situation, exchanging healthcare data among institutions and manually identifying noisy or anomalous data are not trivial tasks. 

During the training phase, only internal data samples and the anomaly detection models (see introductions in Sec.~\ref{anomaly_detection}) are involved. As shown in the left blue part of Fig.~\ref{mura-pipeline1}, a set of anomaly detectors $\mathcal{F}$s for each targeted categories of $D_{A}$ are trained on the internal dataset in an unsupervised fashion, considering the unavailability of external data sources. Each class will then obtain a unique OOD detector $\mathcal{F}_{c}$. The anomaly detector learns to assign each data item with a specific anomaly score, a higher score means more possibility of being an anomalous data. Notably, the anomaly detectors are trained with accessible internal data, and then shared with the external validation sites.

In the test phase, no internal data will be shared but the trained anomaly detector model with shift identification capability will be exchanged. As represented with pink figures and dotted flows in Fig.~\ref{mura-pipeline1}, each trained anomaly detector is evaluated on each corresponding class of dataset $D_{B}$ and assigns anomaly scores for the external dataset. To prepare for the \textit{shiftness} quantification in Sec.~\ref{shift_quantification}, an unsupervised clustering algorithm is subsequently applied to each class and clusters the data items into $k$ groups based on the learnt anomaly scores. For each class, the optimal number of cluster $k$ is determined by the Elbow Method. As observed during our experiments, data collected from the same source usually presents similar distributions. Therefore, we keep $k$ as the same across all the classes.

\subsection{Shiftness Quantification}~\label{shift_quantification}
\begin{figure*}[tp]
\begin{center}
  \includegraphics[width=\linewidth]{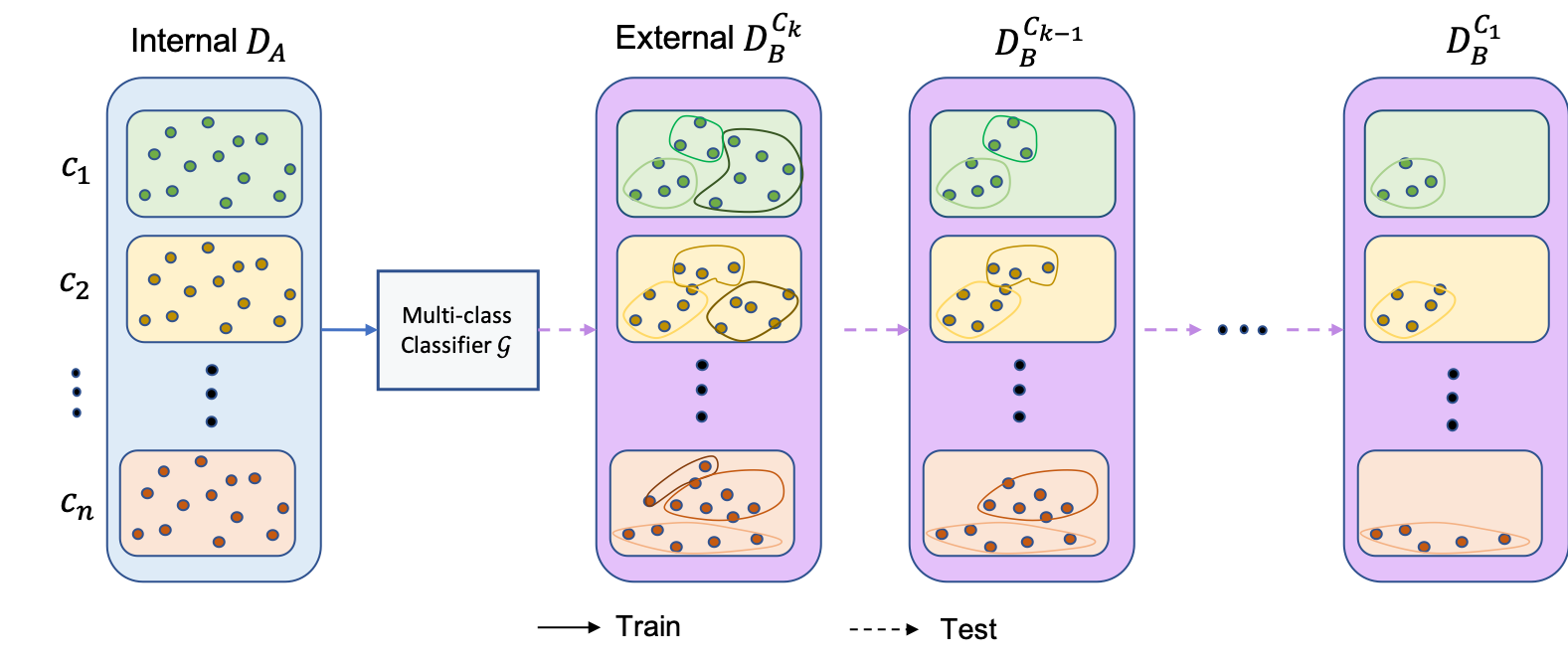}
\end{center}
  \caption{\textit{Shiftness} quantification pipeline} 
\label{mura_pipeline2}
\end{figure*}

The above pipeline can be applied to detect the shift data and assign each data with an anomaly score for indicating its contribution to the dataset shift. Nonetheless, the \textit{shiftness} of the identified data is not simple and straightforward to evaluate in relation with the targeted task. We suggest evaluating them in groups. As prepared in the first stage of the whole pipeline, the clustering has split each class of dataset $D_{B}$ into multiple groups according to the anomaly scores. For simplicity, we assume that each class has $k$ groups. To evaluate the significance of detected outliers, we train a multi-class classifier $\mathcal{G}$ for $D_{A}$ and test on $D_{B}$. As presented in Fig.~\ref{mura_pipeline2}, we gradually drop one group that has the largest anomaly scores among current groups for each class until only one group remains. The corresponding class-wise classification performance is recorded. The performance variation thus is an indicator of the \textit{shiftness} of the specific group.  

\begin{figure*}[tp]
\begin{center}
  \includegraphics[width=\linewidth]{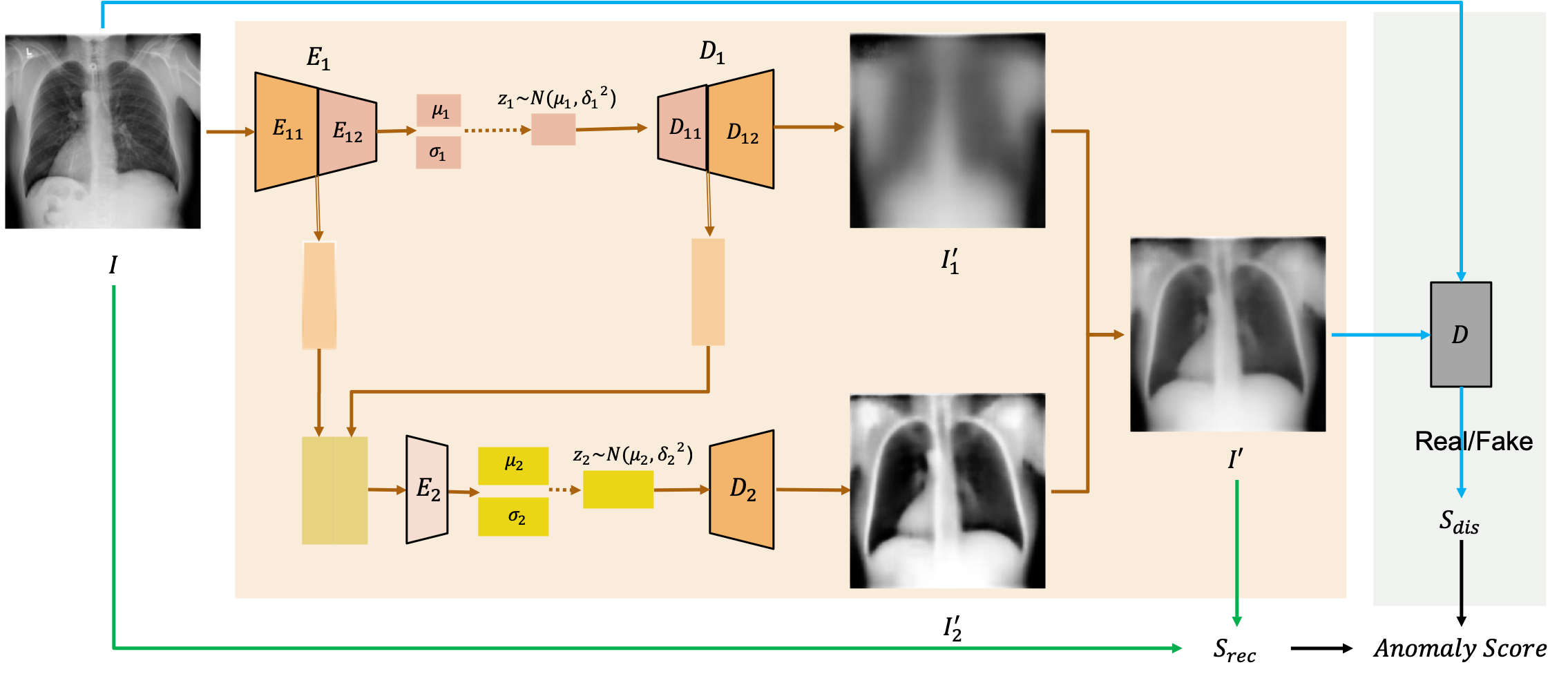}
\end{center}
  \caption{CVAD architecture}
\label{cvad}
\end{figure*}

\subsection{Anomaly Detection}~\label{anomaly_detection}
As claimed in Sec.~\ref{shift_identification}, we propose to utilize anomaly detection models to not only identify distribution shifts in the external dataset but also automated cleaning of the external data without any data sharing. First, we briefly describe our anomaly detection model -  \emph{Cascade Variational autoencoder-based Anomaly Detector (CVAD)}~\cite{guo2021cvad} used in \textit{MedShift}, which was previously been tested on both generic and medical image datasets. As shown in Fig.~\ref{cvad}, CVAD is a self-supervised variational autoencoder-based anomaly detection model which combines latent representation at multiple scales using the cascade architecture of variational autoencoders and thus, can reconstruct the in-distribution image $I$ with high quality. Both the original image $I$ and the reconstruction $I^{'}$ are then fed into a binary discriminator $D$ to separate the synthetic data from the in-distribution ones. The anomaly score includes two parts: the reconstruction error $S_{rec}$ in the first stage and the probability of being the anomaly class $S_{dis}$ in the second stage. To adapt the application of detecting abnormal data for multiple unknown external sources, we modified that anomaly score computation by simply adding the two parts together $S = S_{rec} + S_{dis}$. This gives us the advantage that when dealing with heavy noisy data, the reconstruction error will be the dominant indicator for \textit{shiftness}; when facing the hard distinguished cases the class probability plays the decision role.  

As this method poses no assumption on the input data and the applied situations, we utilize this anomaly detection architecture in our pipeline called \textit{MedShift\_w\_CVAD} across all the experiments. Apart from our anomaly detection model \textit{CVAD}, we also test other anomaly detector f-AnoGAN~\cite{schlegl2019f} in \textit{MedShift} for comparison (\textit{MedShift\_w\_f-AnoGAN} in short).

\subsection{Dataset Quality Measurement}~\label{dataset_quality}
To further quantify the efficacy of identifying the shift data among external datasets, we measure the quality of external datasets compared to the internal dataset and observe the difference after removing the shift data from the external sources in an iterative fashion. We apply the Optimal Transport Dataset Distance~\cite{alvarez2020geometric} (OTDD) measure to calculating similarities, or distances, between classification datasets. It relies on optimal transport\cite{villani2009optimal}, which is a flexible geometric method for comparing probability distributions, and can be used to compare any two datasets, regardless of whether their label sets are directly comparable. Formally, the optimal transport dataset distance is defined as: 

\begin{equation}
    OTDD(\boldsymbol{D}_{A}, \boldsymbol{D}_{B})= min_{\pi \in \prod (P_{A},P_{B}))}\int _{\boldsymbol{Z}\times\boldsymbol{Z}}d(z,z^{'})d\pi(z,z^{'})
\end{equation}
, of which 
\begin{equation}
    d(z, z^{'}) = {(d(x,x^{'})^{2}+W_2(P_y, P_{y^{'}})^{2} )}^{1\over2}
\end{equation}
, where $D_A$, $D_B$ are the two datasets, $x, x^{'}$ and $y, y^{'}$ are their samples and labels respectively,  $W_{p}$ denotes the p-Wassertein distance. Please refer Ref.~\cite{alvarez2020geometric} for more details. 

\section{Experiments}
\subsection{Datasets}
There are two categories of medical datasets used in this paper: (1) \emph{Musculoskeletal radiographs} - Emory MURA dataset (internal) and Stanford MURA dataset~\cite{rajpurkar2017mura} (external); (2) \emph{Chest radiographs} - Emory Chest X-rays (internal, Emory\_CXR in short), CheXpert dataset~\cite{irvin2019chexpert} (external\_1) and MIMIC dataset~\cite{johnson2019mimic} (external\_2).

\begin{figure*}[htp]
     \centering
     \begin{subfigure}[a]{\textwidth}
         \centering
         \includegraphics[width=0.9\linewidth]{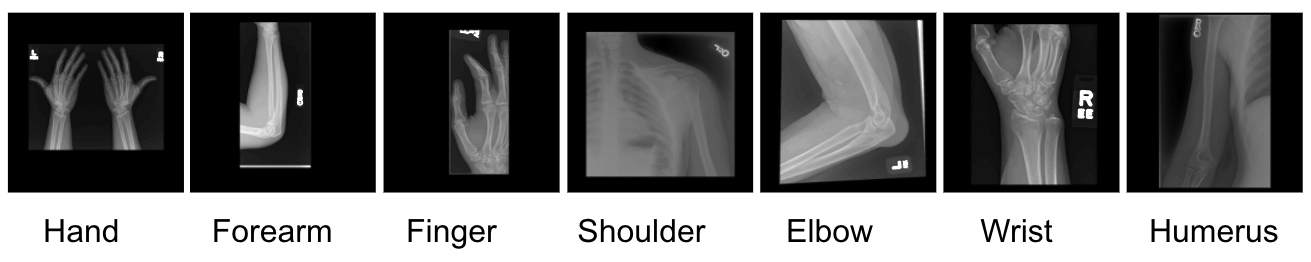}
         \caption{}
         \label{mura}
     \end{subfigure}
     \hfill
     \begin{subfigure}[b]{\textwidth}
         \centering
         \includegraphics[width=\linewidth]{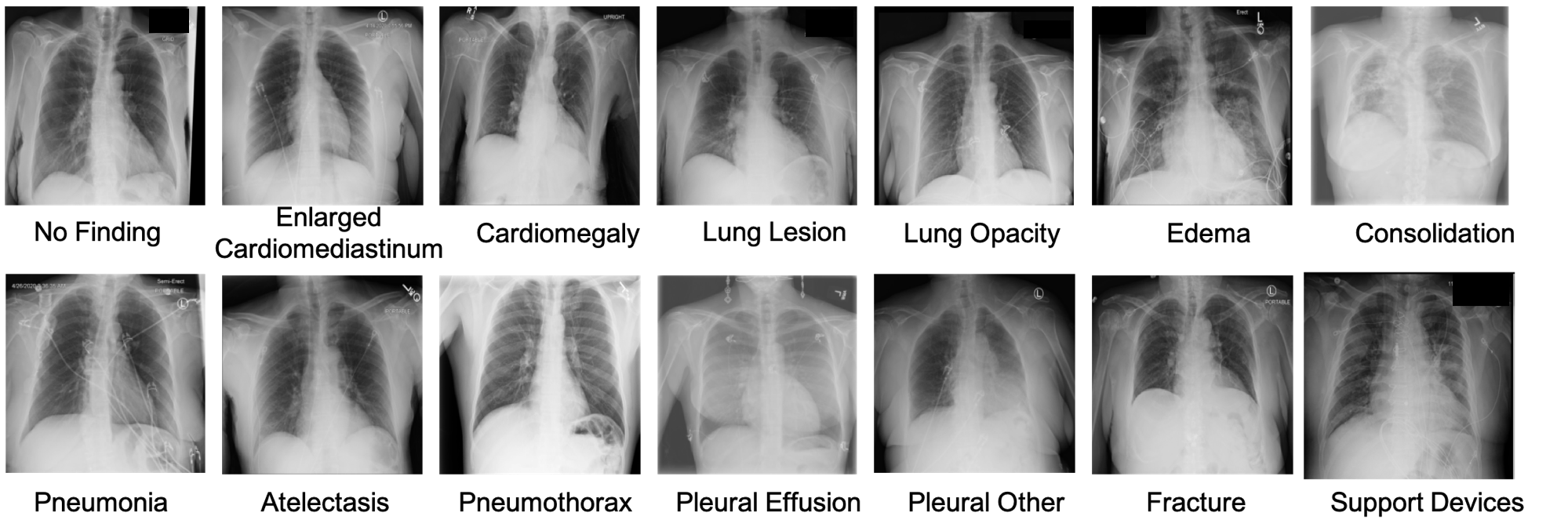}
         \caption{}
         \label{chestxray}
     \end{subfigure}
     \hfill
        \caption{Sample images from the datasets: (a) MURA examples for each anatomical joint class. (Intensity contrasts are changed for better visualization); (b) chest X-ray examples for each class. (Image are resized for better visualization);}
        \label{fig:three graphs}
\end{figure*}

MURA (musculoskeletal radiographs) is a large dataset of bone X-rays. Each MURA dataset has seven classes, \textit{XR\_HAND}, \textit{XR\_FORARM}, \textit{XR\_FIGER}, \textit{XR\_SHOULDER}, \textit{XR\_ELBOW}, \textit{XR\_WRIST}, \textit{XR\_HUMERUS}. Image examples are illustrated in Fig.~\ref{mura} for each class. To demonstrate the effectiveness of detecting shift data, we have Emory MURA and Stanford MURA datasets as a pair and treat Emory MURA as the internal dataset with Stanford MURA as the external one. More class-wise details of the datasets are presented in the upper of Table.~\ref{tab:dataset}.

\begin{table*}[htp]
\resizebox{\textwidth}{!}{%
\begin{tabular}{lcccccccccccccc}
\hline
 & \multicolumn{2}{c}{\textbf{HAND}} & \multicolumn{2}{c}{\textbf{FOREAMR}} & \multicolumn{2}{c}{\textbf{FINGER}} & \multicolumn{2}{c}{\textbf{SHOULDER}} & \multicolumn{2}{c}{\textbf{ELBOW}} & \multicolumn{2}{c}{\textbf{WRIST}} & \multicolumn{2}{c}{\textbf{HUMERUS}} \\
\textbf{Emory\_MURA} & \multicolumn{2}{c}{2,473 (21.33\%)} & \multicolumn{2}{c}{368 (3.17\%)} & \multicolumn{2}{c}{368 (3.17\%)} & \multicolumn{2}{c}{3,451 (29.77\%)} & \multicolumn{2}{c}{1,521 (13.12\%)} & \multicolumn{2}{c}{2,858(24.65\%)} & \multicolumn{2}{c}{553(4.77\%)} \\
\textbf{Stanford\_MURA} & \multicolumn{2}{c}{3,851 (17.94\%)} & \multicolumn{2}{c}{1,097 (5.11\%)} & \multicolumn{2}{c}{3,660 (17.05\%)} & \multicolumn{2}{c}{5,621 (26.18\%)} & \multicolumn{2}{c}{2,397 (11.16\%)} & \multicolumn{2}{c}{3,993(18.60\%)} & \multicolumn{2}{c}{852(3.97\%)} \\ \hline
 & \textbf{\begin{tabular}[c]{@{}c@{}}No \\ Finding\end{tabular}} & \textbf{\begin{tabular}[c]{@{}c@{}}Enlarged\\ Cardiome-\\ diastinum\end{tabular}} & \textbf{\begin{tabular}[c]{@{}c@{}}Cardio-\\ megaly\end{tabular}} & \textbf{\begin{tabular}[c]{@{}c@{}}Lung \\ Lesion\end{tabular}} & \textbf{\begin{tabular}[c]{@{}c@{}}Lung \\ Opacity\end{tabular}} & \textbf{Edema} & \textbf{\begin{tabular}[c]{@{}c@{}}Consoli-\\ dation\end{tabular}} & \textbf{\begin{tabular}[c]{@{}c@{}}Pneu-\\ monia\end{tabular}} & \textbf{\begin{tabular}[c]{@{}c@{}}Atele-\\ ctasis\end{tabular}} & \textbf{\begin{tabular}[c]{@{}c@{}}Pneumo-\\ thorax\end{tabular}} & \textbf{\begin{tabular}[c]{@{}c@{}}Pleural \\ Effusion\end{tabular}} & \textbf{\begin{tabular}[c]{@{}c@{}}Pleural \\ Other\end{tabular}} & \textbf{Fracture} & \textbf{\begin{tabular}[c]{@{}c@{}}Support \\ Devices\end{tabular}} \\
\textbf{Emory\_CXR (train)} & \begin{tabular}[c]{@{}c@{}}57,973\\  (11.35\%)\end{tabular} & \begin{tabular}[c]{@{}c@{}}7,825 \\ (1.53\%)\end{tabular} & \begin{tabular}[c]{@{}c@{}}27,019\\ (5.29\%)\end{tabular} & \begin{tabular}[c]{@{}c@{}}6,157 \\ (1.21\%)\end{tabular} & \begin{tabular}[c]{@{}c@{}}64,439 \\ (12.62\%)\end{tabular} & \begin{tabular}[c]{@{}c@{}}22,540 \\ (4.41\%)\end{tabular} & \begin{tabular}[c]{@{}c@{}}6,906\\  (1.35\%)\end{tabular} & \begin{tabular}[c]{@{}c@{}}9,188 \\ (1.80\%)\end{tabular} & \begin{tabular}[c]{@{}c@{}}66,150 \\ (12.95\%)\end{tabular} & \begin{tabular}[c]{@{}c@{}}11,550 \\ (2.26\%)\end{tabular} & \begin{tabular}[c]{@{}c@{}}51,828\\  (10.15\%)\end{tabular} & \begin{tabular}[c]{@{}c@{}}2,325 \\ (0.46\%)\end{tabular} & \begin{tabular}[c]{@{}c@{}}2,114 \\ (0.41\%)\end{tabular} & \begin{tabular}[c]{@{}c@{}}174,768 \\ (34.22\%)\end{tabular} \\
\textbf{Emory\_CXR (test)} & \begin{tabular}[c]{@{}c@{}}7,962\\  (30.44\%)\end{tabular} & \begin{tabular}[c]{@{}c@{}}523 \\ (2.00\%)\end{tabular} & \begin{tabular}[c]{@{}c@{}}1,256\\ (4.80\%)\end{tabular} & \begin{tabular}[c]{@{}c@{}}397 \\ (1.52\%)\end{tabular} & \begin{tabular}[c]{@{}c@{}}2,141 \\ (8.18\%)\end{tabular} & \begin{tabular}[c]{@{}c@{}}475 \\ (1.82\%)\end{tabular} & \begin{tabular}[c]{@{}c@{}}151\\  (0.58\%)\end{tabular} & \begin{tabular}[c]{@{}c@{}}439 \\ (1.68\%)\end{tabular} & \begin{tabular}[c]{@{}c@{}}1,684 \\ (6.44\%)\end{tabular} & \begin{tabular}[c]{@{}c@{}}150 \\ (0.57\%)\end{tabular} & \begin{tabular}[c]{@{}c@{}}711\\  (2.72\%)\end{tabular} & \begin{tabular}[c]{@{}c@{}}98 \\ (0.37\%)\end{tabular} & \begin{tabular}[c]{@{}c@{}}177 \\ (0.68\%)\end{tabular} & \begin{tabular}[c]{@{}c@{}}9,995 \\ (38.21\%)\end{tabular} \\
\textbf{CheXpert} & \begin{tabular}[c]{@{}c@{}}22,381\\  (4.34\%)\end{tabular} & \begin{tabular}[c]{@{}c@{}}10,798 \\ (2.09\%)\end{tabular} & \begin{tabular}[c]{@{}c@{}}27,000 \\ (5.24\%)\end{tabular} & \begin{tabular}[c]{@{}c@{}}9,186 \\ (1.78\%)\end{tabular} & \begin{tabular}[c]{@{}c@{}}105,581\\  (20.48\%)\end{tabular} & \begin{tabular}[c]{@{}c@{}}52,246\\  (10.13\%)\end{tabular} & \begin{tabular}[c]{@{}c@{}}14,783 \\ (2.87\%)\end{tabular} & \begin{tabular}[c]{@{}c@{}}6,039\\  (1.17\%)\end{tabular} & \begin{tabular}[c]{@{}c@{}}33,376 \\ (6.47\%)\end{tabular} & \begin{tabular}[c]{@{}c@{}}19,448 \\ (3.77\%)\end{tabular} & \begin{tabular}[c]{@{}c@{}}86,187\\ (16.72\%)\end{tabular} & \begin{tabular}[c]{@{}c@{}}3,523 \\ (0.68\%)\end{tabular} & \begin{tabular}[c]{@{}c@{}}9,040 \\ (1.75\%)\end{tabular} & \begin{tabular}[c]{@{}c@{}}116,001 \\ (22.50\%)\end{tabular} \\ 
\textbf{MIMIC} & \begin{tabular}[c]{@{}c@{}}143,352\\  (22.62\%)\end{tabular} & \begin{tabular}[c]{@{}c@{}}84,073 \\ (13.26\%)\end{tabular} & \begin{tabular}[c]{@{}c@{}}76,957 \\ (12.14\%)\end{tabular} & \begin{tabular}[c]{@{}c@{}}76,423 \\ (12.06\%)\end{tabular} & \begin{tabular}[c]{@{}c@{}}65,047\\  (10.26\%)\end{tabular} & \begin{tabular}[c]{@{}c@{}}64,346\\  (10.15\%)\end{tabular} & \begin{tabular}[c]{@{}c@{}}36,564 \\ (5.77\%)\end{tabular} & \begin{tabular}[c]{@{}c@{}}26,222\\  (4.14\%)\end{tabular} & \begin{tabular}[c]{@{}c@{}}14,675 \\ (2.32\%)\end{tabular} & \begin{tabular}[c]{@{}c@{}}14,257 \\ (2.25\%)\end{tabular} & \begin{tabular}[c]{@{}c@{}}10,801\\ (1.70\%)\end{tabular} & \begin{tabular}[c]{@{}c@{}}10,042 \\ (1.58\%)\end{tabular} & \begin{tabular}[c]{@{}c@{}}7,605 \\ (1.20\%)\end{tabular} & \begin{tabular}[c]{@{}c@{}}3,460 \\ (0.55\%)\end{tabular} \\ \hline
\end{tabular}%
}
\caption{Dataset details, with total image number and the percentage (in brackets) of each class presented. Upper part of the table present the MURA datasets and the lower is for Chest X-ray datasets.}
\label{tab:dataset}
\end{table*}

For chest X-ray, we used three datasets - Emory\_CXR (199,029 training and 12,873 test images retrieved from Emory Healthcare system), CheXpert and MIMIC datasets. The bottom part of Table.~\ref{tab:dataset} shows the details of the three datasets. The chest X-ray datasets have 14 classes (or diagnosis) in total. The classes are \textit{No Finding}, \textit{Enlarged Cardiomediastinum}, \textit{Cardiomegaly}, \textit{Lung Lesion}, \textit{Lung Opacity}, \textit{Edema}, \textit{Consolidation}, \textit{Pneumonia}, \textit{Atelectasis}, \textit{Pneumothorax}, \textit{Pleural Effusion}, \textit{Pleural Other}, \textit{Fracture}, \textit{Support Devices}. Image examples are displayed in Fig.~\ref{chestxray}. Different from the MURA dataset where class labels are mutually exclusive, each chest X-ray data may have multiple common diagnoses.  

\subsection{Implementation Details}
We implement the pipeline using Pytorch 1.5.0, Python 3.7.3 and Cuda compilation tools V10.0.130 on a machine with 4 NVIDIA RTX A6000 GPUs with 48 GB memory. More details about the training of anomaly detectors and classifiers are introduced below. 

\subsubsection{Anomaly Detectors}
We resize all the medical images to 256$\times$256$\times$\textit{channel} for simplicity considering the irregular image sizes. To train CVAD, we use the Adam optimizer with a batch size of 256 and 2,048 for MURA and chest X-ray dataset, respectively; we set the learning rate of $1\times{10^{-5}}$ and $1\times{10^{-3}}$ for the generator and the discriminator of CVAD, respectively; we train the generator with 250-500 epochs and the discriminator with 10-20 epochs. 

To train f-AnoGAN, we use the default Adam optimizer with a learning rate of $2\times{10^{-4}}$ and the same batch sizes as CVAD for the corresponding datasets; we run the generative adversarial training for 1000-1500 epochs and the encoder training for 300-500 epochs. 

\subsubsection{Multi-class Classifiers}~\label{multiclass_classifier}
To quantify the \textit{shiftness} of each clustered group for each class of external dataset $D_{B}$, we first train a multi-class classifier $\mathcal{G}$ for the internal dataset $D_{A}$. The classifier learns the class latent features of the internal domain and is able to predict class labels for test data. For MURA data, we train ResNet152~\cite{he2016deep} on the Emory MURA dataset with the publicly available pretrained weights as initialization. We optimize the classifier using the Adam optimizer with a batch size of 512, a learning rate of $1\times{10^{-3}}$ for 50 epochs. For chest X-ray data, we utilize the model proposed by Ref.~\cite{yuan2021large}, which originally aims for multi-label classification of the CheXpert dataset, and modifies it for the Emory\_CXR 14-class classification task. Following the same implementations in Ref.~\cite{yuan2021large}, we use DenseNet121~\cite{huang2017densely} as the feature extraction backbone and initialize it with the public pretrained model weights. We train the classifier with a batch size of 256 for 20 epochs. The corresponding classification performances, including the \textit{Precision}, \textit{Recall}, \textit{F1-score} and \textit{AUC} score are reported in Sec.~\ref{classification_results}.

\subsection{Results}
In this section, we evaluate the performance of our pipeline on three objectives - (i) shift data identification, (ii) shift data partition and (iii) shift data significance evaluation. To increase readability, one representative class is selected for explanation and more results about other classes are supplemented in ~\ref{mura_more} and ~\ref{chest_more}. 

\begin{figure*}[tp]
\centering
\begin{subfigure}{.5\textwidth}
  \centering
  \includegraphics[width=\linewidth]{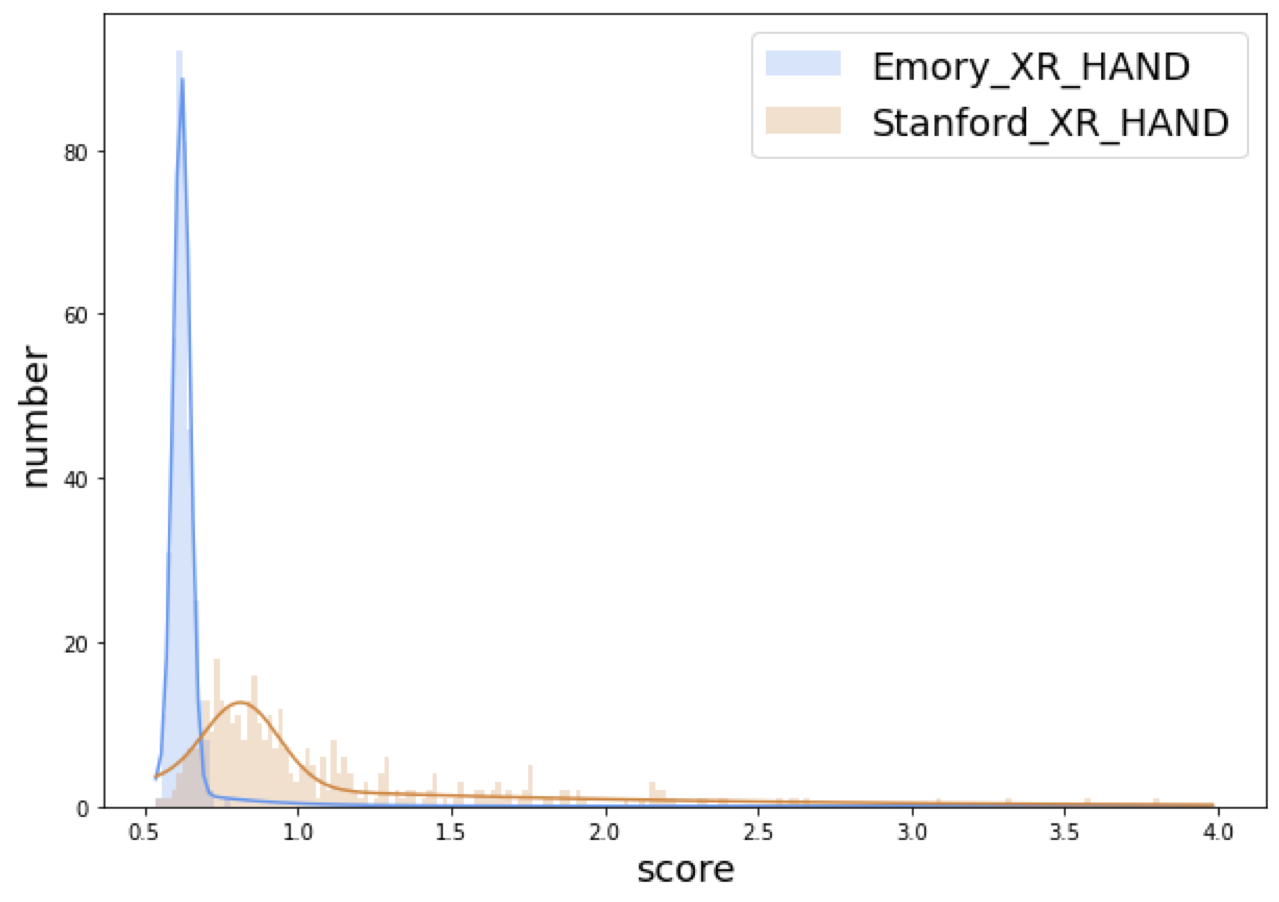}
  \label{fig:mura_hand}
\end{subfigure}%
\begin{subfigure}{.5\textwidth}
  \centering
  \includegraphics[width=\linewidth]{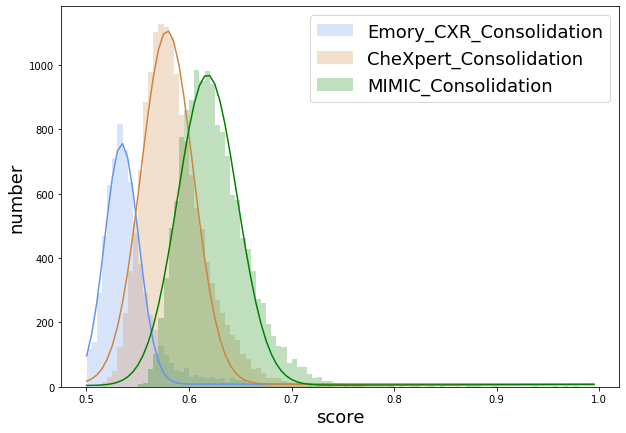}
  \label{fig:chest_consolidataion}
\end{subfigure}
\caption{\textit{MedShift\_w\_CVAD} example results of shift identification with anomaly detection - (left) anomaly score distributions on MURA \textit{HAND}; (right) anomaly score distributions for chest X-ray \textit{Consolidation}. Distributions are truncated on samples with large anomaly scores for better visualization.}
\label{fig:anomaly-detection-res}
\end{figure*}

\subsubsection{Shift Identification with Anomaly Detection}
In the process of identifying the shift data from the external source, each class of the internal dataset will obtain its own anomaly detector. Figure~\ref{fig:anomaly-detection-res} presents the anomaly score distributions of the representative class from both MURA and Chest X-ray obtained by \textit{MedShift\_w\_CVAD} architecture. The X-axis represents the anomaly score and Y-axis stands for the number of images that have anomaly scores in the corresponding range. In both cases, Emory data is considered as internal data.

For \textbf{MURA} dataset, the anomaly score distribution for \textit{XR\_HAND} is shown in the left of Fig.~\ref{fig:anomaly-detection-res}, with the blue curve for Emory \textit{XR\_HAND} and the orange distribution curve for Stanford \textit{XR\_HAND} data. As can be observed, the peaks of the two distributions are clearly separated, the Stanford data generally gets higher OOD scores than the internal Emory data. The difference between the internal and external anomaly score distributions can be easily observed. The closer and more similar the two distributions are, the less shift the external dataset has. 

The similar phenomenon can also be seen in chest X-ray data when being tested on two external datasets. For \textbf{chest X-ray} dataset, the OOD detection for \textit{Consolidation} is shown in the right of Fig.~\ref{fig:anomaly-detection-res}, with the blue histogram and curve for internal Emory\_CXR dataset, the orange for CheXpert dataset and the green for MIMIC dataset. The differences in the distributions reflect how different the external chest X-ray data is from the internal domain. MIMIC \textit{Consolidation} has less overlapping with the internal Emory\_CXR compared with the CheXpert distribution, which indicates that MIMIC contains more shift \textit{Consolidation} data than the CheXpert dataset.

\begin{figure*}[tp]
\centering
\includegraphics[width=\linewidth]{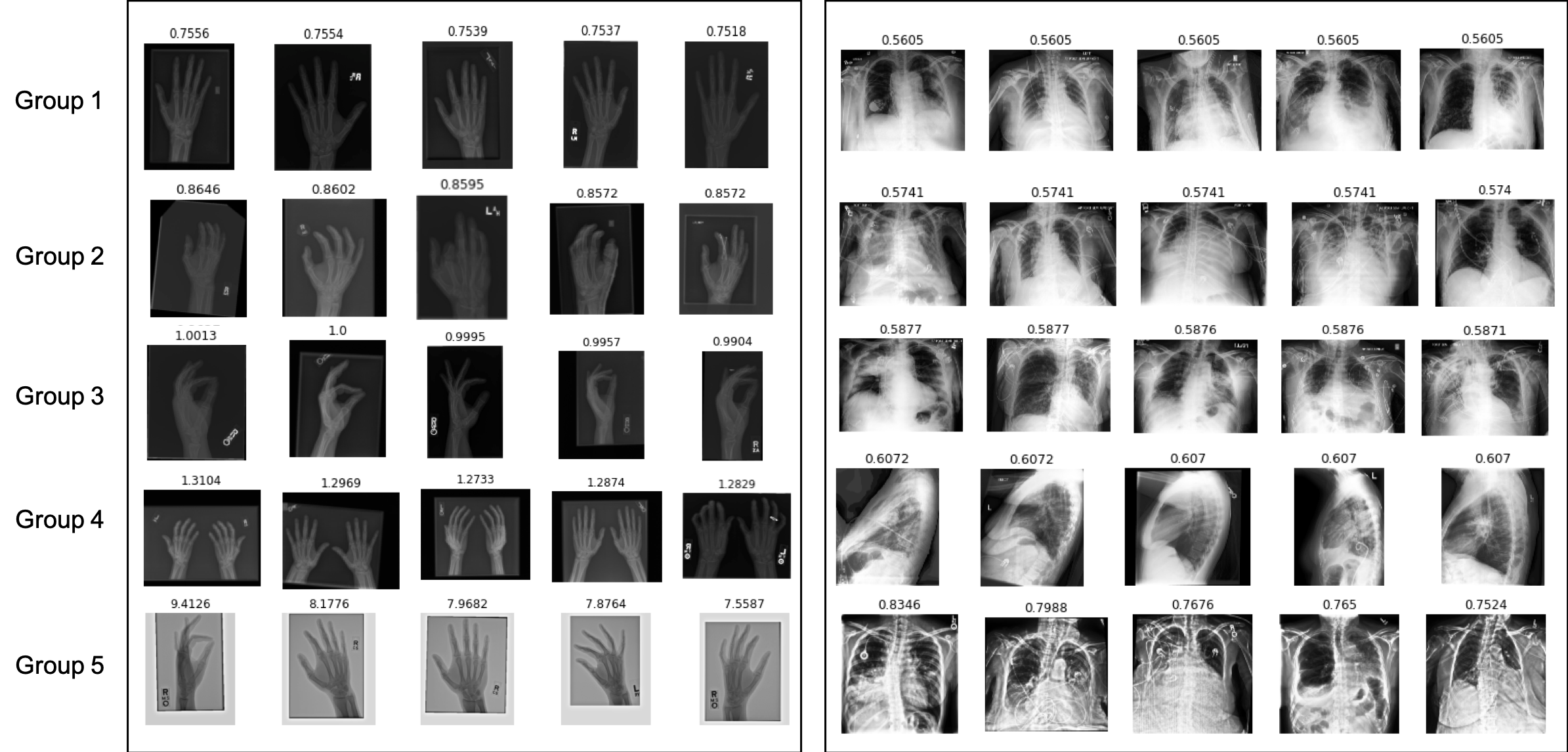}
\caption{\textit{MedShift\_w\_CVAD} examples of clustering results - (left) clustering results on Stanford\_MURA \textit{HAND} data; and (right) clustering results on CheXpert \textit{Consolidation} data. Each row represents one group with five example images. The groups are illustrated in ascending order based on the anomaly scores from top to bottom. The corresponding anomaly score is on top of each image.}
\label{fig:cluster_results_example}
\end{figure*}

\subsubsection{Shift Data Clustering Results}
In this section, we showcase the clustering results based on anomaly scores for both MURA and chest X-ray datasets. Specifically, Stanford MURA dataset, CheXpert and MIMIC data are clustered into different groups according to their anomaly scores obtained in the previous step. The selection of group numbers is decided by the Elbow distortion curves, please refer to \ref{mura_distortion_more} and \ref{mura_distortion_more_fanogan} for the curve details.

An \textit{XR\_HAND} example of \textbf{MURA} dataset is shown in the left of Fig.~\ref{fig:cluster_results_example}. There are 5 cluster groups in total, with each row representing one cluster. The groups are sorted in ascending order, namely, the top row is with the lowest anomaly scores and the bottom has the largest anomaly scores. For better understanding, their corresponding scores are labelled on top of each example item. As can be observed, the hand data gradually shows more and more variations in terms of image quality, positioning, and noise, as the anomaly score becomes large, especially when comparing the group 1 (first row with lowest anomaly score) to group 5 (last row with highest anomaly score). The variance exhibiting in the abnormal data indicates the existence of distribution shift in the external dataset. Nonetheless, the significance of the detected under-represented/shift data samples in affecting deep learning models' prediction/classification remains to be explored. Similarly, an example of \textbf{chest X-ray} \textit{Consolidation} is presented in the right of Fig.~\ref{fig:cluster_results_example}. Following the same arrange order, the difference for each group can be clearly captured.

\subsubsection{Classification Results for Shiftness Evaluation}~\label{classification_results}
As introduced in Sec.~\ref{multiclass_classifier}, a multi-class classifier has to be trained on the internal dataset to quantify the effect of removing the \textit{shiftness} of external datasets for the two targeted classification tasks. In this section, we report the classification training and testing performance on the internal dataset, and the performance on the external datasets after dropping the highest anomaly score group gradually. The external group-wise \textit{shiftness} is thus revealed by the performance variation. An evident decrease suggests a significant distribution shift in the dropped group. For comparison, we report the classification outcomes on external dataset based on the clustering results obtained with both anomaly scores computed with CVAD~\cite{guo2021cvad} and f-AnoGAN~\cite{schlegl2019f} architectures.

\begin{table*}[tp]
\caption{MURA classification class-wise results with CVAD (left) and f-AnoGAN (right). The best classification values are in bold for each method.}
\resizebox{\textwidth}{!}{%
\begin{tabular}{l|lccccccc|cc}
\hline
\multirow{2}{*}{\textbf{Dataset}} & \multirow{2}{*}{\textbf{Metric}} & \multirow{2}{*}{\textbf{HAND}} & \multirow{2}{*}{\textbf{FOREARM}} & \multirow{2}{*}{\textbf{SHOULDER}} & \multirow{2}{*}{\textbf{FINGER}} & \multirow{2}{*}{\textbf{ELBOW}} & \multirow{2}{*}{\textbf{WRIST}} & \multirow{2}{*}{\textbf{HUMERUS}} & \multicolumn{2}{c}{\textbf{Average}} \\ \cline{10-11} 
 &  &  &  &  &  &  &  &  & \textbf{Macro} & \textbf{Weighted} \\ \hline


\multirow{5}{*}{\textbf{Emory\_test}} & \textit{\#images} & 495 & 74 & 691 & 74 & 305 & 572 & 111 & \multicolumn{2}{c}{2,322} \\
& \textit{Precision} & 0.842 & 0.704 & 0.979 & 0.312 & 0.929 & 0.957 & 0.875 & 0.800 & 0.903 \\
& \textit{Recall} & 0.970 & 0.770 & 0.999 & 0.068 & 0.862 & 0.942 & 0.820 & 0.776 & 0.915 \\
& \textit{F1-score} & 0.901 & 0.735 & 0.989 & 0.111 & 0.895 & 0.950 & 0.847 & 0.775 & 0.905 \\ 
& \textit{AUC} & 0.960 & 0.880 & 0.995 & 0.531 & 0.926 & 0.964 & 0.907 & 0.984 & 0.992 \\\hline

\multicolumn{1}{l|}{\multirow{4}{*}{\textbf{Stanford\_TOP 5}}} & \textit{\#images} & 3,851 & 1,097 & 5,621 & 3,660 & 2,397 & 3,993 & 852 & \multicolumn{2}{c}{21,471} \\
& \textit{Precision} & \textbf{0.921} & 0.758 & 0.977 & 0.765 & 0.695 & 0.380 & 0.395 & 0.699 & 0.754 \\
& \textit{Recall} & 0.450 & 0.160 & 0.746 & 0.188 & 0.701 & 0.983 & 0.664 & 0.556 & 0.604 \\
& \textit{F1-score} & 0.605 & 0.264 & 0.846 & 0.301 & 0.698 & 0.548 & 0.496 & 0.537 & 0.594 \\ 
& \textit{AUC} & 0.721 & 0.578 & 0.870 & 0.588 & 0.831 & 0.808 & 0.811 & 0.902 & 0.915 \\ \hline

\multicolumn{1}{l|}{\multirow{4}{*}{\textbf{Stanford\_TOP 4}}} & \textit{\#images} & 3,098 / 3,838 & 880 / 1,091 & 4,499 / 5,584 & 2,904 / 3,658 & 1,923 / 2,387 & 3,182 / 3,933 & 686 / 848 & \multicolumn{2}{c}{17,172 / 21,339} \\
& \textit{Precision} & 0.921 / 0.921 & 0.758 / 0.758 & 0.986 / 0.978 & 0.768 / \textbf{0.765} & \textbf{0.695} / 0.695 & 0.426 / \textbf{0.379} & 0.545 / 0.404 & 0.728 / 0.700 & 0.772 / 0.755 \\
& \textit{Recall} & 0.558 / 0.452 & 0.195 / 0.160 & \textbf{0.827} / 0.750 & 0.233 / 0.188 & 0.777 / 0.704 & \textbf{0.990} / 0.987 & 0.691 / 0.665 & 0.610 / 0.558 & 0.665 / \textbf{0.605} \\
& \textit{F1-score} & 0.695 / 0.606 & 0.311 / 0.265 & \textbf{0.899} / 0.849 & 0.358 / 0.302 & 0.734 / 0.700 & 0.596 / \textbf{0.548} & 0.609 / 0.503 & 0.600 / 0.539 & 0.654 / \textbf{0.596} \\ 
& \textit{AUC} & 0.774 / 0.722 & 0.596 / 0.579 & \textbf{0.911} / 0.872 & 0.609 / 0.588 & 0.867 / 0.832 & 0.843 / 0.811 & 0.833 / 0.812 & 0.938 / 0.903 & 0.949 / \textbf{0.916} \\\hline

\multicolumn{1}{l|}{\multirow{4}{*}{\textbf{Stanford\_TOP 3}}} & \textit{\#images} & 2,331 / 3,814 & 661 / 1,079 & 3,368 / 5,419 & 2,159 / 3,648 & 1,443 / 2,367 & 2,380 / 3,066 & 517 / 808 & \multicolumn{2}{c}{12,859 / 20,201} \\
& \textit{Precision} & 0.913 / 0.920 & 0.759 / 0.758 & 0.986 / \textbf{0.981} & 0.789 / 0.765 & 0.690 / 0.698 & 0.471 / 0.329 & 0.589 / 0.414 & 0.743 / 0.695 & 0.784 / 0.764 \\
& \textit{Recall} & 0.661 / 0.455 & 0.253 / 0.162 & 0.821 / 0.769 & 0.279 / 0.188 & 0.831 / 0.710 & 0.988 / 0.991 & 0.747 / 0.666 & 0.654 / 0.563 & 0.701 / 0.595 \\
& \textit{F1-score} & 0.767 / 0.609 & 0.379 / 0.267 & 0.896 / 0.862 & 0.413 / 0.302 & \textbf{0.754} / 0.704 & 0.638 / 0.494 & 0.659 / 0.511 & 0.644 / 0.535 & 0.692 / 0.593 \\ 
& \textit{AUC} & 0.823 / 0.723 & 0.624 / 0.580 & 0.908 / 0.882 & 0.632 / 0.588 & 0.891 / 0.834 & 0.867 / \textbf{0.814} & 0.862 / 0.813 & 0.953 / 0.905 & 0.961 / 0.916 \\\hline

\multicolumn{1}{l|}{\multirow{4}{*}{\textbf{Stanford\_TOP 2}}} & \textit{\#images} & 1,553 / 3,761 & 440 / 1,068 & 2,234 / 3,839 & 1,429 / 3,483 & 959 / 2,335 & 1,587 / 2,048 & 345 / 717 & \multicolumn{2}{c}{8,547 / 17,251} \\
& \textit{Precision} & 0.894 / 0.921 & 0.763 / 0.771 & 0.984 / 0.979 & 0.801 / 0.765 & 0.666 / 0.724 & 0.520 / 0.262 & 0.592 / 0.483 & \textbf{0.746} / \textbf{0.701} & \textbf{0.788} / 0.770 \\
& \textit{Recall} & 0.748 / 0.461 & 0.359 / 0.164 & 0.795 / \textbf{0.818} & 0.324 / 0.195 & 0.842 / 0.719 & 0.986 / 0.991 & \textbf{0.754} / 0.658 & 0.687 / 0.572 & 0.724 / 0.575 \\
& \textit{F1-score} & 0.815 / 0.614 & 0.488 / 0.270 & 0.879 / \textbf{0.891} & 0.461 / 0.311 & 0.744 / 0.722 & 0.681 / 0.414 & \textbf{0.663} / 0.557 & 0.676 / \textbf{0.540} & 0.717 / 0.582 \\
& \textit{AUC} & 0.864 / 0.725 & 0.677 / 0.580 & 0.895 / \textbf{0.907} & 0.654 / 0.590 & 0.894 / 0.838 & 0.889 / 0.808 & \textbf{0.866} / 0.814 & 0.963 / 0.904 & 0.968 / 0.908 \\\hline

\multirow{4}{*}{\textbf{Stanford\_TOP 1}} & \textit{\#images} & 773 / 3,697 & 219 / 1,042 & 1,110 / 1,921 & 711 / 2,417 & 477 / 2,236 & 795 / 1,023 & 172 / 463 & \multicolumn{2}{c}{4,257 / 12,799} \\
& \textit{Precision} & 0.855 / \textbf{0.925} & \textbf{0.779} / \textbf{0.799} & \textbf{0.989} / 0.966 & \textbf{0.816} / 0.751 & 0.612 / \textbf{0.767} & \textbf{0.575} / 0.184 & 0.559 / 0.467 & 0.741 / 0.694 & 0.788 / \textbf{0.784}  \\
& \textit{Recall} & \textbf{0.814} / \textbf{0.468} & 0.434 / 0.168 & 0.730 / 0.791 & \textbf{0.368} / \textbf{0.257} & \textbf{0.881} / \textbf{0.743} & 0.974 / \textbf{0.997} & 0.738 / 0.590 & \textbf{0.705} / \textbf{0.573} & \textbf{0.732} / 0.547 \\
& \textit{F1-score} & \textbf{0.834} / \textbf{0.622} & \textbf{0.557} / \textbf{0.278} & 0.840 / 0.869 & \textbf{0.508} / \textbf{0.382} & 0.722 / \textbf{0.755} & \textbf{0.723} / 0.310 & 0.637 / 0.521 & \textbf{0.689} / 0.534 & \textbf{0.726} / 0.580 \\ 
& \textit{AUC} & \textbf{0.892} / \textbf{0.726} & \textbf{0.714} / \textbf{0.582} & 0.863 / 0.893 & \textbf{0.676} / \textbf{0.618} & \textbf{0.905} / \textbf{0.848} & \textbf{0.904} / 0.806 & 0.857 / 0.782 & \textbf{0.969} / \textbf{0.909} & \textbf{0.971} / 0.907 \\\hline
\end{tabular}%
}
\label{MURA-cls-tab-all-in-one}
\end{table*}

Table.~\ref{MURA-cls-tab-all-in-one} shows the classification results for the \textbf{MURA} data, including the test results of Emory MURA and evaluation on Stanford MURA groups. Both the class-wise and average performances are reported, including \textit{Precision}, \textit{Recall}, \textit{F1-score} and \textit{AUC} scores. As the classification is evaluated in the order of TOP\_k, TOP\_k-2, ..., TOP\_1 order, which is TOP\_5, TOP\_4, TOP\_3, TOP\_2, TOP\_1 for our experiments, meaning that we gradually drop the group that with the highest anomaly scores and evaluate the classification performance on the remaining data. There are five groups being clustered for each class. Therefore, the TOP 5 clusters constitute the whole external dataset and the corresponding classification results for CVAD version and f-AnoGAN version are the same. For simplicity, only one version is present (see Table.~\ref{MURA-cls-tab-all-in-one} \textit{Row} \textbf{Stanford\_ MURA\_TOP 5}). The total number of images being evaluated on is listed in the row \#\textit{images} for each class. The amount of data samples in the dropped group is the number difference between the adjacent groups. Take \textit{XR\_HAND} for example, group 5 of \textit{MedShift\_w\_CVAD} has 753 samples by calculating the difference of total image number of TOP 5 clusters (3851) and TOP 4 clusters (3098), (i.e., $753 = 3851-3098$) and group 5 of \textit{MedShift\_w\_f-AnoGAN} has 13 samples ($13 = 3851-3838$). As can be observed in the table, the classifier's predictions become more and more accurate as the groups are discarded gradually based on their anomaly score order. Look into the \textit{AUC} scores of \textit{XR\_HAND} from TOP 5 to TOP 1, the values of both CVAD and f-AnoGAN are growing, which means the removed group contains data with certain \textit{shiftness} and will affect the in-domain model's ability. The extent of \textit{shiftness} can be inferred via the change of classification measurements for a notable improvement indicates a severe shifting exists in the dropped group. Although the same trend is noted for both CVAD and f-AnoGAN versions in general, the CVAD version can get more increase after expelling the most anomalous group than the f-AnoGAN version, which demonstrates the effectiveness of our \textit{MedShift} framework in determining shift data among external datasets.  

\begin{figure*}[t]
    \centering 
\begin{subfigure}{0.33\textwidth}
  \includegraphics[width=\linewidth]{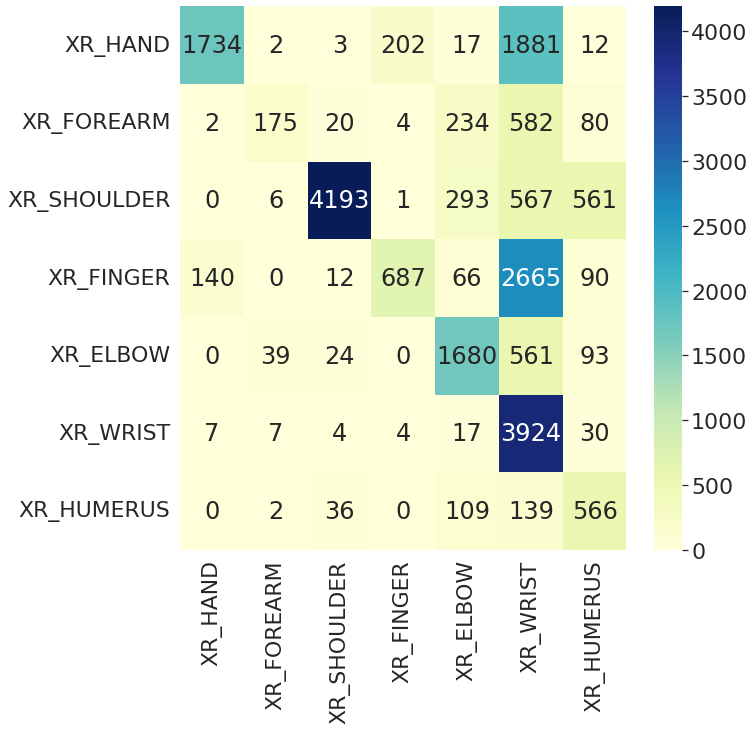}
  \caption{}
  \label{fig:mura-matrix5}
\end{subfigure}\hfil 
\begin{subfigure}{0.33\textwidth}
  \includegraphics[width=\linewidth]{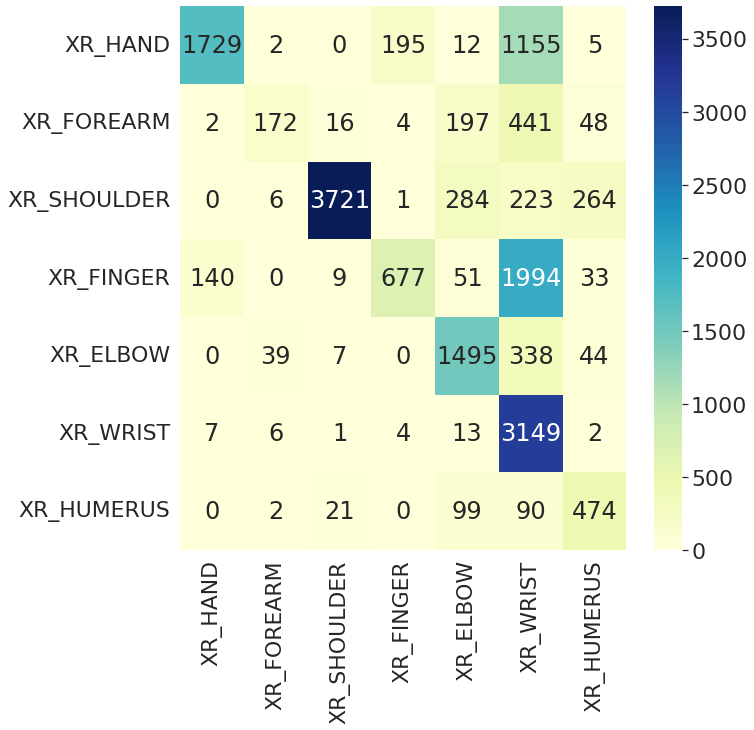}
  \caption{}
  \label{fig:mura-matrix4}
\end{subfigure}\hfil 
\begin{subfigure}{0.33\textwidth}
  \includegraphics[width=\linewidth]{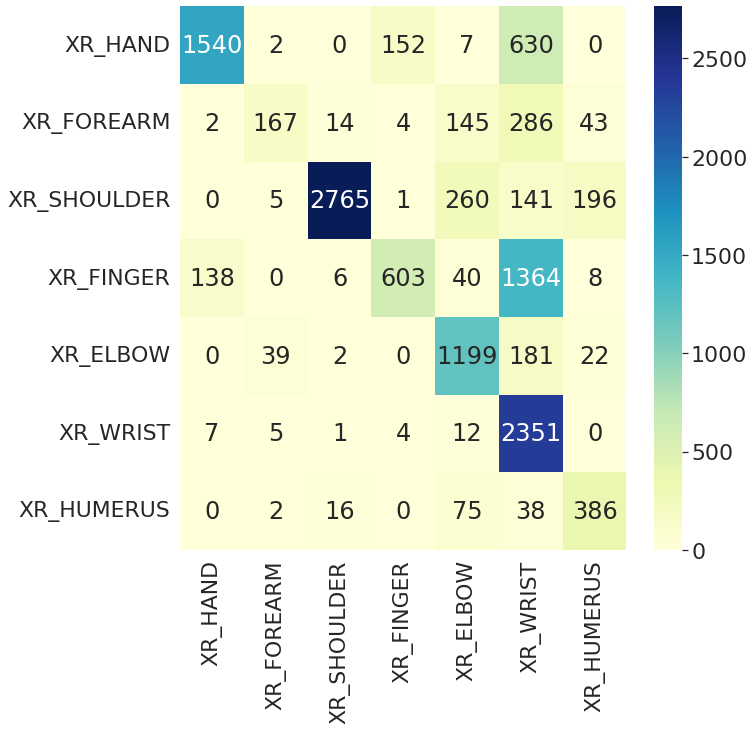}
  \caption{}
  \label{fig:mura-matrix3}
\end{subfigure}

\medskip
\begin{subfigure}{0.33\textwidth}
  \includegraphics[width=\linewidth]{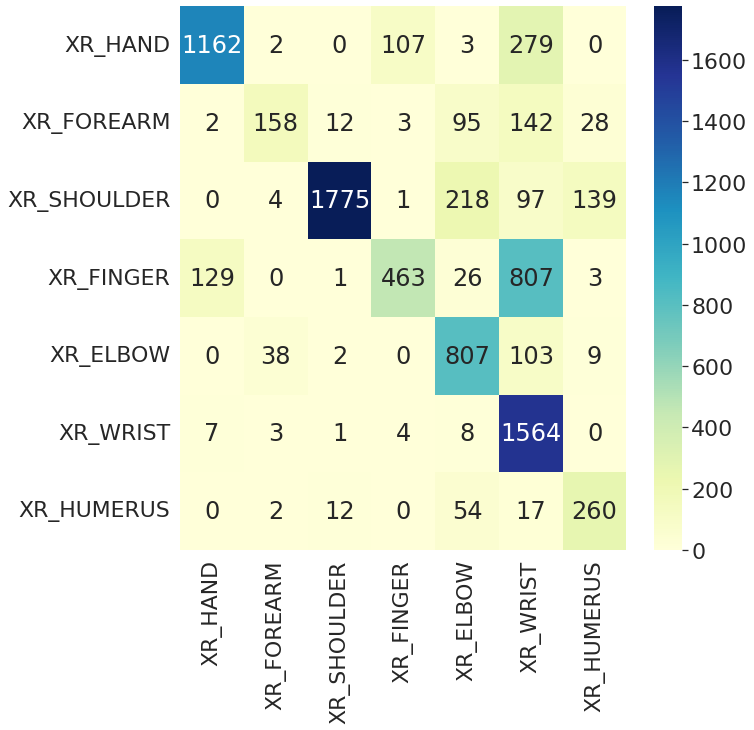}
  \caption{}
  \label{fig:mura-matrix2}
\end{subfigure}\hfil 
\begin{subfigure}{0.33\textwidth}
  \includegraphics[width=\linewidth]{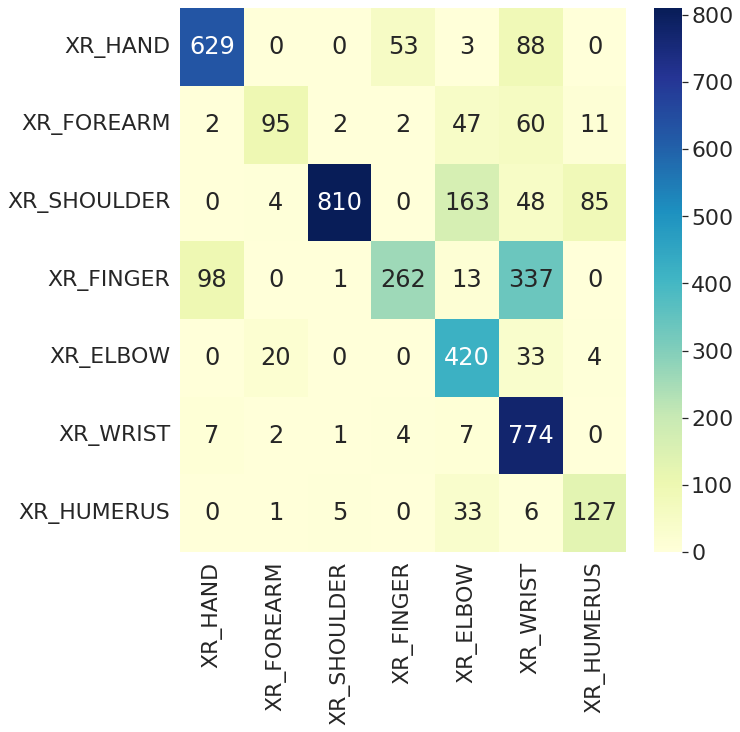}
  \caption{}
  \label{fig:mura-matrix1}
\end{subfigure}\hfil 
\begin{subfigure}{0.33\textwidth}
  \includegraphics[width=\linewidth]{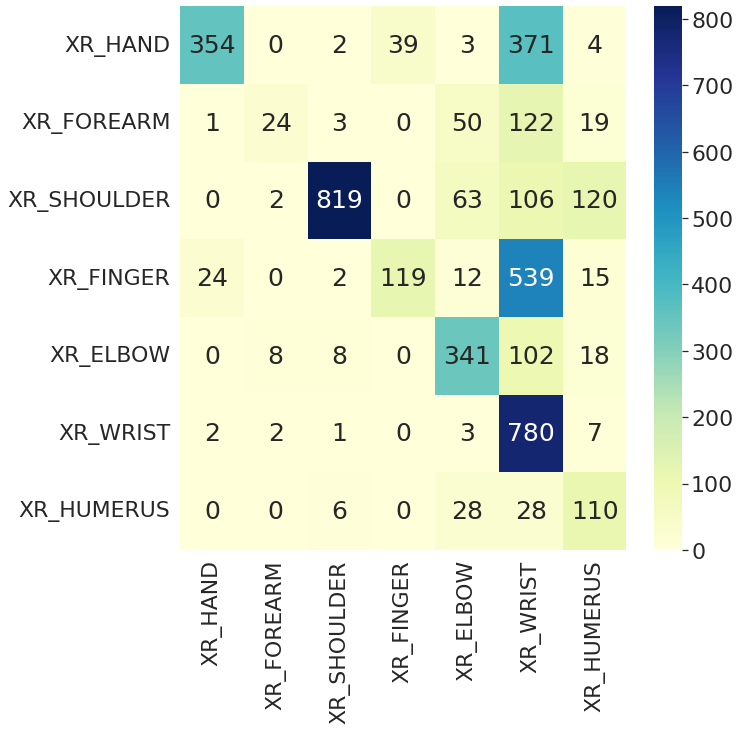}
  \caption{}
  \label{fig:mura-matrix0}
\end{subfigure}
\caption{\textit{MedShift\_w\_CVAD} confusion matrices for Stanford MURA with different numbers of clusters - (a) confusion matrix of TOP 5 clusters; (b) confusion matrix of TOP 4 clusters; (c) confusion matrix of TOP 3 clusters; (d) confusion matrix of TOP 2 clusters; (e) confusion matrix of TOP 1 cluster; (f) confusion matrix with random sampling the same number of images as the TOP 1 cluster.}
\label{fig:mura-matrix}
\end{figure*}

Aside from the quantification results above, we show the confusion matrices in Fig.~\ref{fig:mura-matrix} to report more details about the classification performance. Each confusion matrix represents a particular situation of using different cluster data during evaluation. To be clear, we use the clustering results based on the pipeline with our CVAD architecture. From left to right, top to bottom of Fig.~\ref{fig:mura-matrix}, the confusion matrices are for TOP 5, TOP 4, TOP 3, TOP 2, TOP 1 clusters, respectively. Since the groups that are dropped contain shift data, the classification accuracy is gradually rising after removing the shift data items, which can be observed from the confusion matrix differences.

Furthermore, to demonstrate our pipeline's capability of separating shift data that are with high anomaly scores, we random sample the same number of images as the \textit{MedShift\_w\_CVAD}'s TOP 1 cluster of Stanford MURA dataset for each class and run the classifier to evaluate the classification accuracy. The random sampling confusion matrix is displayed in Fig~\ref{fig:mura-matrix0}. Compared to the confusion matrix of TOP 1 in Fig.~\ref{fig:mura-matrix1}, there are more misclassified cases, especially for \textit{XR\_HAND} class, where there are 629 samples correctly predicted and 144 wrong predictions after applying our pipeline whereas 354 correct predictions and 419 misclassified cases for the random sampling situation. The improvement of classification accuracy manifests that our \textit{MedShift} can identify the shift data that will degrade the performance of an in-domain model. The same observation holds for the \textbf{chest X-ray} experiments, please refer to Appendix ~\ref{chest_classification} for the classification details.

\begin{figure*}[tp]
\centering
\begin{subfigure}{.5\textwidth}
  \centering
  \includegraphics[width=0.9\linewidth]{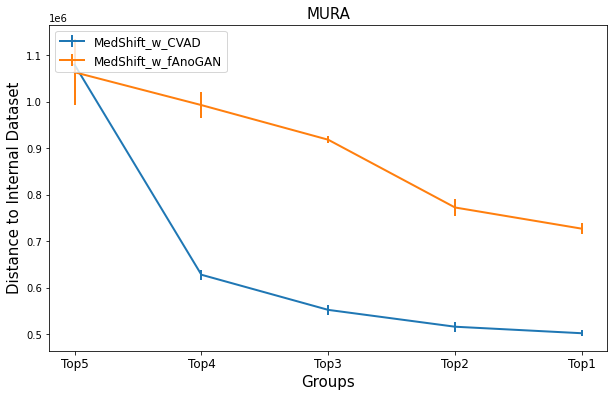}
\end{subfigure}%
\begin{subfigure}{.5\textwidth}
  \centering
  \includegraphics[width=0.9\linewidth]{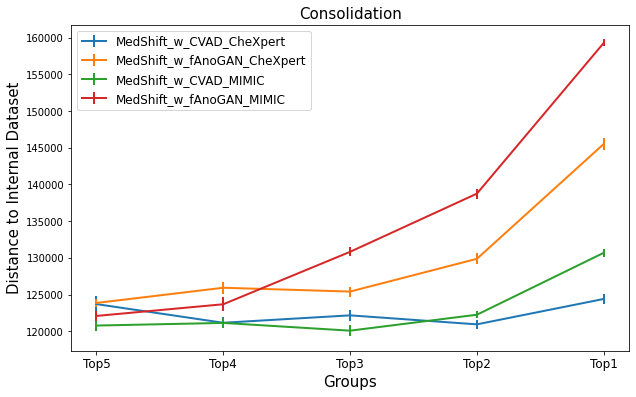}
\end{subfigure}
\caption{Dataset quality measurement results - (left) Stanford MURA whole dataset's quality; (right) CheXpert and MIMIC Consolidation class quality. X-axis values represent situations of the groups in use, and Y-axis values indicate the distance between the internal and external datasets (the lower the better). Distance mean and stdev values of ten rounds of evaluations are present in the plots. }
\label{fig:qual_results}
\end{figure*}

\subsubsection{Dataset Quality Measurement Results}
We report the Stanford \textbf{MURA} dataset quality calculated via the OTDD metric proposed above in the left of Fig.~\ref{fig:qual_results}. We respectively evaluate the quality for TOP\_5, TOP\_4, TOP\_3,  TOP\_2, TOP\_1 cases as indicated by the X-axis values of the plots. To compare, we test our pipeline with both CVAD and f-AnoGAN anomaly detection architectures. As can be seen, the distance between Stanford MURA and Emory MURA datasets is decreasing when the anomalous groups with shift data are removed gradually. Nevertheless, our CVAD version (in blue) shortens the distance more and faster than the f-AnoGAN (in orange) version. And the general external dataset quality achieves the best when it is composed by the group with the lowest anomaly scores, which follows the same conclusion as the average classification performance in Tab.~\ref{MURA-cls-tab-all-in-one}.    

For the reason that the OTDD method computes the distance values with label-data pairs, it was not designed for multi-label datasets. To adapt for the \textbf{chest X-ray} scenario, we report the class quality instead of the whole dataset. In the right of Fig.~\ref{fig:qual_results}, we show CheXpert and MIMIC Consolidation class quality obtained by both the CVAD and f-AnoGAN versions. Generally, the distances between the internal and external are shortened in a limited way with \textit{MedShift\_w\_CVAD} model, but the distance values are enlarged by the f-AnoGAN version. Since the distance represents the dissimilarity between the evaluated dataset pair, an increase of distance indicates a failure of identifying shift data in the external domain. Here, the CVAD version shows better performance than the \textit{MedShift\_w\_f-AnoGAN} model. 

Moreover, an increase of distance is also an indicator of stop sign for detecting shift data of a well-performed shift identification model. From the anomaly score distribution plots of Fig.~\ref{fig:qual_results}, it is clear that external MURA \textit{HAND} has more variance than the external chest X-ray \textit{Consolidation} data. Thus, shift data identification is relatively difficult for the chest X-ray dataset, and the quality improvement is limited when little \textit{shiftness} exists in the external dataset. Depending on the quality expectations, users can decide to remain the original \textit{Consolidation} class or remove one or two top groups from \textit{Consolidation}. Due to the space limitation, only one chest X-ray class case is illustrated, please refer to \ref{chest_qual_more} for more class quality results.

\section{Discussion}
In this paper, we have designed an automated pipeline - \emph{MedShift}, for medical dataset curation based on anomaly score. Under-the-hood, \emph{MedShift} identifies image data distribution shift based on anomaly detection and unsupervised clustering to discriminate the poor-quality, noisy and under-represented samples from the external data. The anomaly detection architecture involves two separate implementation phases - (1) internal training - time consuming and needs to be trained for each targeted class label, and (2) test phase - quick, only forward pass which needs minimal data pre-processing and cleaning from the external sites. Once trained, the anomaly detectors should be able to identify unknown anomalous patterns from an external dataset without ever seeing such anomalous data examples in training. This quality makes the proposed pipeline particularly suitable for medical image dataset curation since exchanging healthcare data among institutions and manually identifying noisy or anomalous data are both extremely challenging in the current healthcare situation.  

Our pipeline is flexible towards the particular anomaly detector architectures. We evaluated two use-cases - diagnosis from chest X-ray and classifying anatomical joints from MURA and applied two different anomaly detectors CVAD and f-AnoGAN. Even though our CVAD version efficiently shortens the data quality matrix (OTDD) faster than f-AnoGAN and reaches convergence for the shift data removal by dropping lower number of cases from external data, the targeted final classification performance stays similar for both architectures. 

Our experiments showed that being trained only on internal Emory datasets, deep learning models  classification accuracy is gradually rising on the external dataset after removing the shift data items via \emph{MedShift} and ultimately achieved performance close to the internal data. The improvement of classification accuracy represents the fact that the \textit{MedShift} can identify relevant shift data that will degrade the performance of an in-domain model and be able to reproduce the internal performance on unseen external data. Moreover, the brief cluster exploration on the external dataset showed that higher anomaly cluster groups contain more variations in terms of image quality, positioning, noise, and the pipeline correctly identified the shift data. As an immediate future study, we plan to conduct a reader study with expert radiologists to interactively evaluate the proposed platform and quantify the performance based on user-feedback matrices. 

In its current state, the proposed pipeline \textit{MedShift} can serve domain-specific quality checks and derive powerful and actionable insights. The suggested workflow will be beneficial in future non-shareable healthcare collaboration where the \textit{MedShift} pipeline will be set up as a browser-based service within the local firewall for automated dataset curation with  multi-class labels.

\clearpage

\appendix
\onecolumn
\section{MURA Results}~\label{mura_more}
\subsection{Anomaly Score Distribution Results with CVAD}~\label{mura_anomaly_score_more}
\begin{figure*}[htp]
\centering
\begin{subfigure}{.49\textwidth}
\includegraphics[width=\linewidth, height=0.6\linewidth]{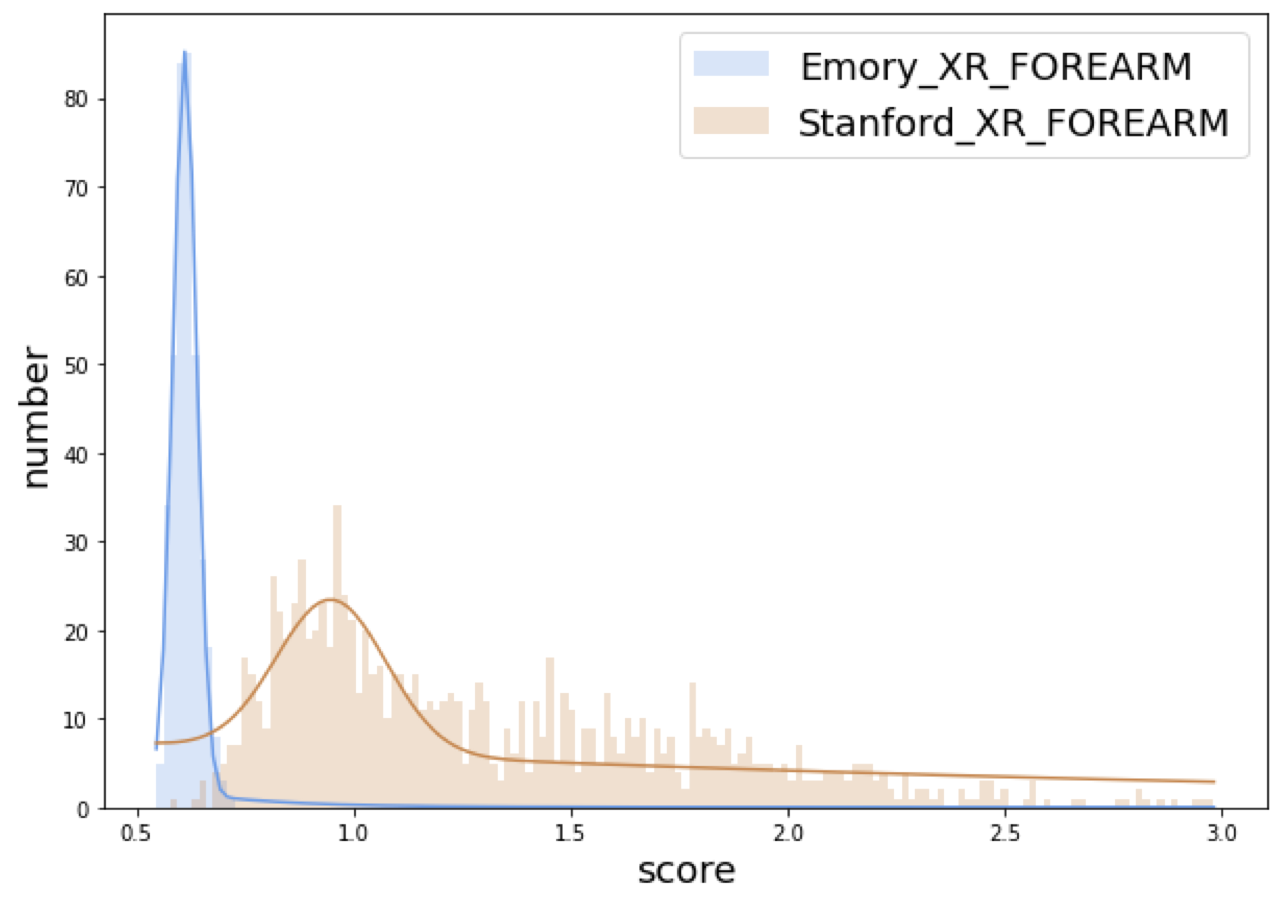}
\end{subfigure}
\begin{subfigure}{.49\textwidth}
\includegraphics[width=\linewidth, height=0.6\linewidth]{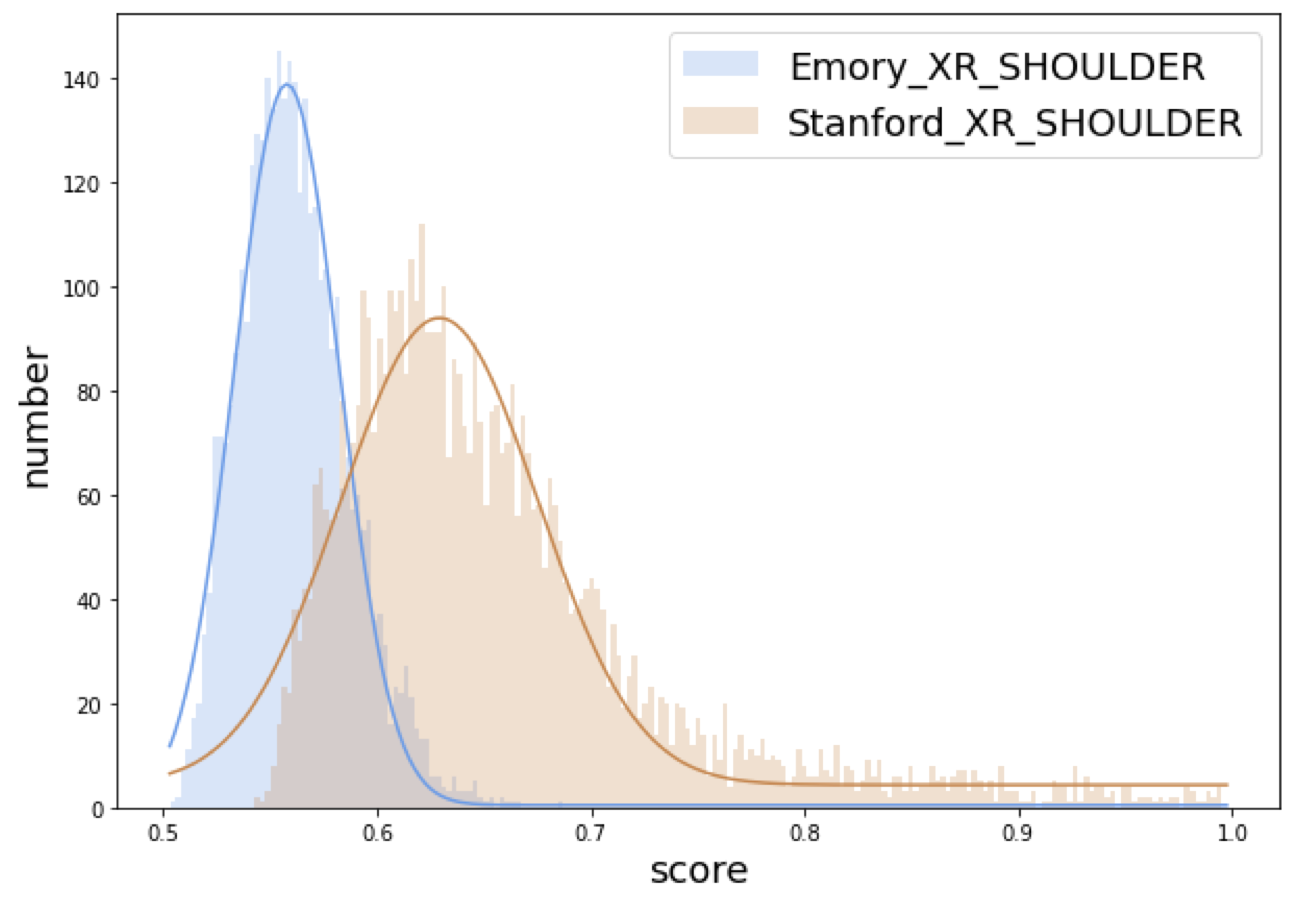}
\end{subfigure}
\begin{subfigure}{.49\textwidth}
\includegraphics[width=\linewidth, height=0.6\linewidth]{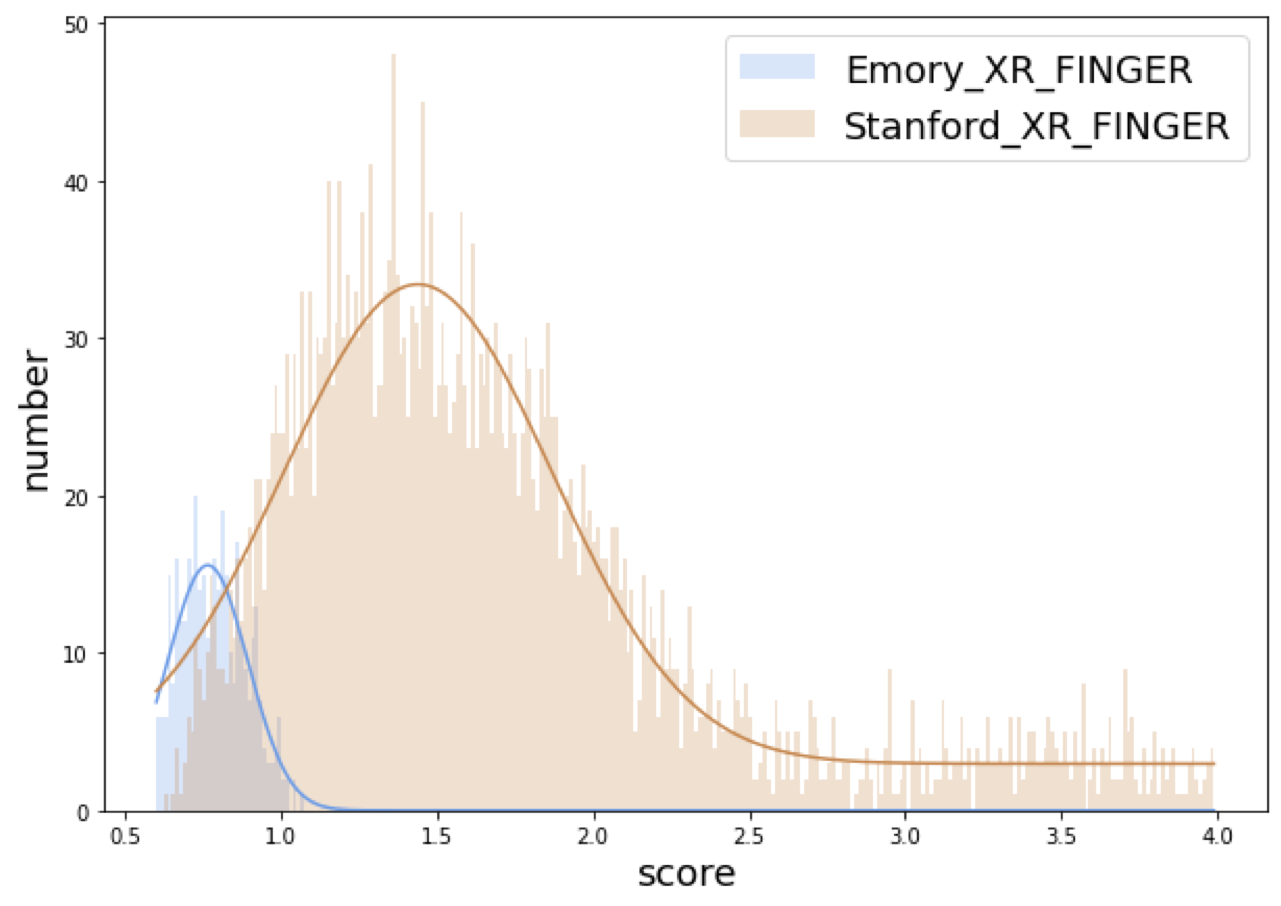}
\end{subfigure}
\begin{subfigure}{.49\textwidth}
\includegraphics[width=\linewidth, height=0.6\linewidth]{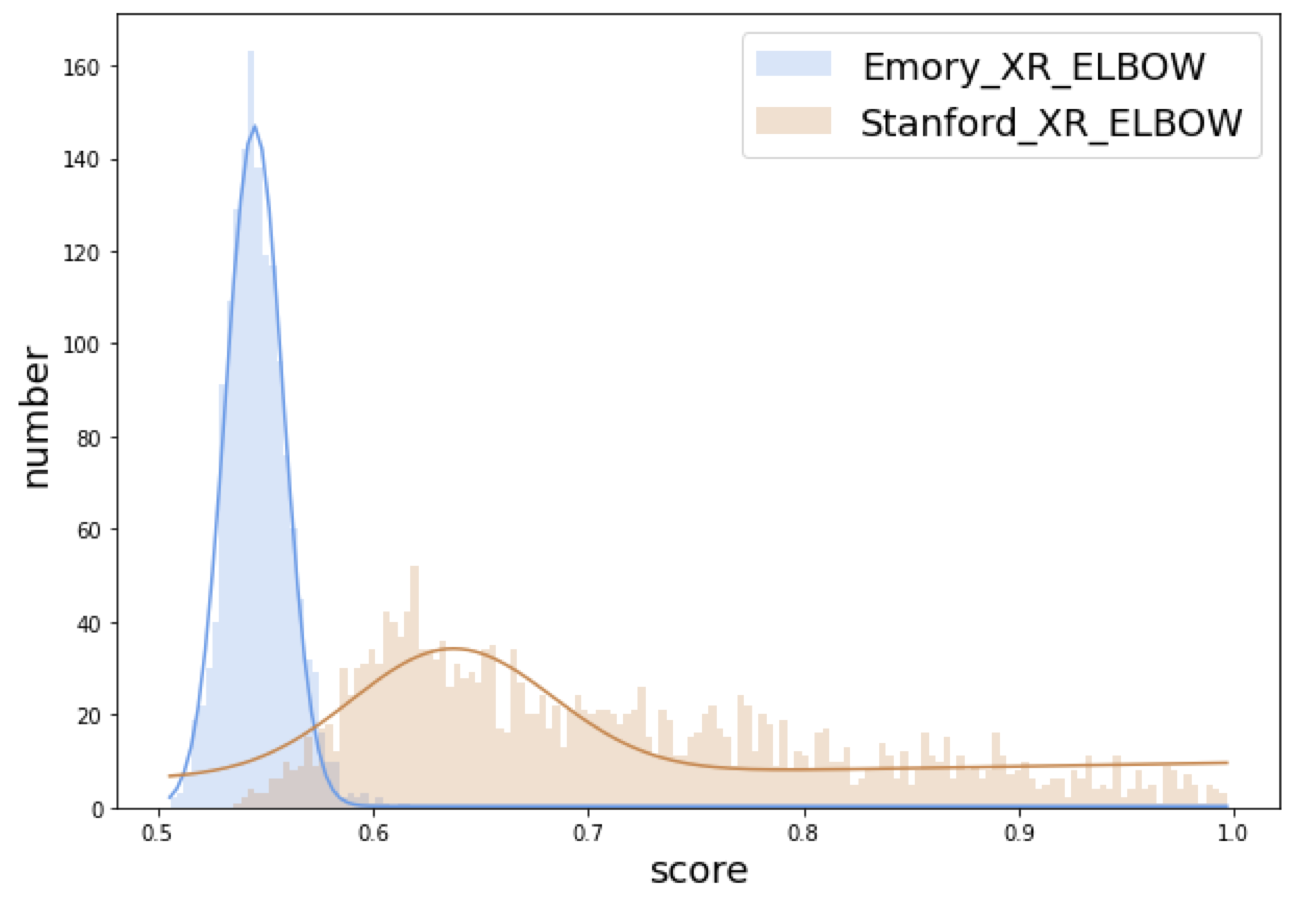}
\end{subfigure}
\begin{subfigure}{.49\textwidth}
\includegraphics[width=\linewidth, height=0.6\linewidth]{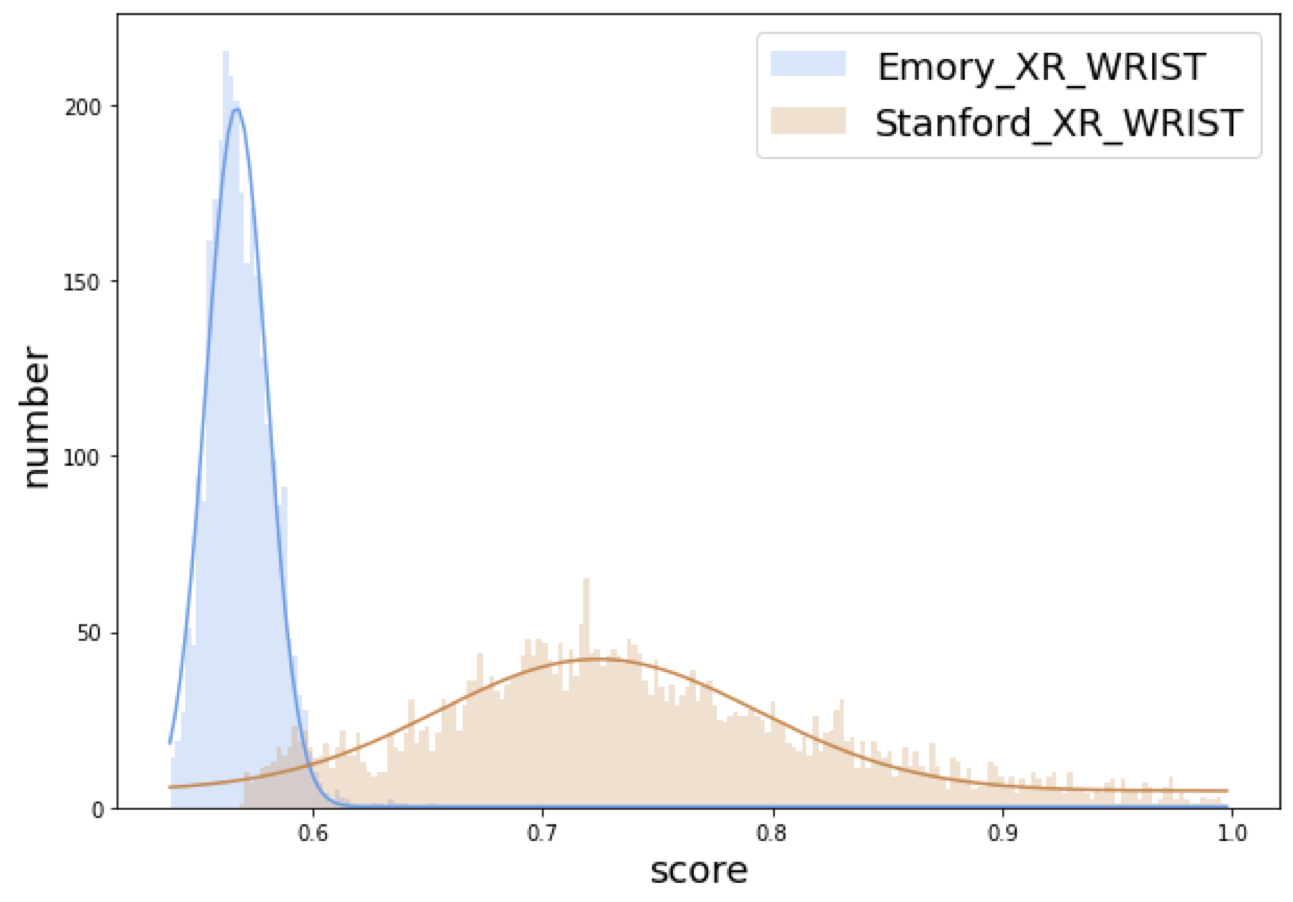}
\end{subfigure}
\begin{subfigure}{.49\textwidth}
\includegraphics[width=\linewidth, height=0.6\linewidth]{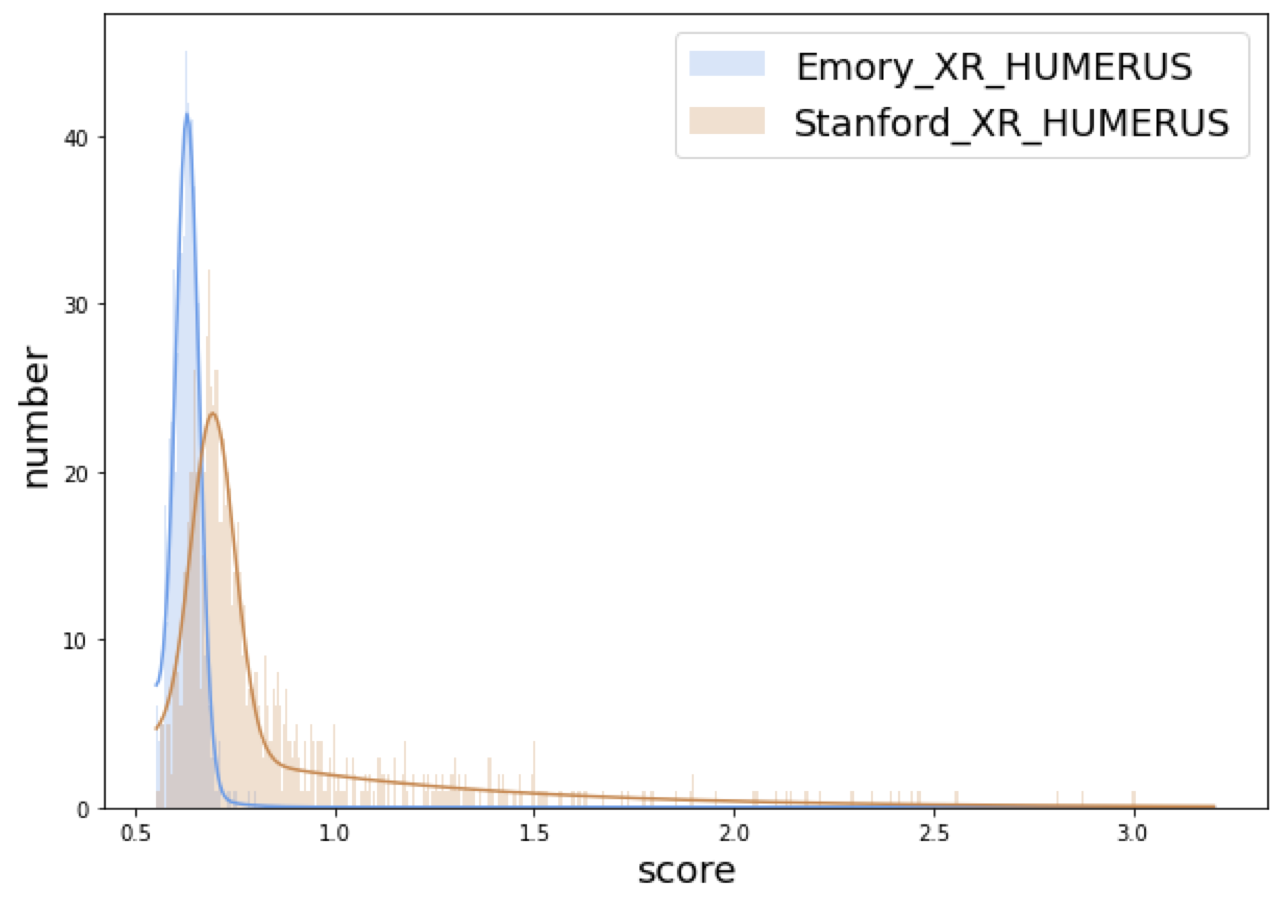}
\end{subfigure}
\caption{\textit{MedShift\_w\_CVAD}'s MURA anomaly score distribution results for classes \textit{FOREARM}, \textit{SHOULDER}, \textit{FINGER}, \textit{ELBOW}, \textit{WRIST} and \textit{HUMERUS} from left to right, top to bottom.}
\vspace{-20mm}
\label{fig:mura-intra-all}
\end{figure*}

\clearpage
\subsection{Anomaly Score Distribution Results with f-AnoGAN}~\label{mura_anomaly_score_more_fanogan}
\begin{figure*}[htp]
\vspace{-7mm}
\centering
\includegraphics[width=0.85\linewidth, height=0.98\linewidth]{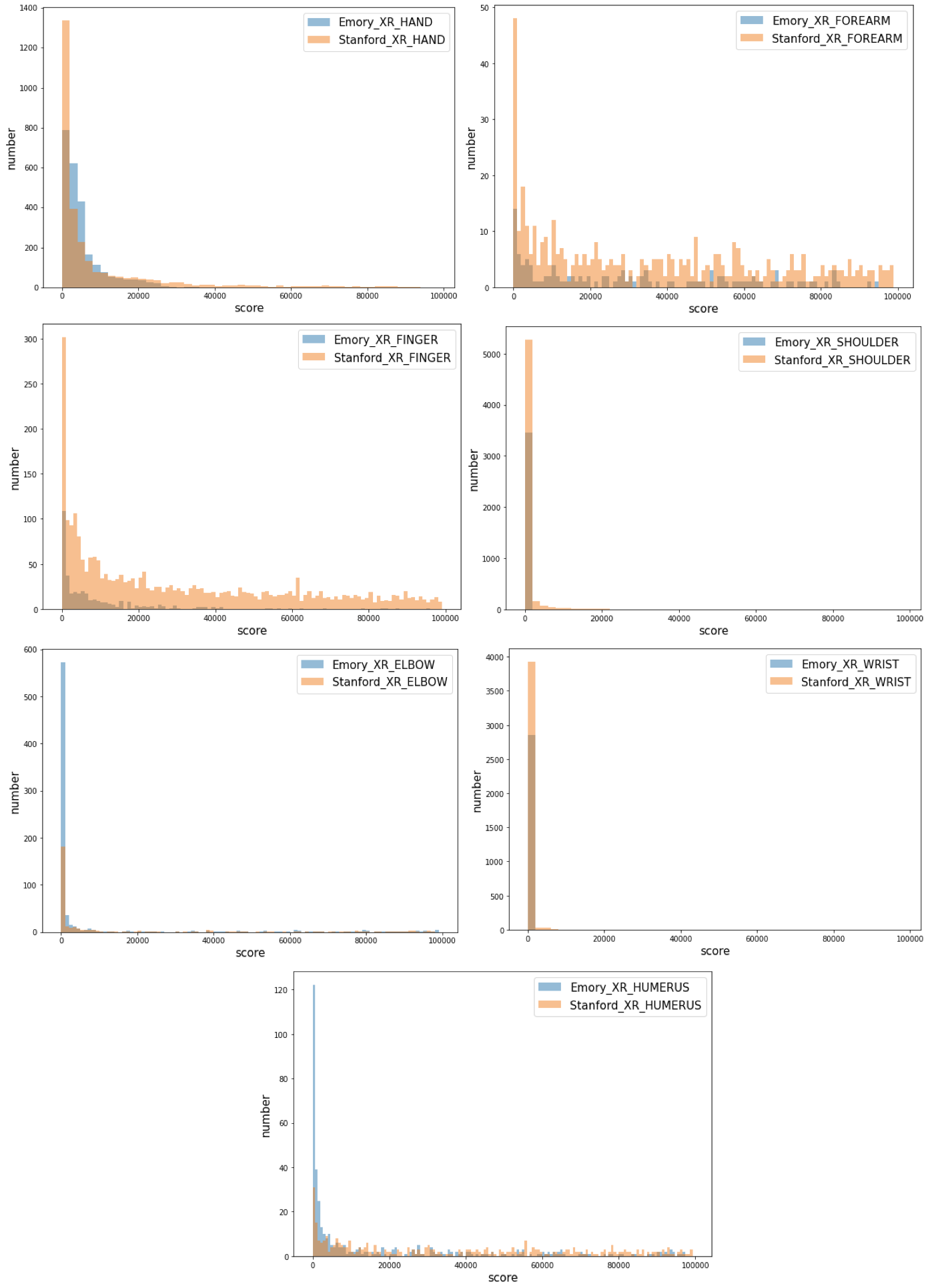}
\label{fig:mura-intra-all-fanogan}
\caption{\textit{MedShift\_w\_f-AnoGAN}'s MURA anomaly score distribution results for classes \textit{HAND}, \textit{FOREARM}, \textit{SHOULDER}, \textit{FINGER}, \textit{ELBOW}, \textit{WRIST} and \textit{HUMERUS} from left to right, top to bottom.}
\vspace{-15mm}
\end{figure*}

\clearpage
\subsection{\textit{MedShift\_w\_CVAD} Elbow Distortion Curve Results}~\label{mura_distortion_more}
\begin{figure*}[htp]
\vspace{-7mm}
\centering
\includegraphics[width=0.85\linewidth, height=0.95\linewidth]{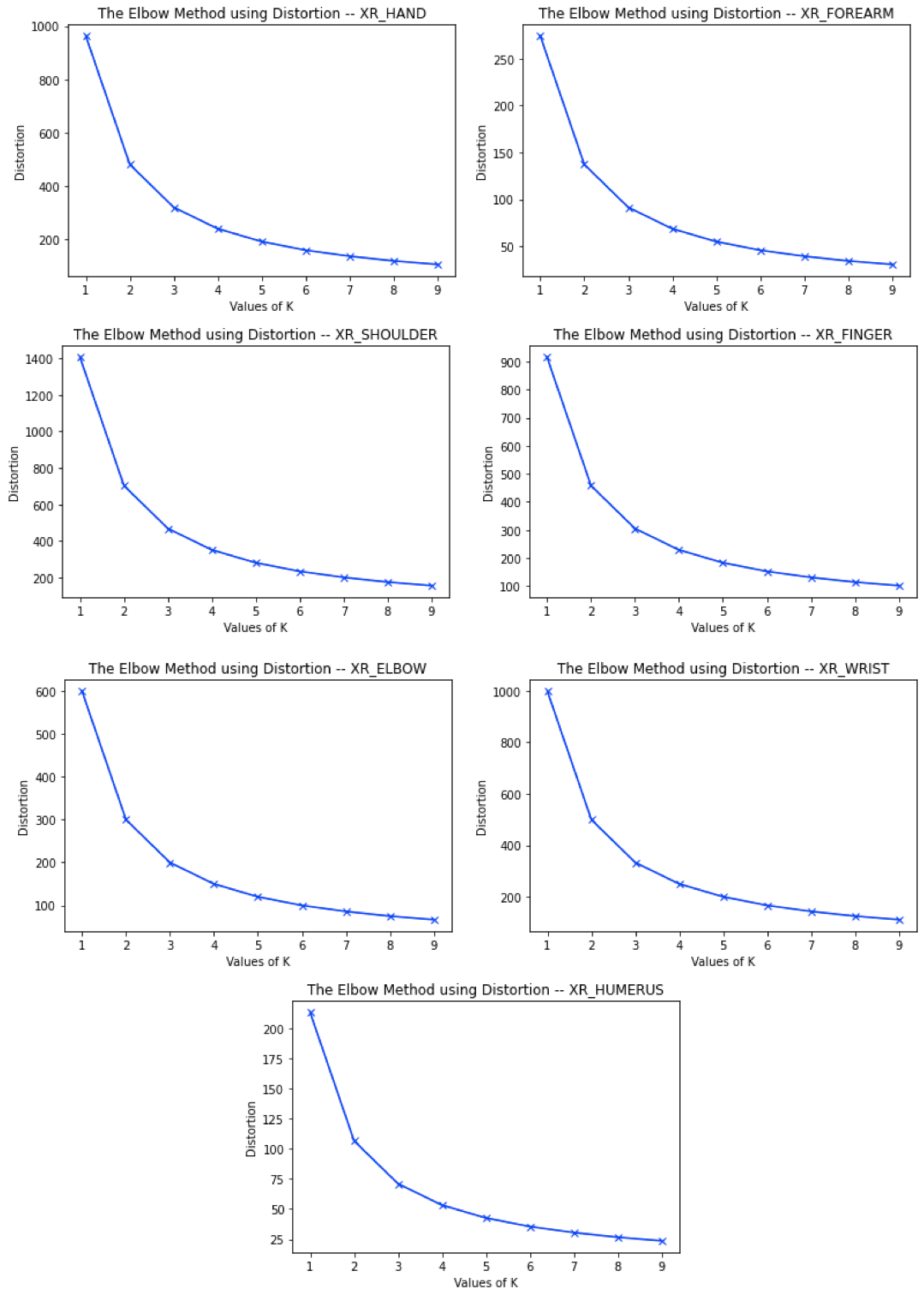}
\caption{\textit{MedShift\_w\_CVAD}'s elbow distortion curves on Stanford MURA dataset. From left to right, top to bottom, there are plots for \textit{XR\_HAND}, \textit{XR\_FOREARM}, \textit{XR\_SHOULDER}, \textit{XR\_FINGER}, \textit{XR\_ELBOW}, \textit{XR\_WRIST}, \textit{XR\_HUMERUS}, respectively. X-axis values represent the selection of $K$, the number of groups to be clustered into, and Y-axis values indicate the distortion. }
\vspace{-15mm}
\label{fig:mura_distortion}
\end{figure*}

\clearpage
\subsection{\textit{MedShift\_w\_fAnoGAN} Elbow Distortion Curve Results}~\label{mura_distortion_more_fanogan}
\begin{figure*}[htp]
\vspace{-7mm}
\centering
\includegraphics[width=0.85\linewidth, height=0.95\linewidth]{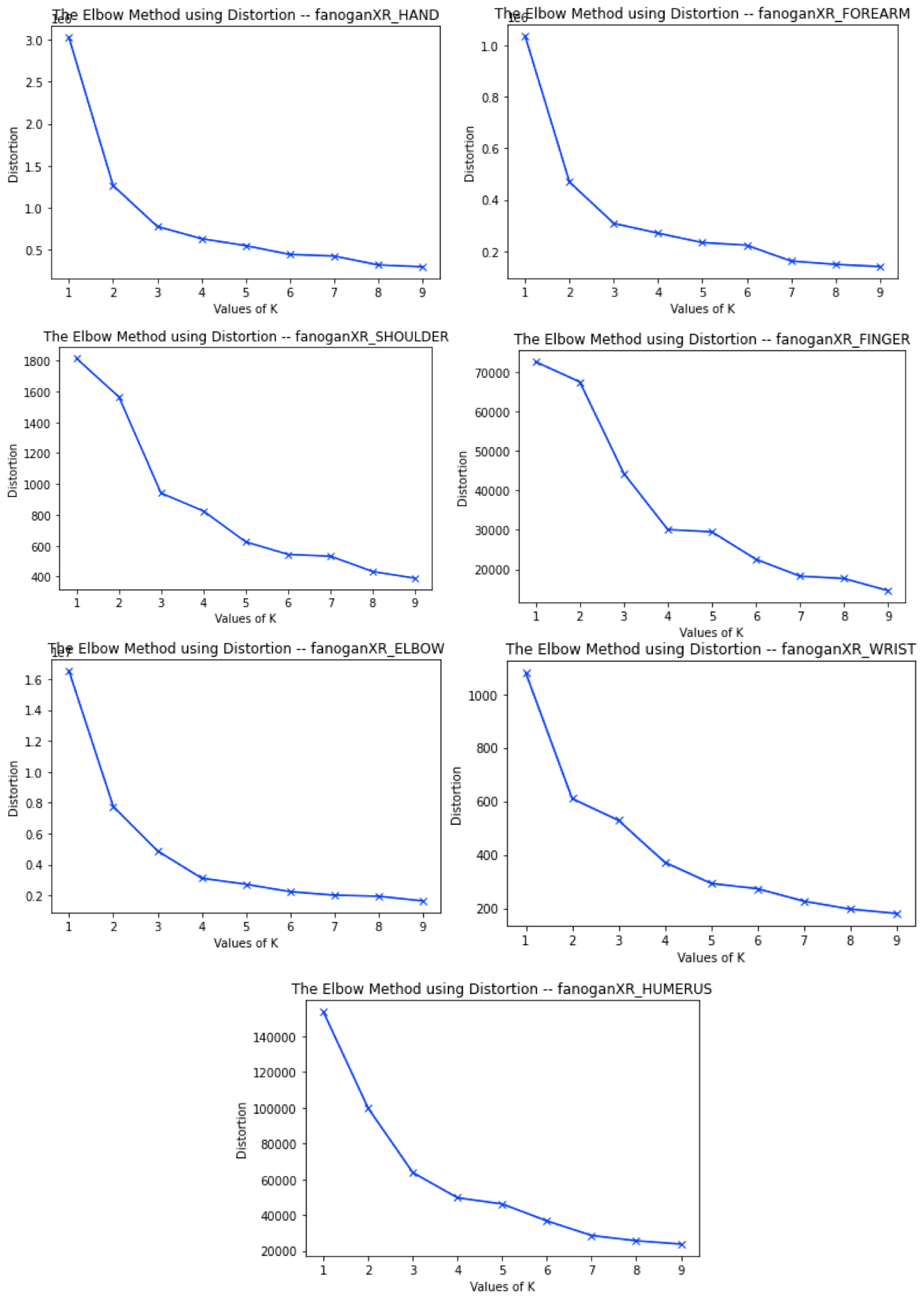}
\caption{\textit{MedShift\_w\_fAnoGAN}'s elbow distortion curves on Stanford MURA dataset. From left to right, top to bottom, there are plots for \textit{XR\_HAND}, \textit{XR\_FOREARM}, \textit{XR\_SHOULDER}, \textit{XR\_FINGER}, \textit{XR\_ELBOW}, \textit{XR\_WRIST}, \textit{XR\_HUMERUS}, respectively. X-axis values represent the selection of $K$, the number of groups to be clustered into, and Y-axis values indicate the distortion. }
\vspace{-15mm}
\label{fig:mura_distortion_fanogan}
\end{figure*}

\clearpage
\subsection{Clustering Results with CVAD}~\label{mura_clustering_more}

\begin{figure*}[htp]
\vspace{-6mm}
\centering
\begin{subfigure}{.5\textwidth}
  \includegraphics[width=0.95\linewidth, height=0.88\linewidth, frame]{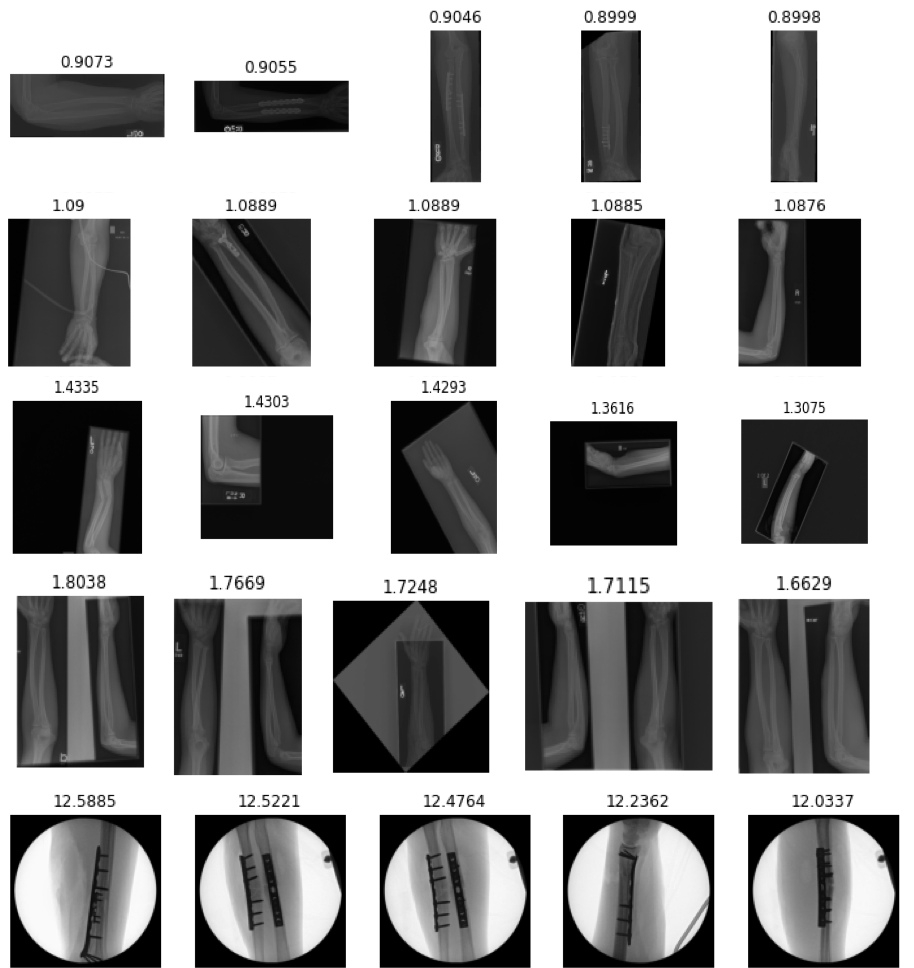}
  \caption{FOREARM}
\end{subfigure}%
\hspace*{\fill}
\begin{subfigure}{.5\textwidth}
  \includegraphics[width=0.95\linewidth, height=0.88\linewidth, frame]{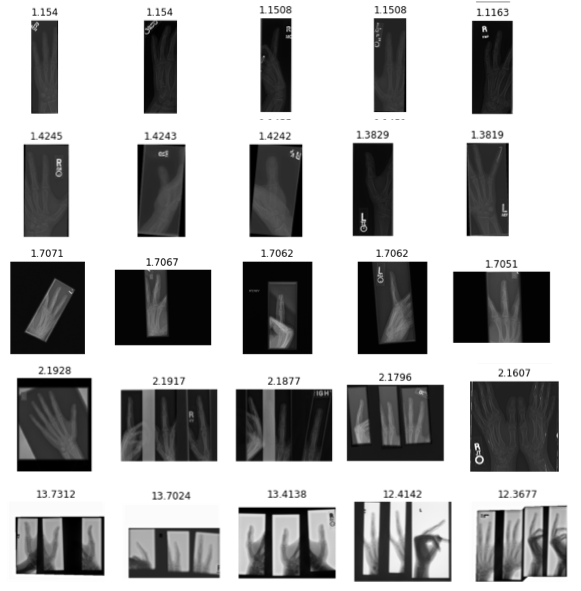}
  \caption{FINGER}
\end{subfigure}
\begin{subfigure}{.5\textwidth}
  \includegraphics[width=0.95\linewidth, height=0.88\linewidth,frame]{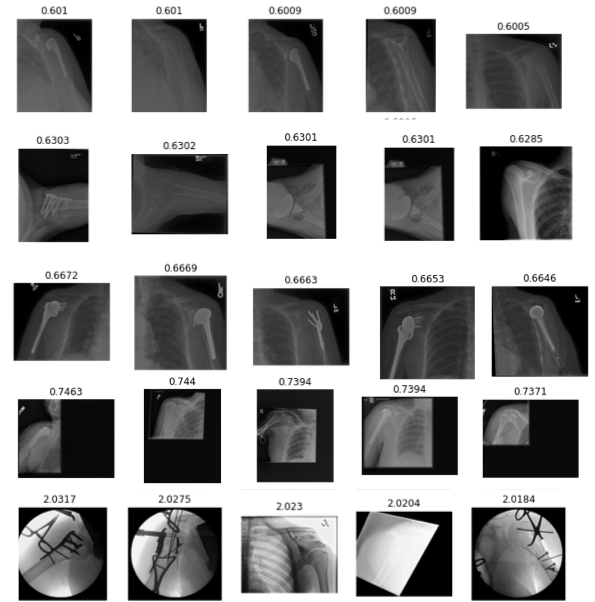}
  \caption{SHOULDER}
\end{subfigure}%
\hspace*{\fill}
\begin{subfigure}{.5\textwidth}
  \includegraphics[width=0.95\linewidth, height=0.88\linewidth,frame]{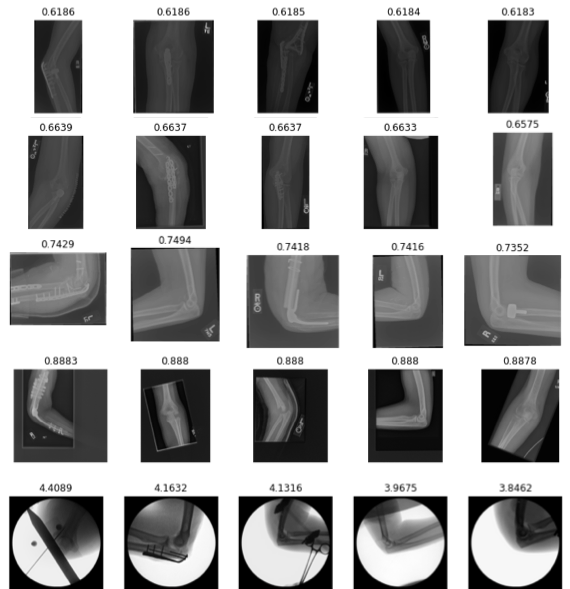}
  \caption{ELBOW}
\end{subfigure}
\caption{\textit{MedShift\_w\_CVAD}'s clustering results of Stanford MURA classes (a) \textit{FOREARM}; (b) \textit{FINGER}; (c) \textit{SHOULDER}; and (d) \textit{ELBOW}. Each row represents one group with five example images. The groups are illustrated in ascending order based on the anomaly scores from top to bottom. The corresponding anomaly score is on top of each image.}
\label{fig:mura_clustering_all_1}
\vspace{-15mm}
\end{figure*}

\begin{figure*}[htp]
\centering
\begin{subfigure}{.49\textwidth}
\includegraphics[width=0.95\linewidth, height=0.95\linewidth,frame]{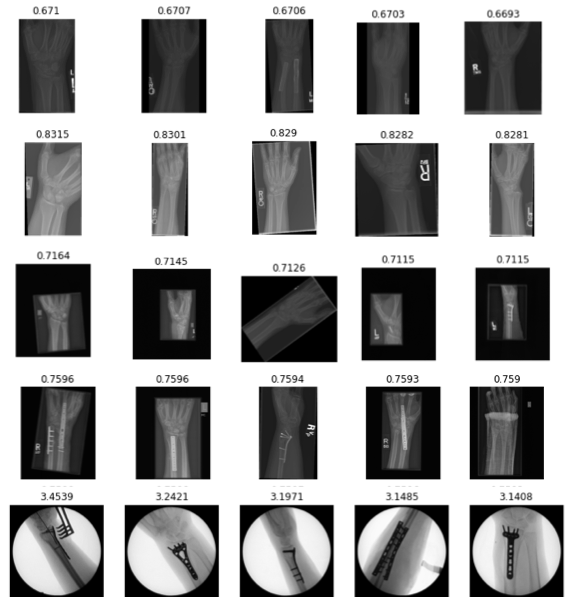}
  \caption{WRIST}
\end{subfigure}
\hspace*{\fill}
\begin{subfigure}{.49\textwidth}
  \includegraphics[width=0.95\linewidth, height=0.95\linewidth,frame]{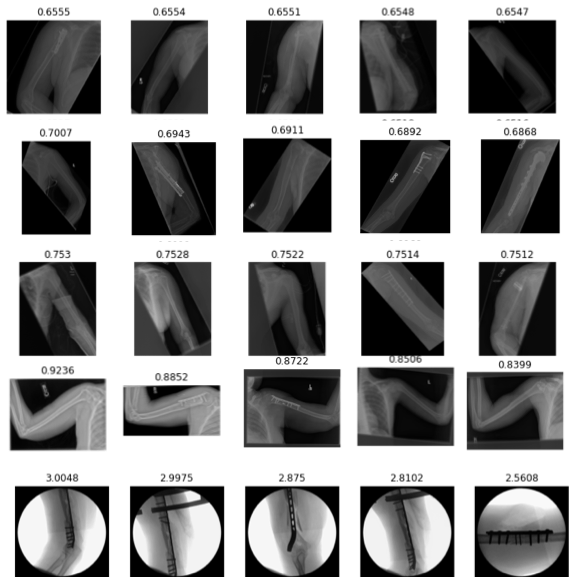}
  \caption{HUMERUS}
\end{subfigure}
\caption{More \textit{MedShift\_w\_CVAD}'s clustering results of Stanford MURA classes (a) \textit{WRIST}; (b) \textit{HUMERUS} following the same arrangement style of Fig.~\ref{fig:mura_clustering_all_1}.}
\vspace{-10mm}
\label{fig:mura_clustering_all_2}
\end{figure*}

\clearpage
\subsection{Clustering Results with f-AnoGAN}~\label{mura_clustering_more_fanogan}

\begin{figure*}[htp]
\vspace{-6mm}
\centering
\begin{subfigure}{.5\textwidth}
  \includegraphics[width=0.95\linewidth, height=0.88\linewidth, frame]{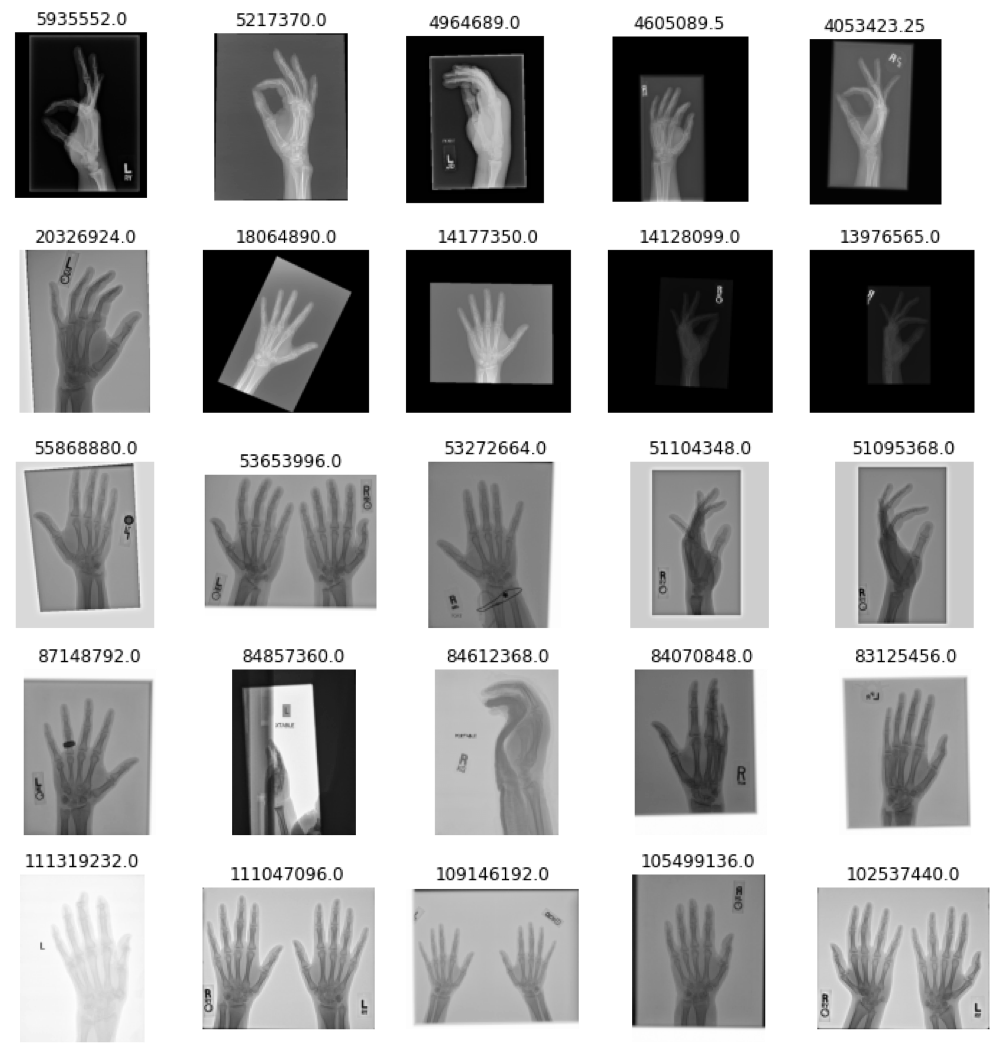}
  \caption{HAND\_fanogan}
\end{subfigure}%
\hspace*{\fill}
\begin{subfigure}{.5\textwidth}
  \includegraphics[width=0.95\linewidth, height=0.88\linewidth, frame]{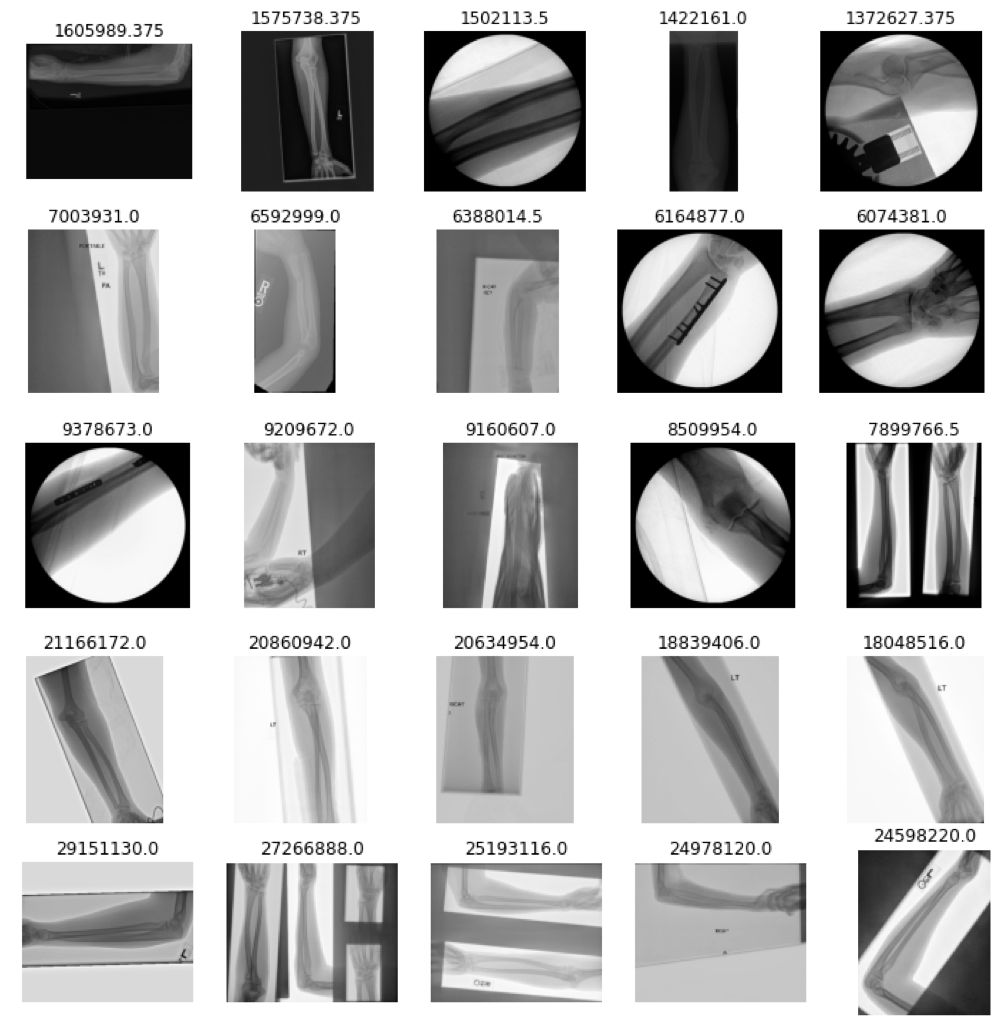}
  \caption{FOREARM\_fanogan}
\end{subfigure}
\begin{subfigure}{.5\textwidth}
  \includegraphics[width=0.95\linewidth, height=0.88\linewidth, frame]{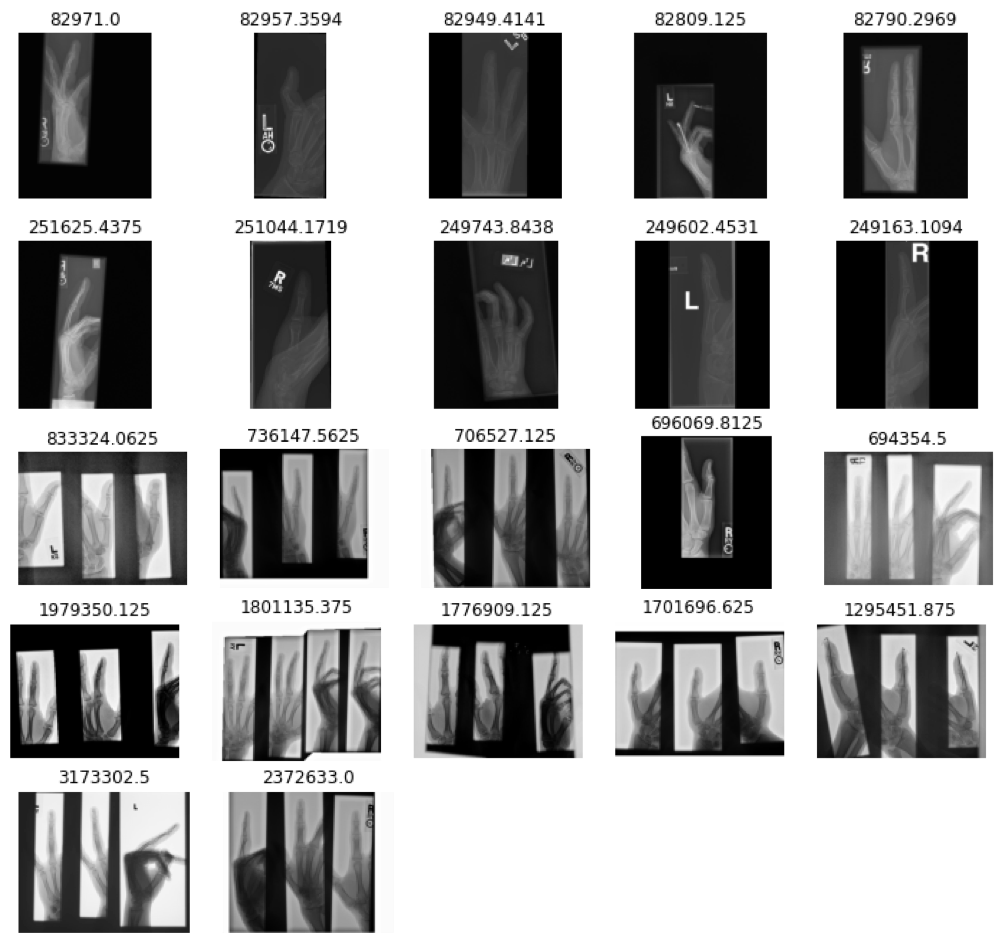}
  \caption{FINGER\_fanogan}
\end{subfigure}%
\hspace*{\fill}
\begin{subfigure}{.5\textwidth}
  \includegraphics[width=0.95\linewidth, height=0.88\linewidth,frame]{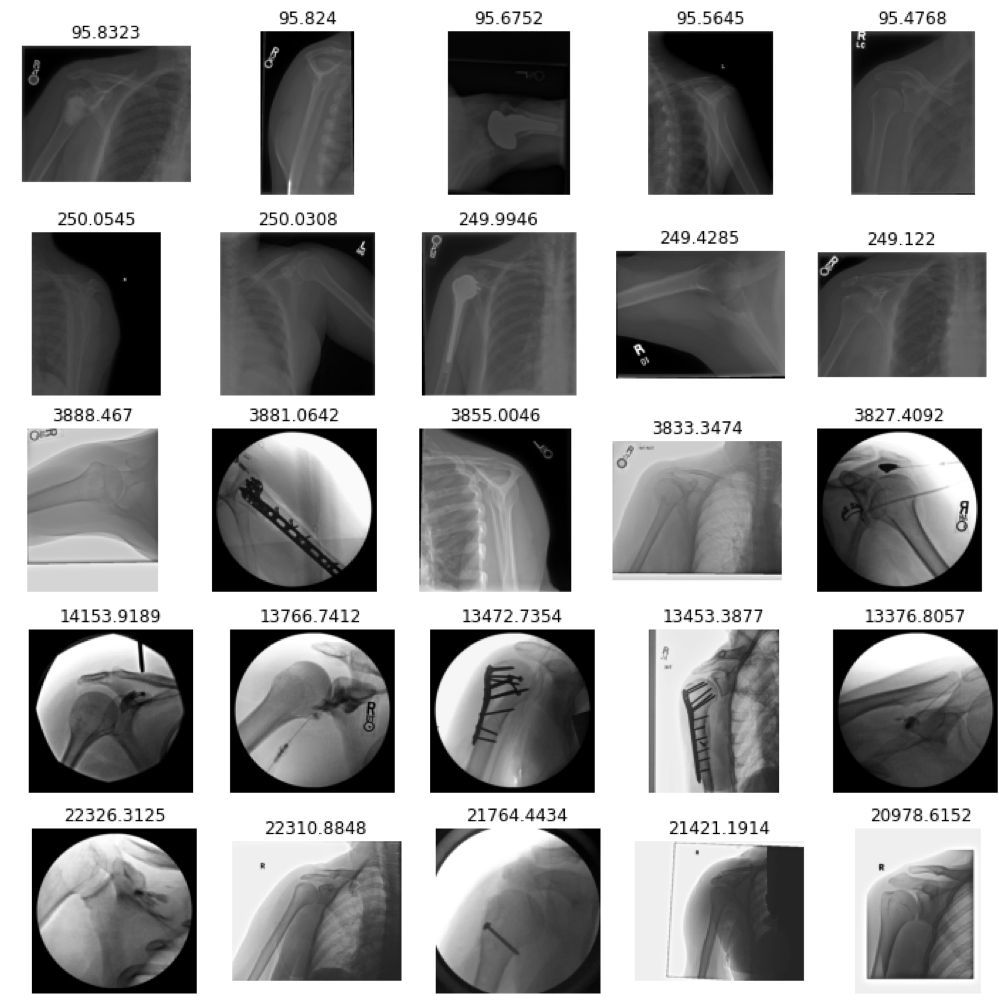}
  \caption{SHOULDER\_fanogan}
\end{subfigure}
\caption{\textit{MedShift\_w\_f-AnoGAN}'s clustering results of Stanford MURA classes (a) \textit{HAND}; (b) \textit{FOREARM}; (c) \textit{FINGER} and (d) \textit{SHOULDER}. Each row represents one group with five example images. The groups are illustrated in ascending order based on the anomaly scores from top to bottom. The corresponding anomaly score is on top of each image.}
\label{fig:mura_clustering_all_fanogan_1}
\vspace{-15mm}
\end{figure*}

\begin{figure*}[htp]
\centering
\begin{subfigure}{.49\textwidth}
  \includegraphics[width=0.95\linewidth, height=\linewidth,frame]{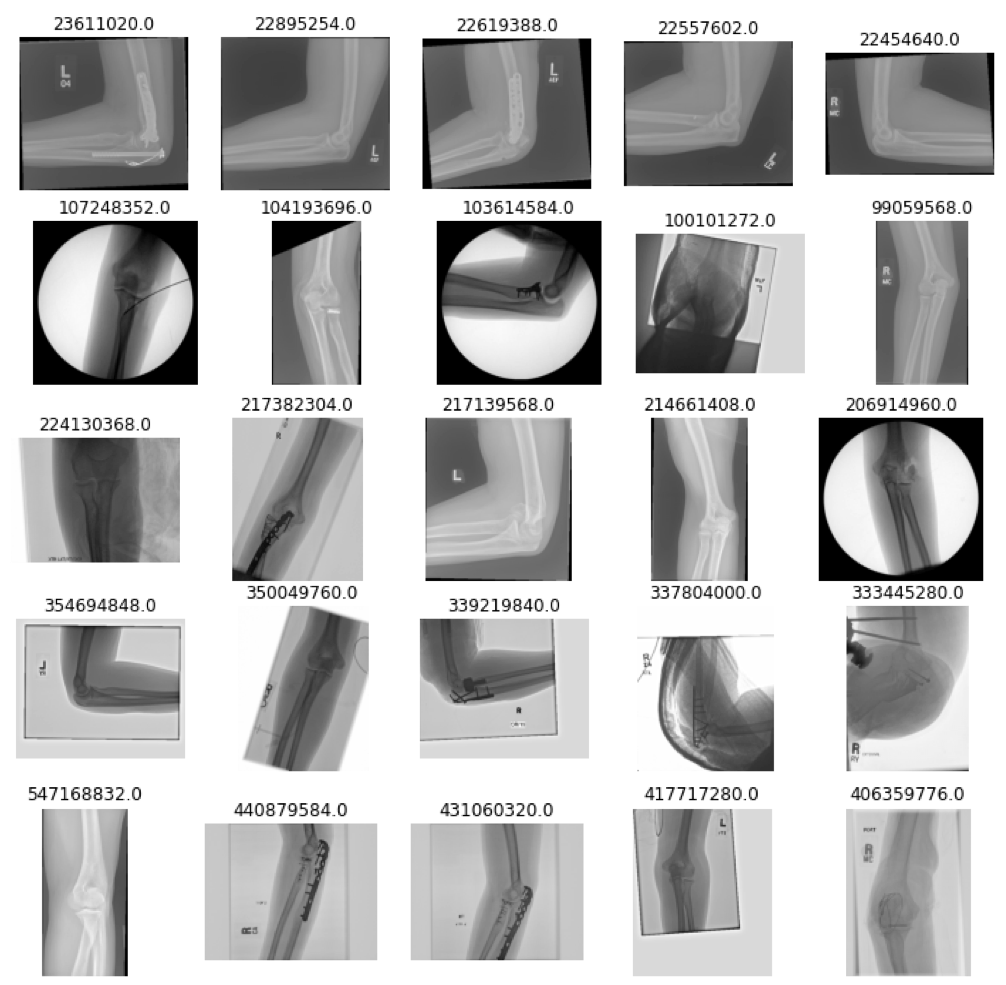}
  \caption{ELBOW\_fanogan}
\end{subfigure}
\hspace*{\fill}
\begin{subfigure}{.49\textwidth}
\includegraphics[width=0.95\linewidth, height=\linewidth,frame]{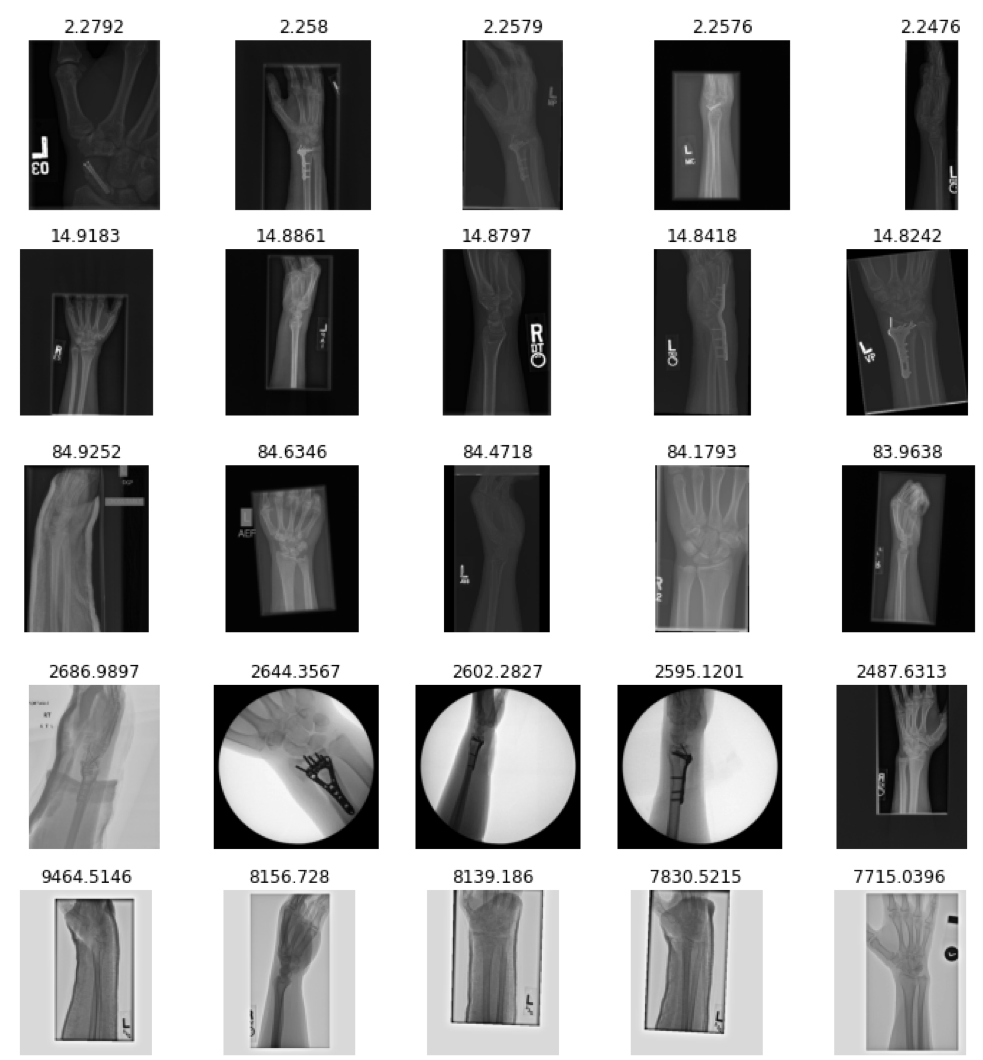}
  \caption{WRIST\_fanogan}
\end{subfigure}
\centering
\begin{subfigure}{.49\textwidth}
  \includegraphics[width=0.95\linewidth, height=\linewidth,frame]{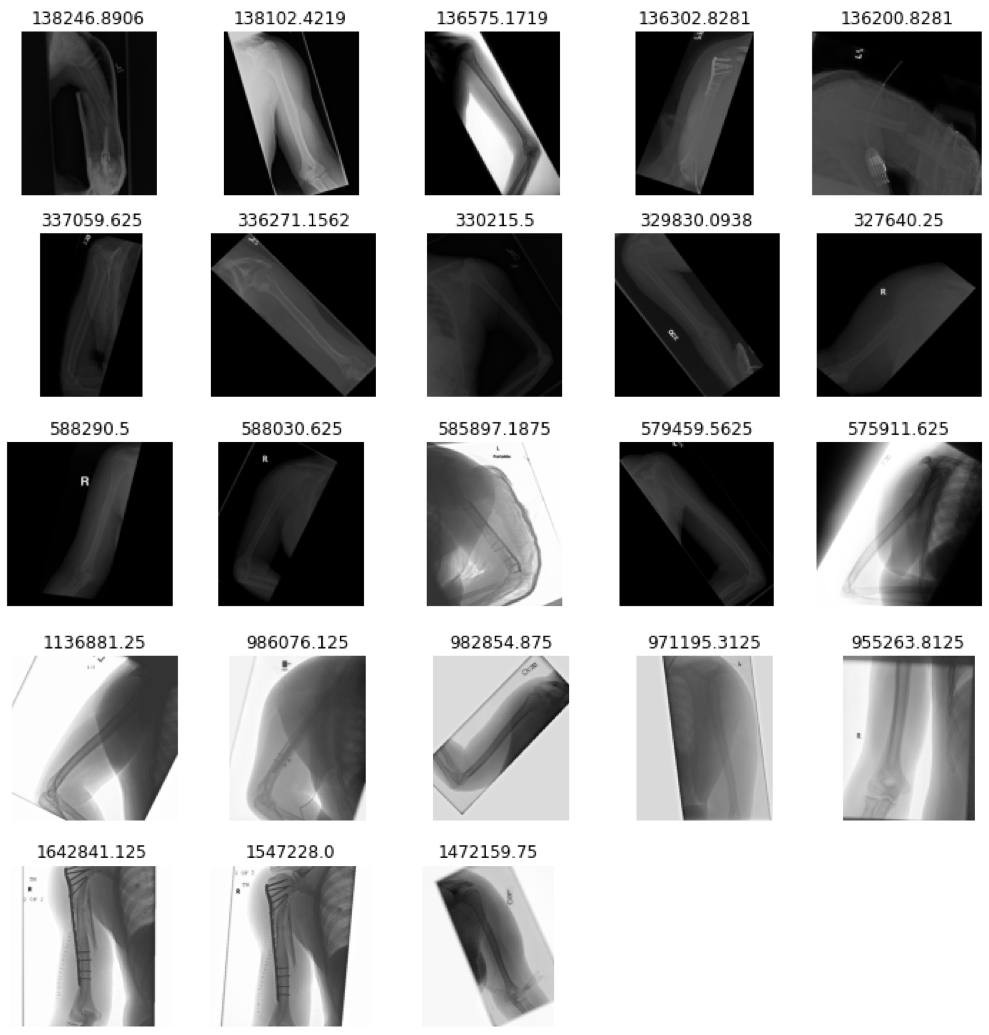}
  \caption{HUMERUS\_fanogan}
\end{subfigure}
\caption{More \textit{MedShift\_w\_f-AnoGAN}'s clustering results of Stanford MURA classes (a) \textit{ELBOW};  (b) \textit{WRIST} and (c) \textit{HUMERUS} following the same arrangement style of Fig.~\ref{fig:mura_clustering_all_1}.}
\label{fig:mura_clustering_all_fanogan_2}
\end{figure*}

\clearpage
\section{Chest X-ray Results}~\label{chest_more}

\subsection{Anomaly Score Distribution Results}~\label{chest_anomaly_score_more}
\begin{figure*}[htp]
\vspace{-6mm}
\centering
\begin{subfigure}{.45\textwidth}
  \includegraphics[width=\linewidth]{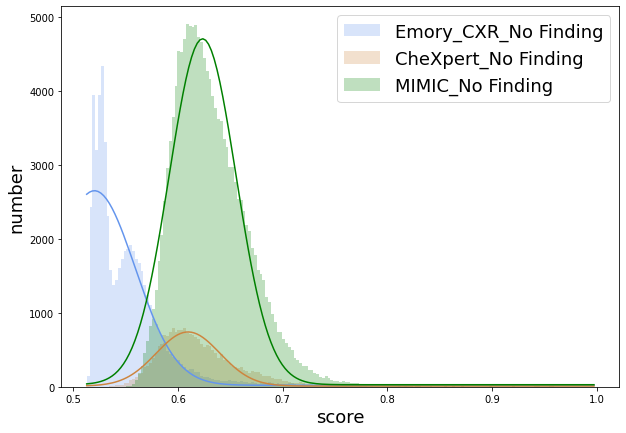}
\end{subfigure}%
\begin{subfigure}{.45\textwidth}
  \includegraphics[width=\linewidth]{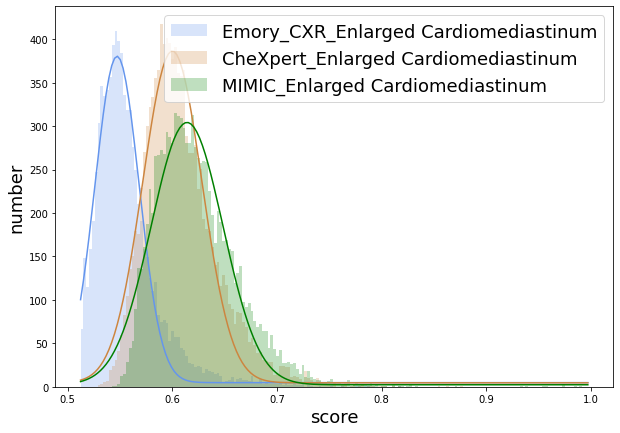}
\end{subfigure}
\begin{subfigure}{.45\textwidth}
  \includegraphics[width=\linewidth]{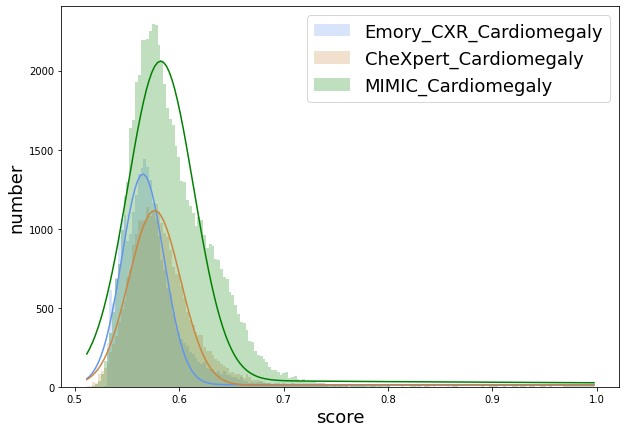}
\end{subfigure}
\begin{subfigure}{.45\textwidth}
  \includegraphics[width=\linewidth]{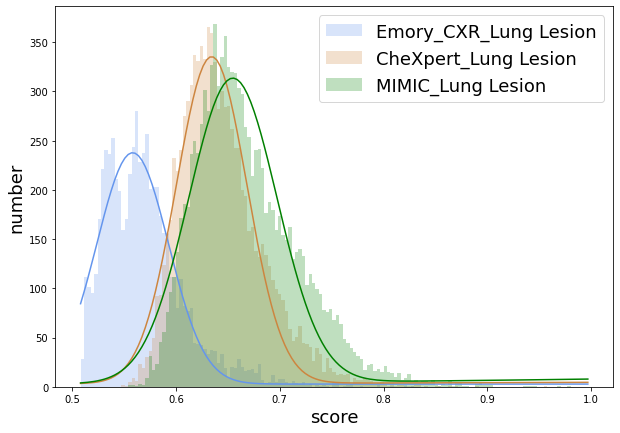}
\end{subfigure}
\begin{subfigure}{.45\textwidth}
  \includegraphics[width=\linewidth]{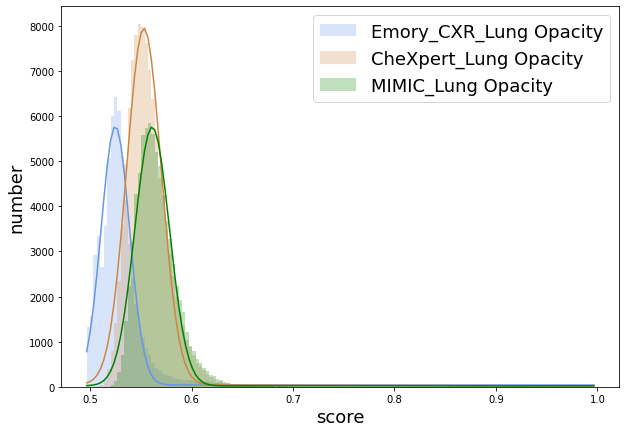}
\end{subfigure}
\begin{subfigure}{.45\textwidth}
  \includegraphics[width=\linewidth]{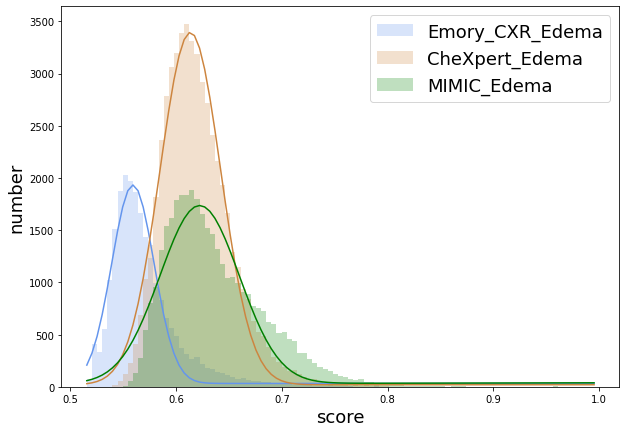}
\end{subfigure}
\caption{\textit{MedShift\_w\_CVAD} class-wise chest X-ray anomaly score distributions for Emory\_CXR (blue), CheXpert (orange) and MIMIC (green) datasets. Distributions may be truncated for better visualization.} 
\vspace{-15mm}
\label{fig:chestchexpertmimic1}
\end{figure*}

\begin{figure*}[htp]
\centering
\begin{subfigure}{.45\textwidth}
  \includegraphics[width=\linewidth]{Consolidation_distribution_all.png}
\end{subfigure}
\begin{subfigure}{.45\textwidth}
  \includegraphics[width=\linewidth]{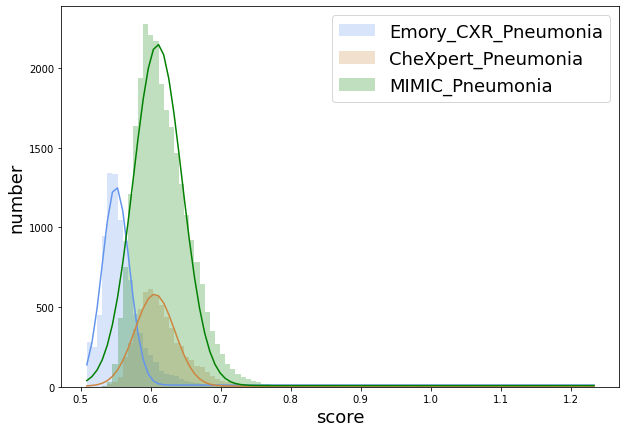}
\end{subfigure}
\begin{subfigure}{.45\textwidth}
  \includegraphics[width=\linewidth]{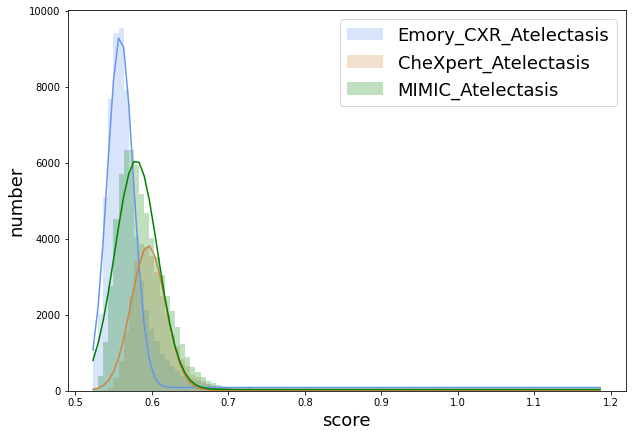}
\end{subfigure}
\begin{subfigure}{.45\textwidth}
  \includegraphics[width=\linewidth]{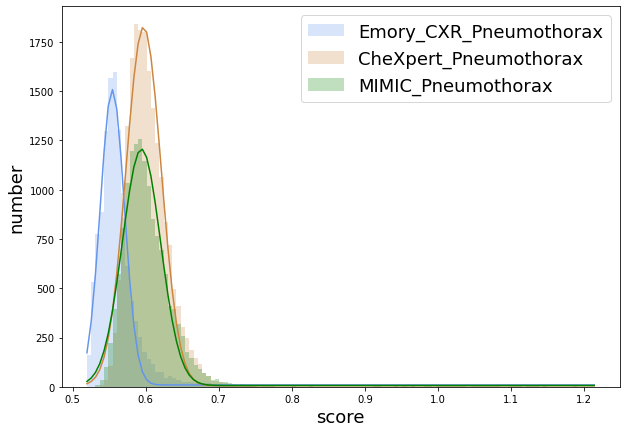}
\end{subfigure}
\begin{subfigure}{.45\textwidth}
  \includegraphics[width=\linewidth]{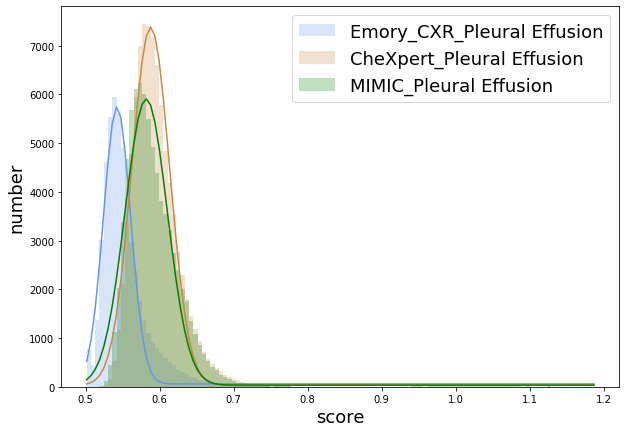}
\end{subfigure}
\begin{subfigure}{.45\textwidth}
  \includegraphics[width=\linewidth]{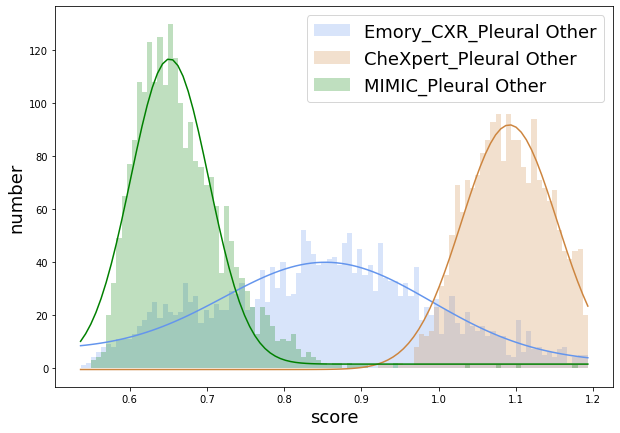}
\end{subfigure}
\begin{subfigure}{.45\textwidth}
  \includegraphics[width=\linewidth]{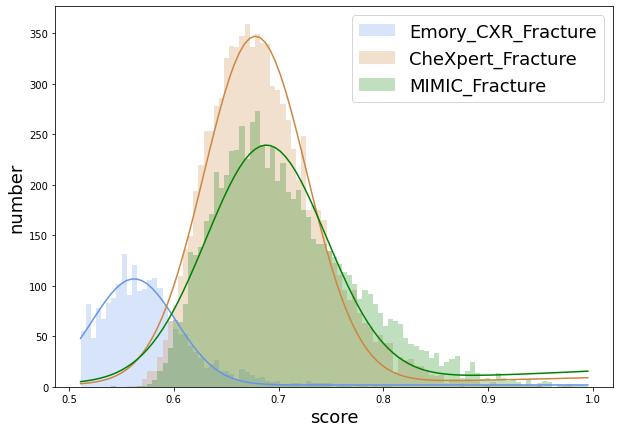}
\end{subfigure}
\begin{subfigure}{.45\textwidth}
  \includegraphics[width=\linewidth]{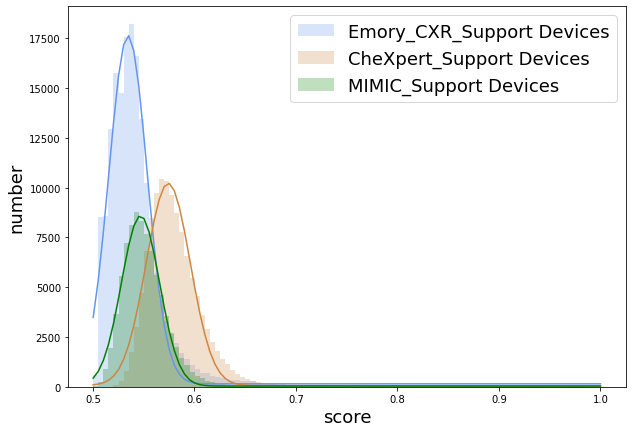}
\end{subfigure}
\caption{More \textit{MedShift\_w\_CVAD} class-wise chest X-ray anomaly score distributions for Emory\_CXR (blue), CheXpert (orange) and MIMIC (green) datasets. Distributions may be truncated for better visualization.}
\label{fig:chestchexpertmimic2}
\end{figure*}
\clearpage

\subsection{Clustering Results}~\label{chest_clustering_more}
\vspace{-8mm}
\subsubsection{Clustering Results for CheXpert}
\begin{figure*}[htp]
\vspace{-2mm}
\begin{subfigure}{.5\textwidth}
  \includegraphics[width=0.95\linewidth, height=0.88\linewidth, frame]{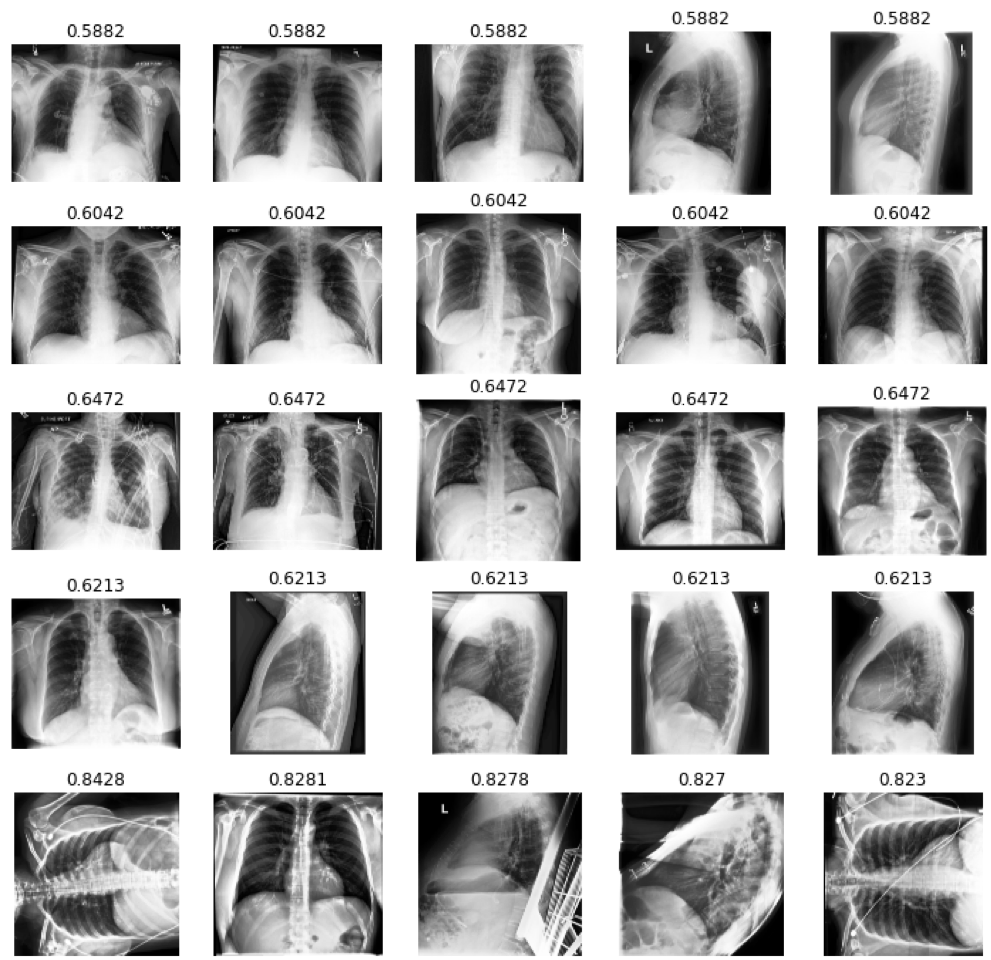}
  \caption{No Finding}
\end{subfigure}%
\hspace*{\fill}
\begin{subfigure}{.5\textwidth}
  \includegraphics[width=0.95\linewidth, height=0.88\linewidth, frame]{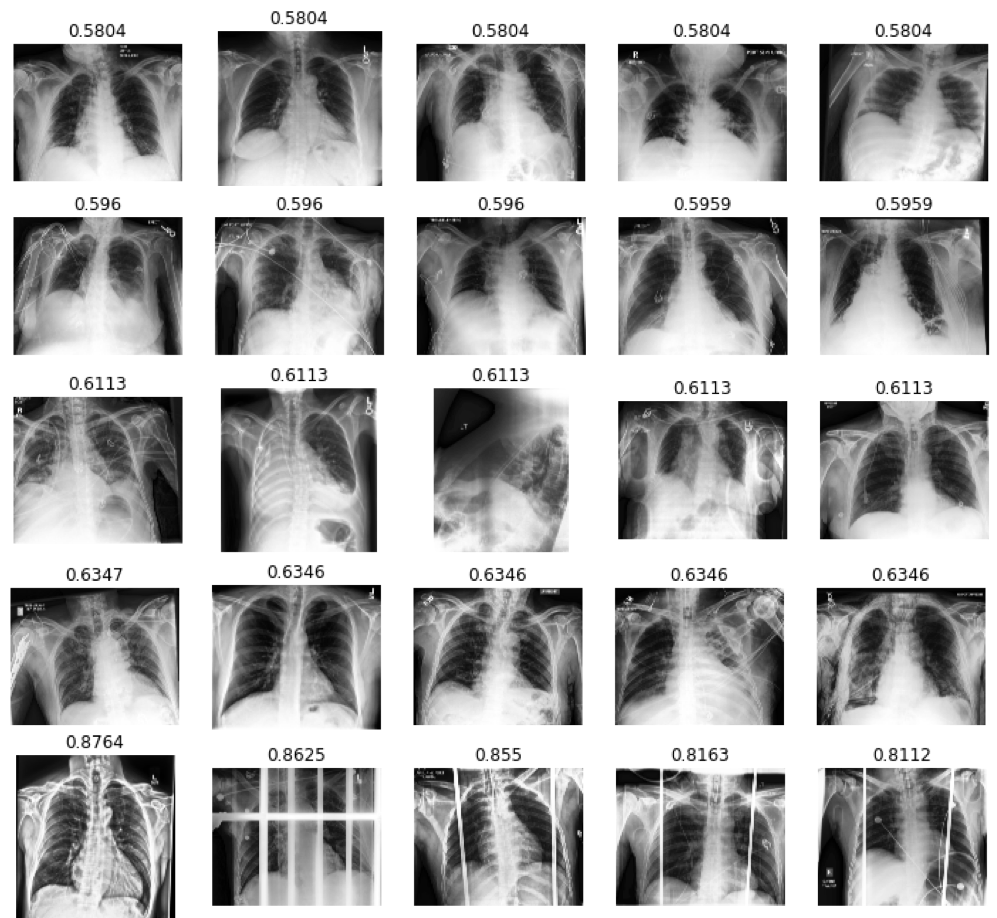}
  \caption{Enlarge Cardiomediastinum}
\end{subfigure}
\begin{subfigure}{.5\textwidth}
  \includegraphics[width=0.95\linewidth, height=0.88\linewidth, frame]{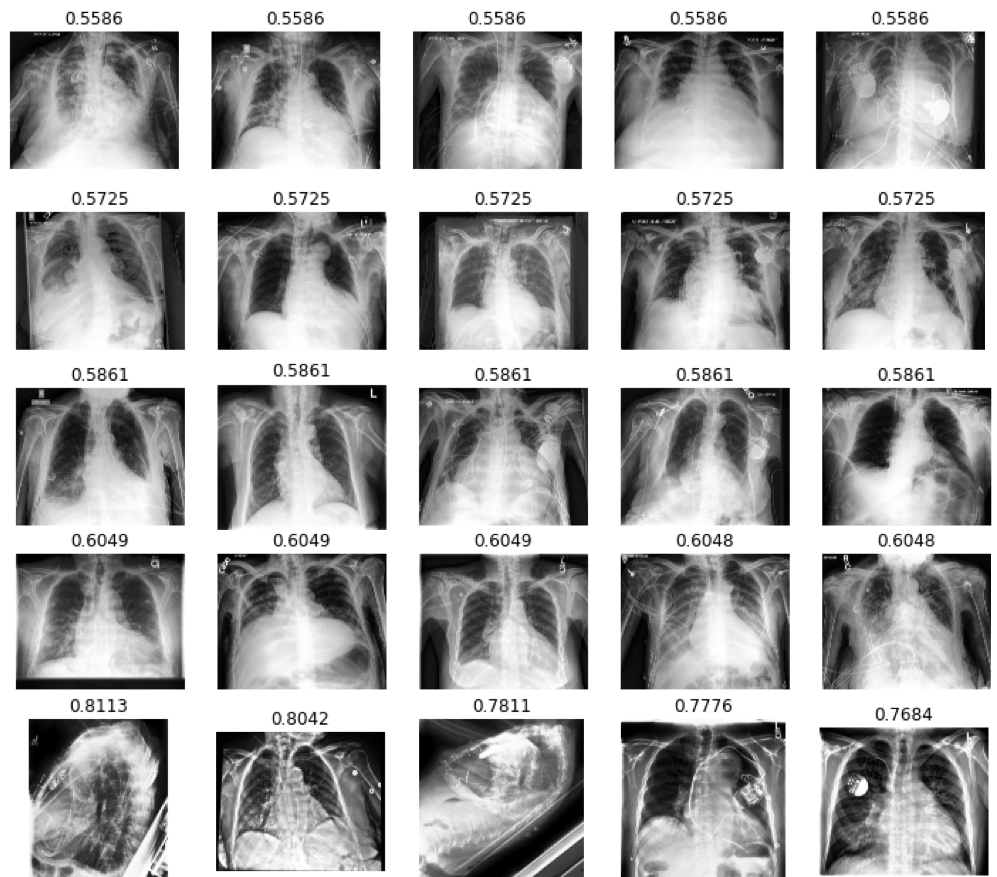}
  \caption{Cardiomegaly}
\end{subfigure}%
\hspace*{\fill}
\begin{subfigure}{.5\textwidth}
  \includegraphics[width=0.95\linewidth, height=0.88\linewidth, frame]{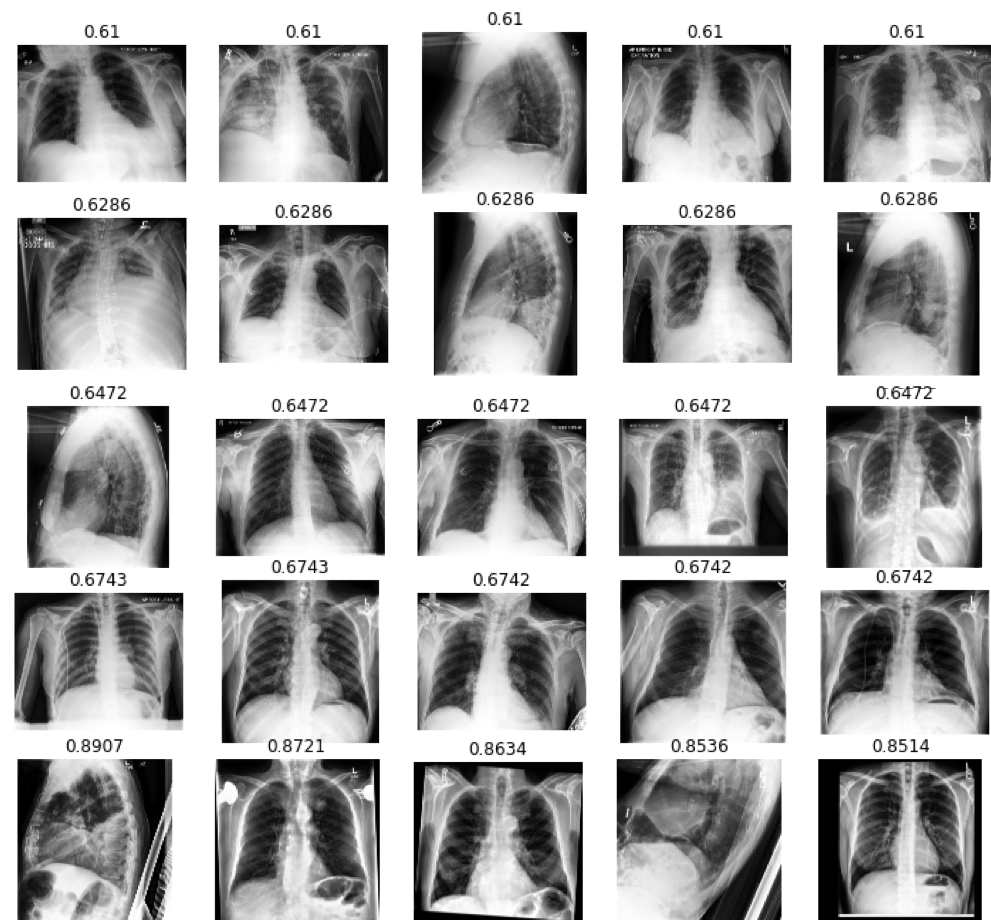}
  \caption{Lung Lesion}
\end{subfigure}
\caption{\textit{MedShift\_w\_CVAD} CheXpert clustering results for classes (a) \textit{No Finding}; (b) \textit{Enlarge Cardiomediastinum}; (c) \textit{Cardiomegaly} and (d) \textit{Lung Lesion} following the same arrangement style of Fig.~\ref{fig:mura_clustering_all_1}. }
\vspace{-15mm}
\label{fig:chest_clustering1}
\end{figure*}

\begin{figure*}[htp]
\begin{subfigure}{.5\textwidth}
    \includegraphics[width=0.95\linewidth, height=\linewidth, frame]{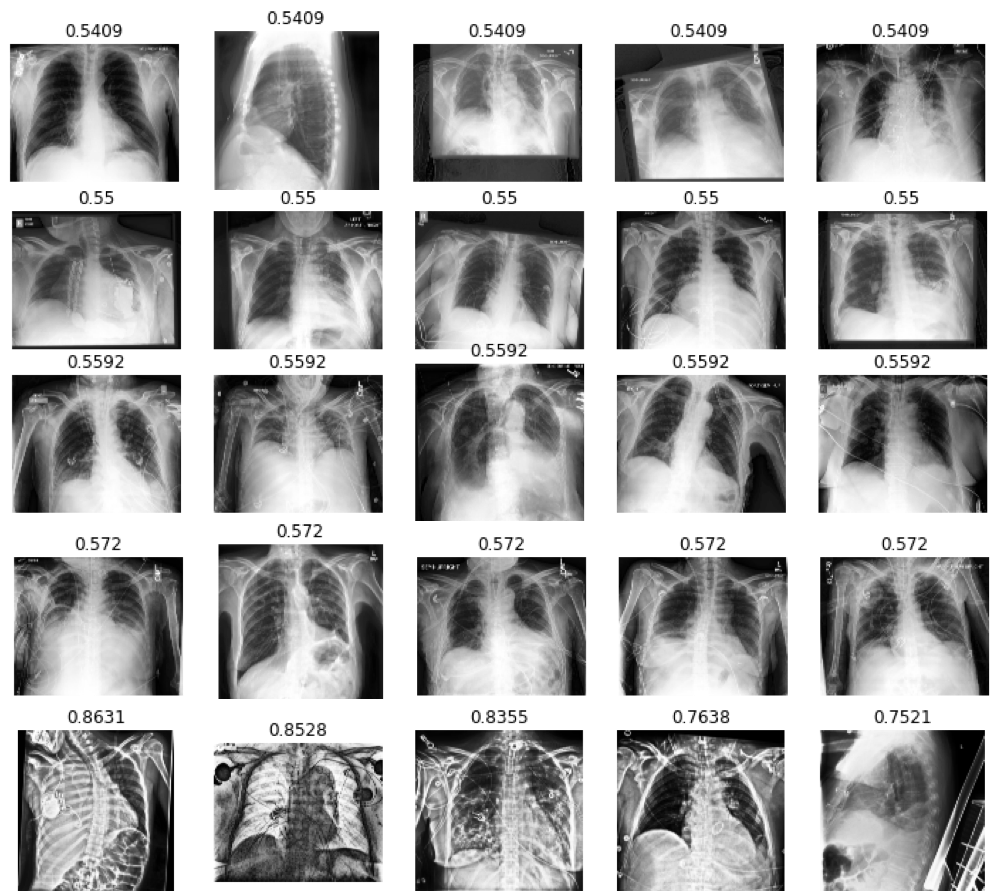}
    \caption{Lung Opacity}
\end{subfigure}
\hspace*{\fill}
\begin{subfigure}{.5\textwidth}
  \includegraphics[width=0.95\linewidth, height=\linewidth, frame]{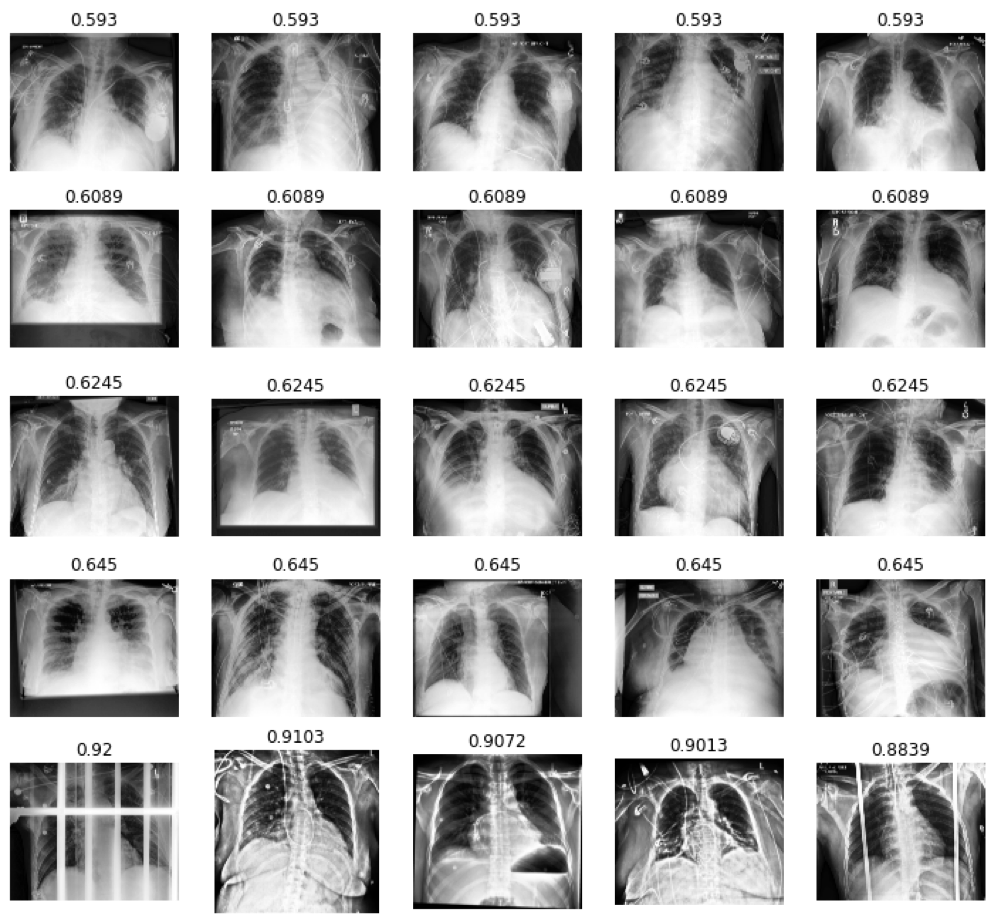}
  \caption{Edema}
\end{subfigure}
\begin{subfigure}{.5\textwidth}
  \includegraphics[width=0.95\linewidth, height=\linewidth, frame]{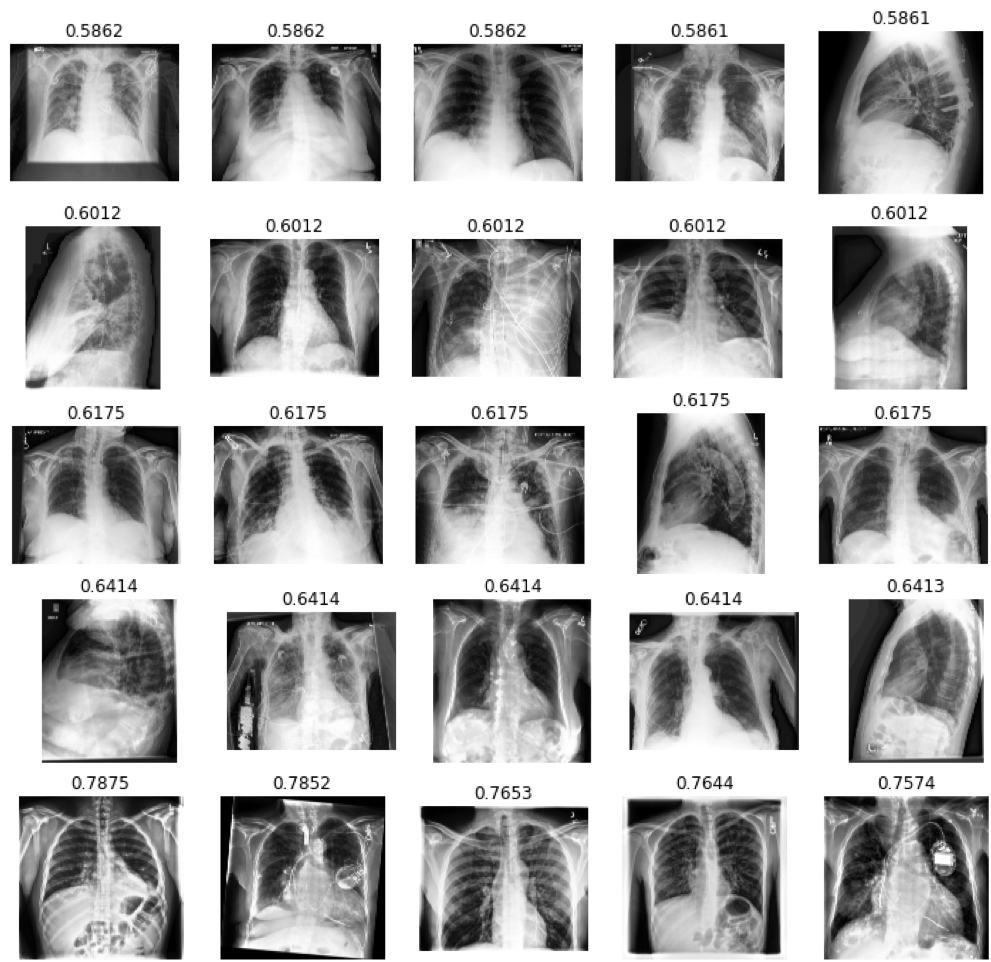}
  \caption{Pneumonia}
\end{subfigure} 
\hspace*{\fill}
\begin{subfigure}{.5\textwidth}
  \includegraphics[width=0.95\linewidth, height=\linewidth, frame]{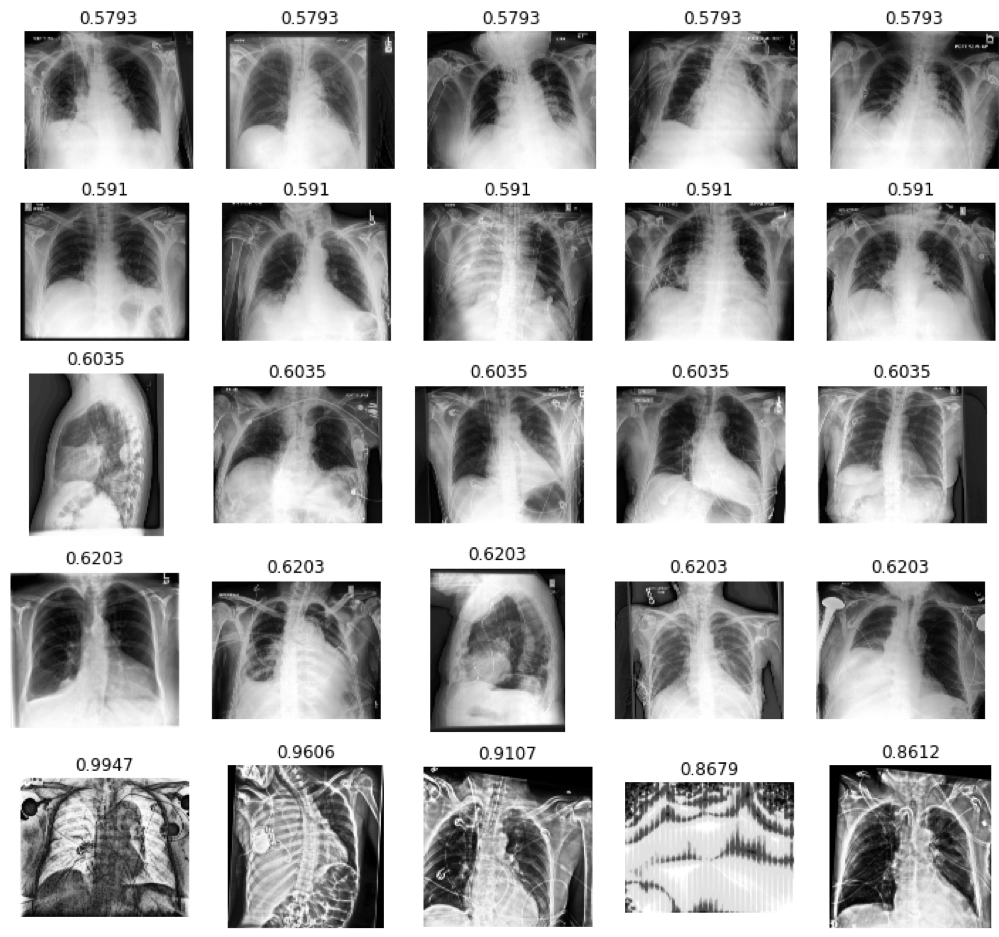}
  \caption{Atelectasis}
\end{subfigure}%
\caption{More \textit{MedShift\_w\_CVAD} CheXpert clustering results for classes (a) \textit{Lung Opacity}; (b) \textit{Edema}; (c) \textit{Pneumonia} and (d) \textit{Atelectasis} following the same arrangement style of Fig.~\ref{fig:mura_clustering_all_1}. }
\label{fig:chest_clustering2}
\end{figure*}

\begin{figure*}[htp]
\begin{subfigure}{.5\textwidth}
  \includegraphics[width=0.95\linewidth, height=\linewidth, frame]{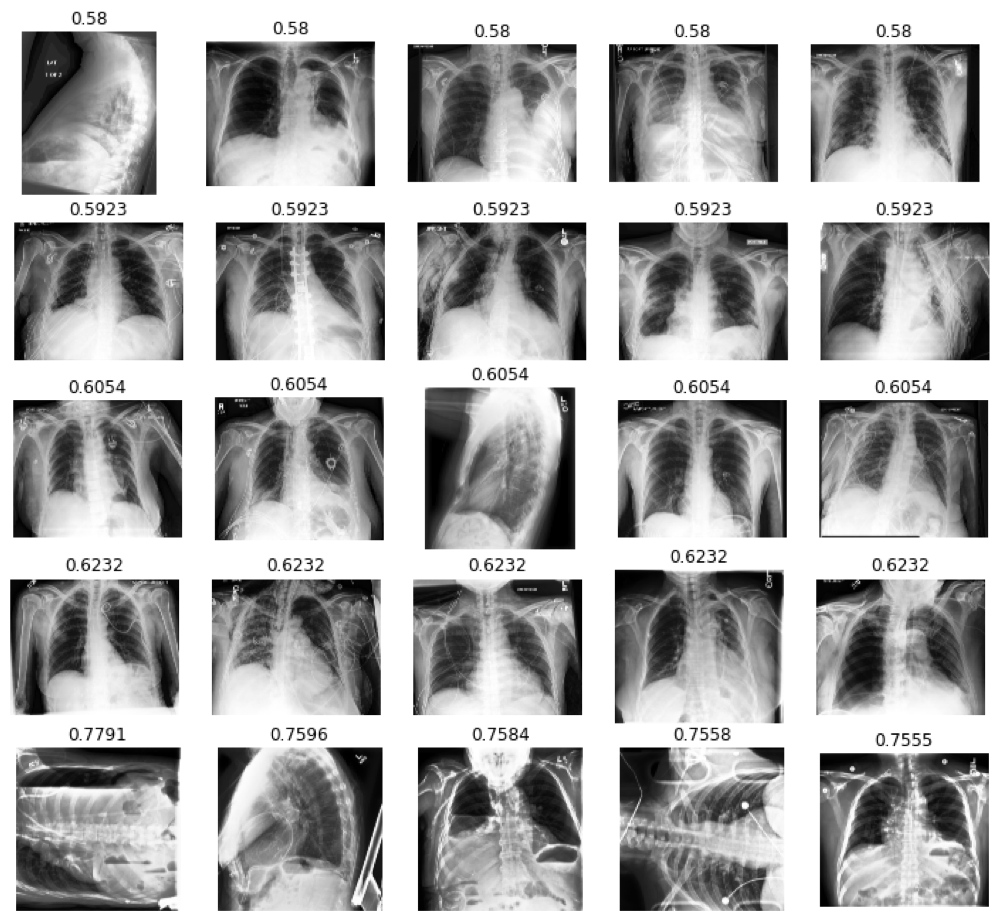}
  \caption{Pneumothorax}
\end{subfigure}
\hspace*{\fill}
\begin{subfigure}{.5\textwidth}
\includegraphics[width=0.95\linewidth, height=\linewidth, frame]{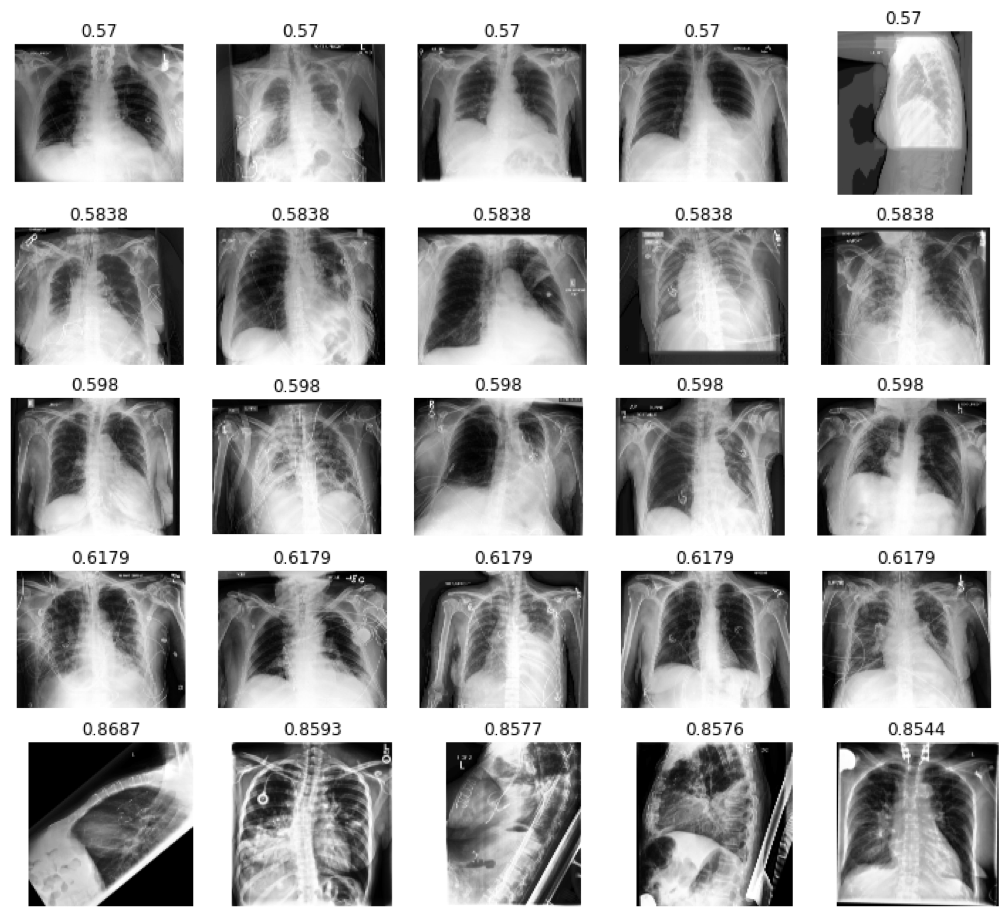}
\caption{Pleural Effusion}
\end{subfigure}
\begin{subfigure}{.5\textwidth}
  \includegraphics[width=0.95\linewidth, height=\linewidth, frame]{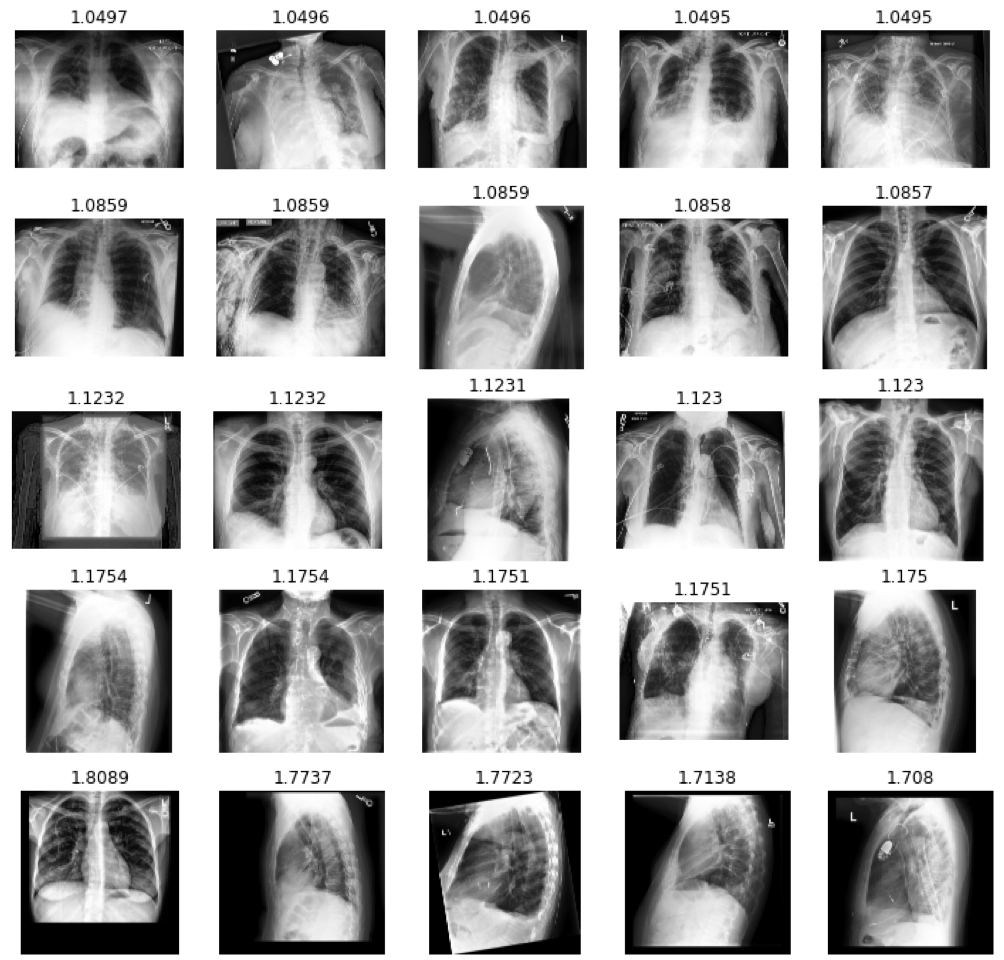}
  \caption{Pleural Other}
\end{subfigure}
\hspace*{\fill}
\begin{subfigure}{.5\textwidth}
\includegraphics[width=0.95\linewidth, height=\linewidth, frame]{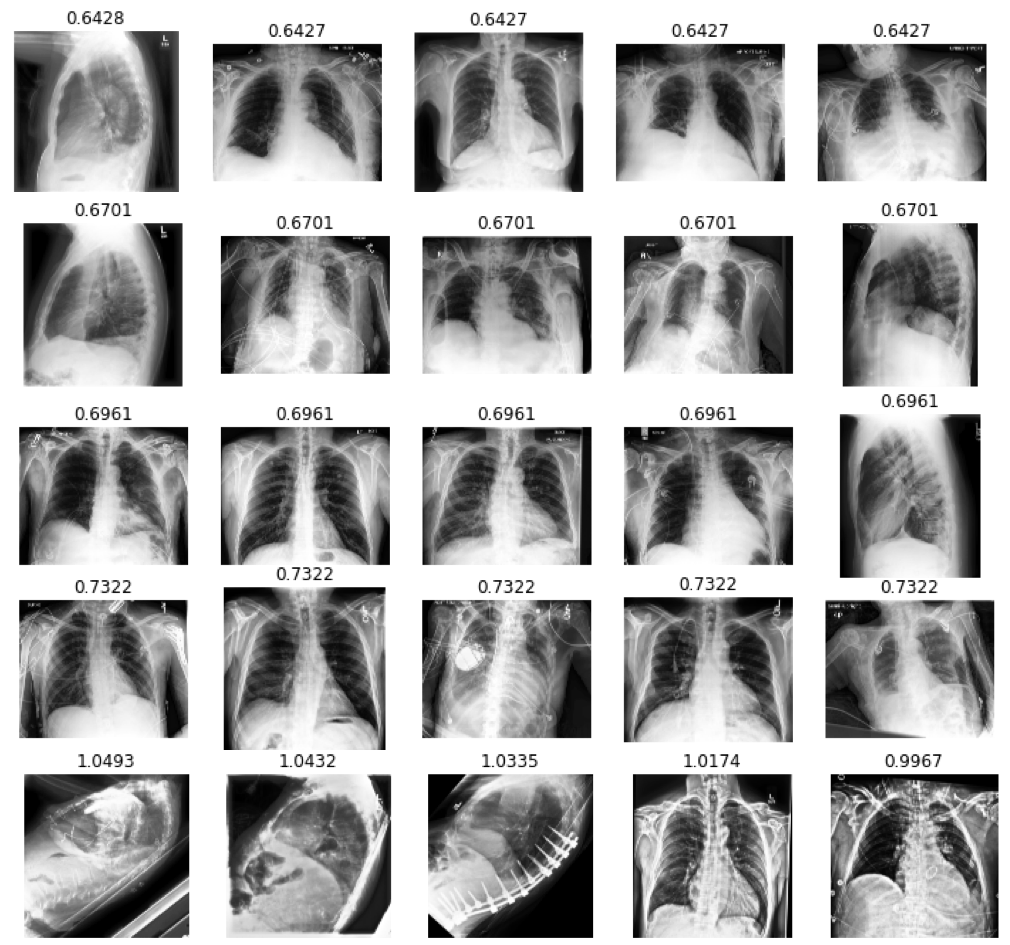}
\caption{Fracture}
\end{subfigure}
\caption{More \textit{MedShift\_w\_CVAD} CheXpert clustering results for classes (a) \textit{Pneumothorax}; (b) \textit{Pleural Effusion}; (c) \textit{Pleural Other} and (d) \textit{Fracture} following the same arrangement style of Fig.~\ref{fig:mura_clustering_all_1}. }
\label{fig:chest_clustering3}
\end{figure*}

\begin{figure*}[tp]
\begin{subfigure}{.5\textwidth}
  \includegraphics[width=0.95\linewidth, height=\linewidth, frame]{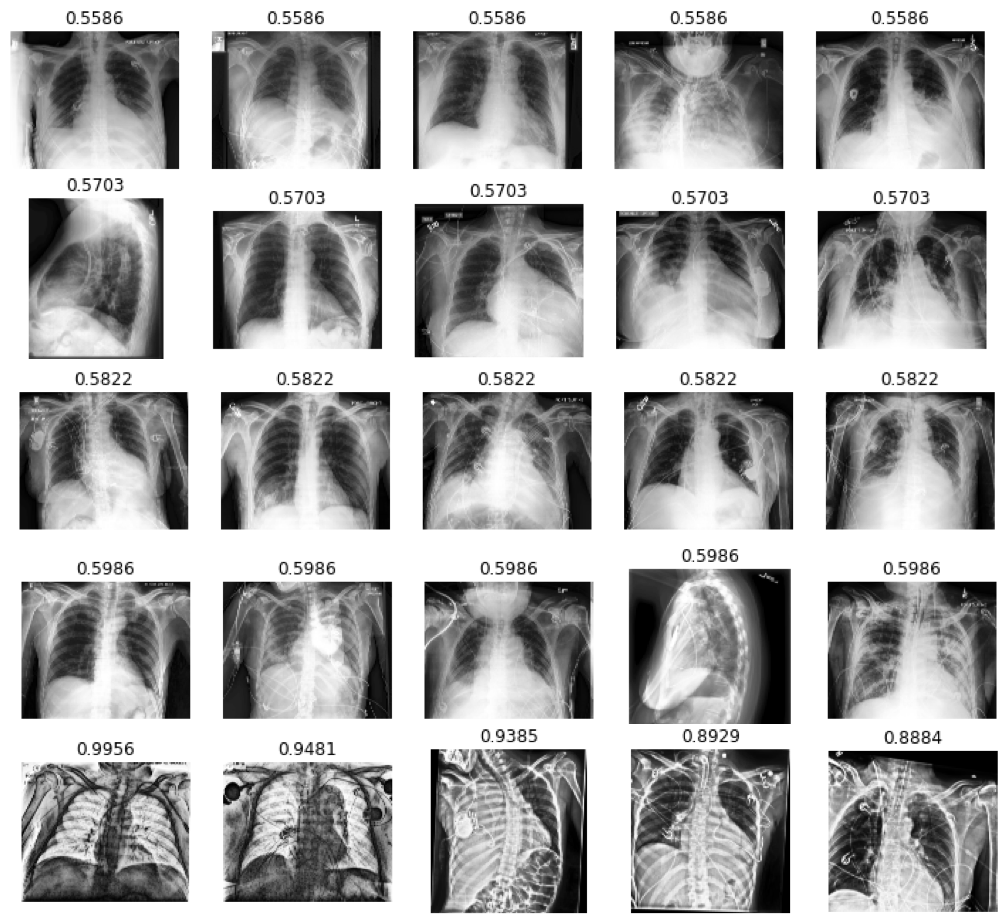}
  \caption{Support Devices}
\end{subfigure}
\caption{More \textit{MedShift\_w\_CVAD} CheXpert clustering results for classes (a) \textit{Support Devices} following the same arrangement style of Fig.~\ref{fig:mura_clustering_all_1}. }
\label{fig:chest_clustering4}
\end{figure*}

\clearpage
\subsubsection{Clustering Results for MIMIC}
\begin{figure*}[htp]
\vspace{-2mm}
\begin{subfigure}{0.5\textwidth}
  \includegraphics[width=0.95\linewidth, height=0.88\linewidth, frame]{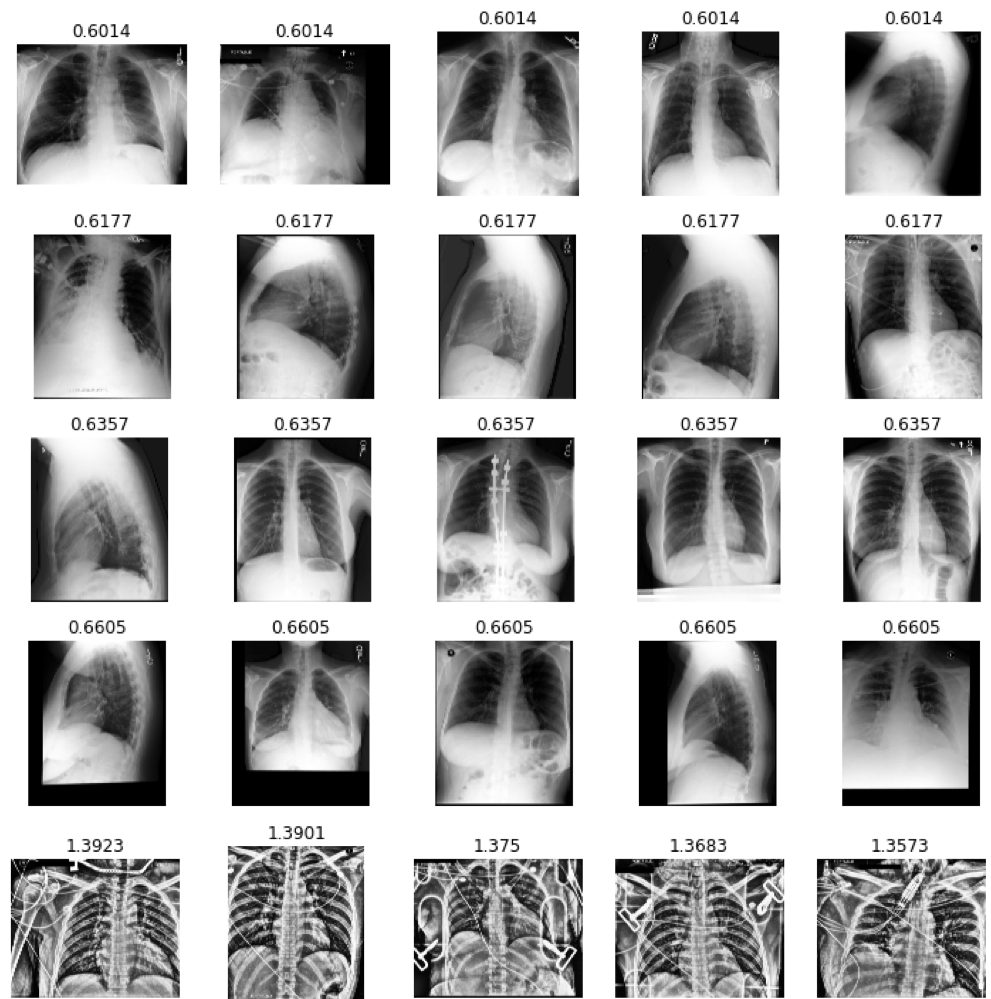}
  \caption{No Finding}
\end{subfigure}%
\hspace*{\fill}
\begin{subfigure}{0.5\textwidth}
  \includegraphics[width=0.95\linewidth, height=0.88\linewidth, frame]{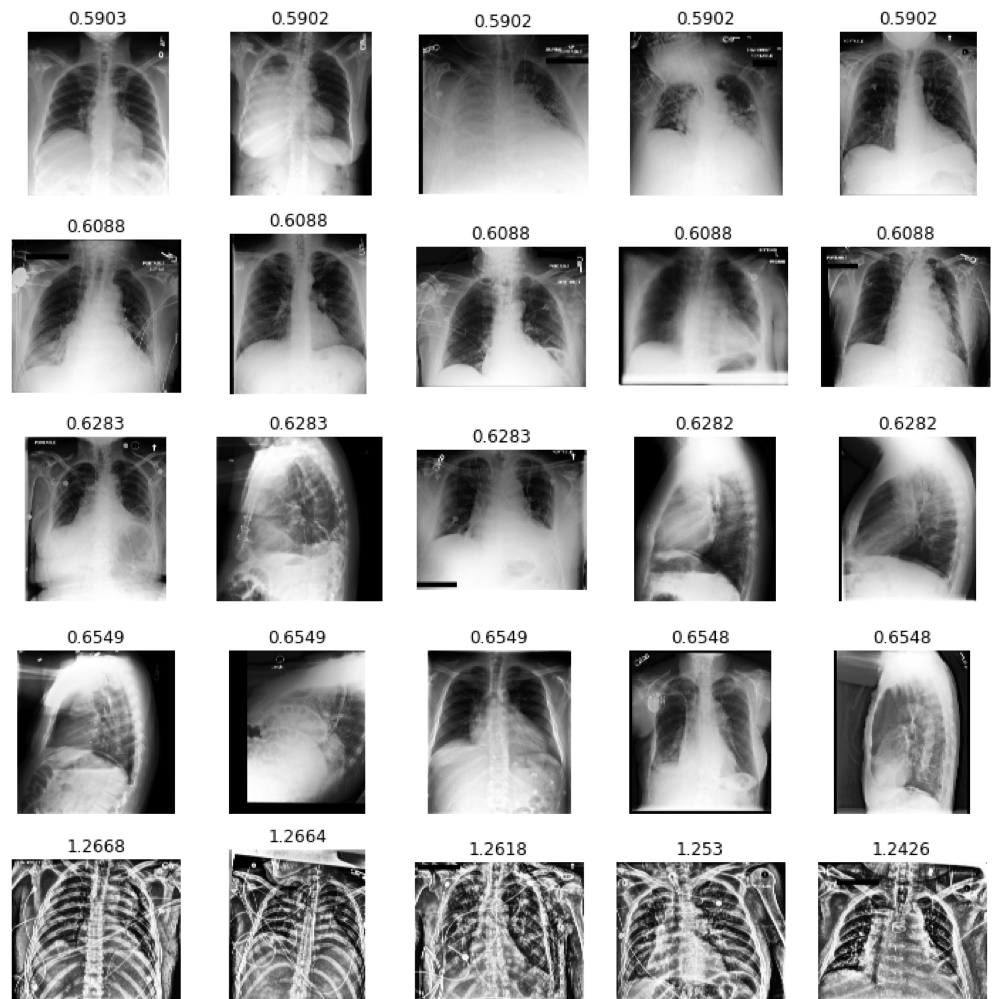}
  \caption{Enlarge Cardiomediastinum}
\end{subfigure}
\begin{subfigure}{0.5\textwidth}
  \includegraphics[width=0.95\linewidth, height=0.88\linewidth, frame]{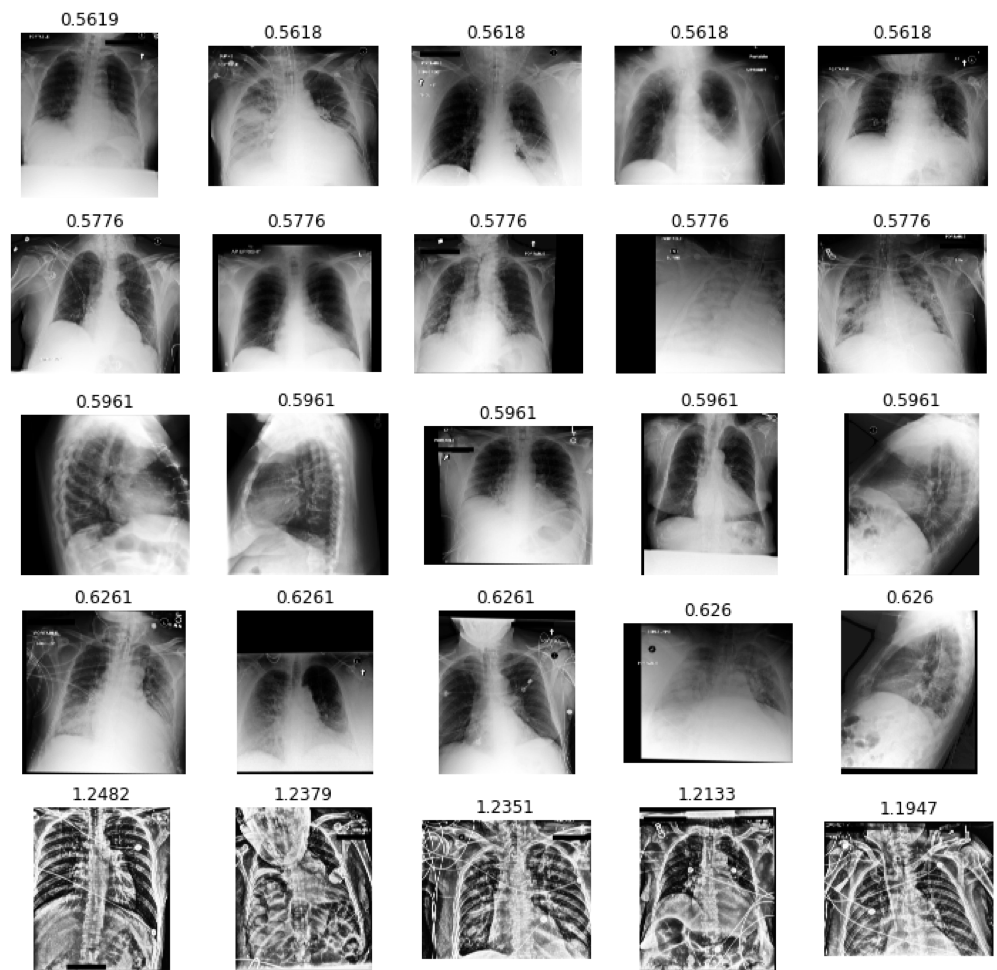}
  \caption{Cardiomegaly}
\end{subfigure}%
\hspace*{\fill}
\begin{subfigure}{0.5\textwidth}
  \includegraphics[width=0.95\linewidth, height=0.88\linewidth, frame]{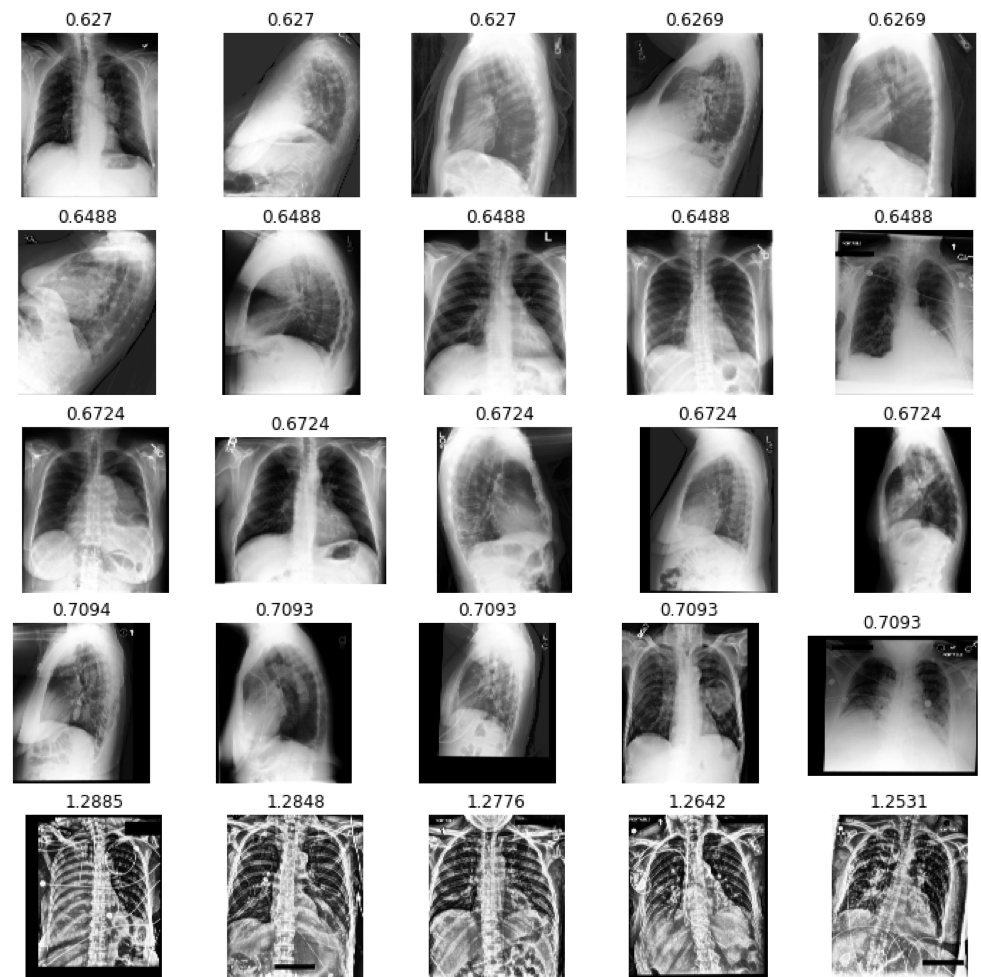}
  \caption{Lung Lesion}
\end{subfigure}
\caption{\textit{MedShift\_w\_CVAD} MIMIC clustering results for classes (a) \textit{No Finding}; (b) \textit{Enlarge Cardiomediastinum}; (c) \textit{Cardiomegaly} and (d) \textit{Lung Lesion} following the same arrangement style of Fig.~\ref{fig:mura_clustering_all_1}. }
\vspace{-15mm}
\label{fig:mimic_clustering1}
\end{figure*}


\begin{figure*}[htp]
\begin{subfigure}{0.5\textwidth}
\includegraphics[width=0.95\linewidth, height=\linewidth, frame]{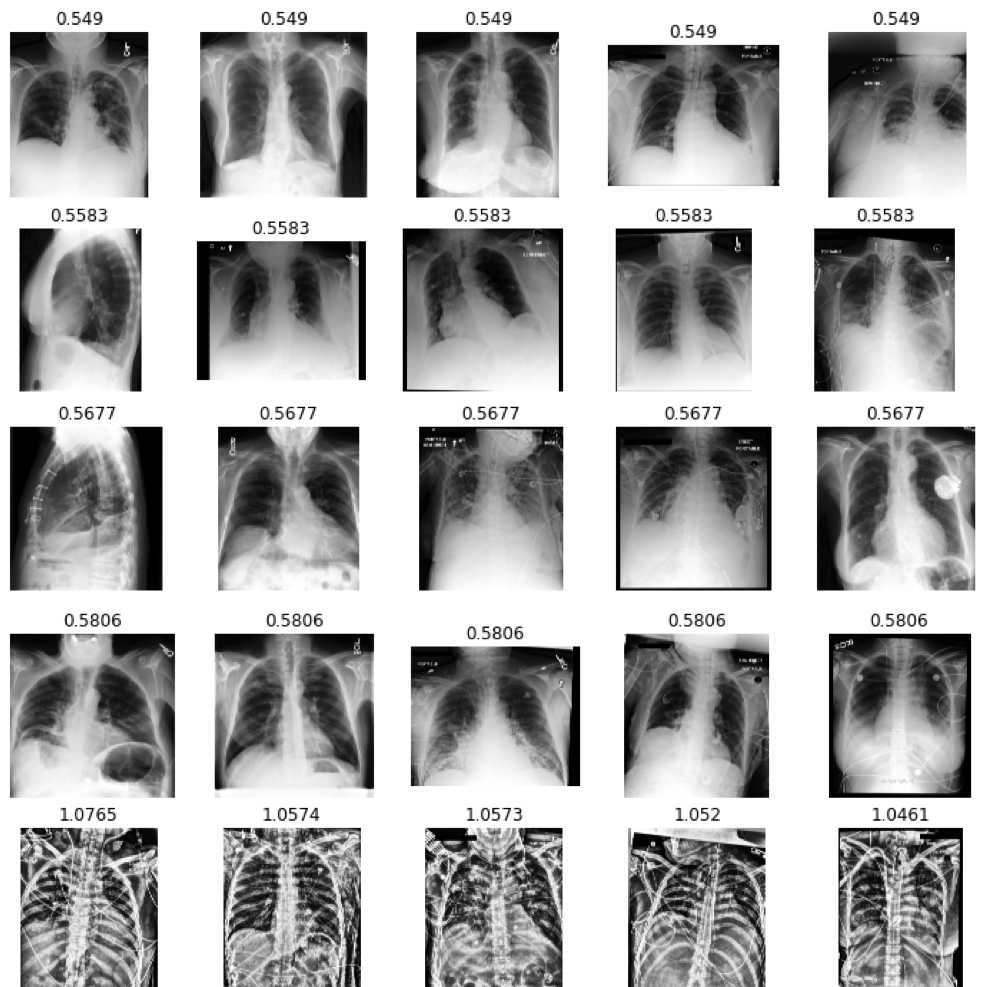}
\caption{Lung Opacity}
\end{subfigure}
\hspace*{\fill}
\begin{subfigure}{0.5\textwidth}
  \includegraphics[width=0.95\linewidth, height=\linewidth, frame]{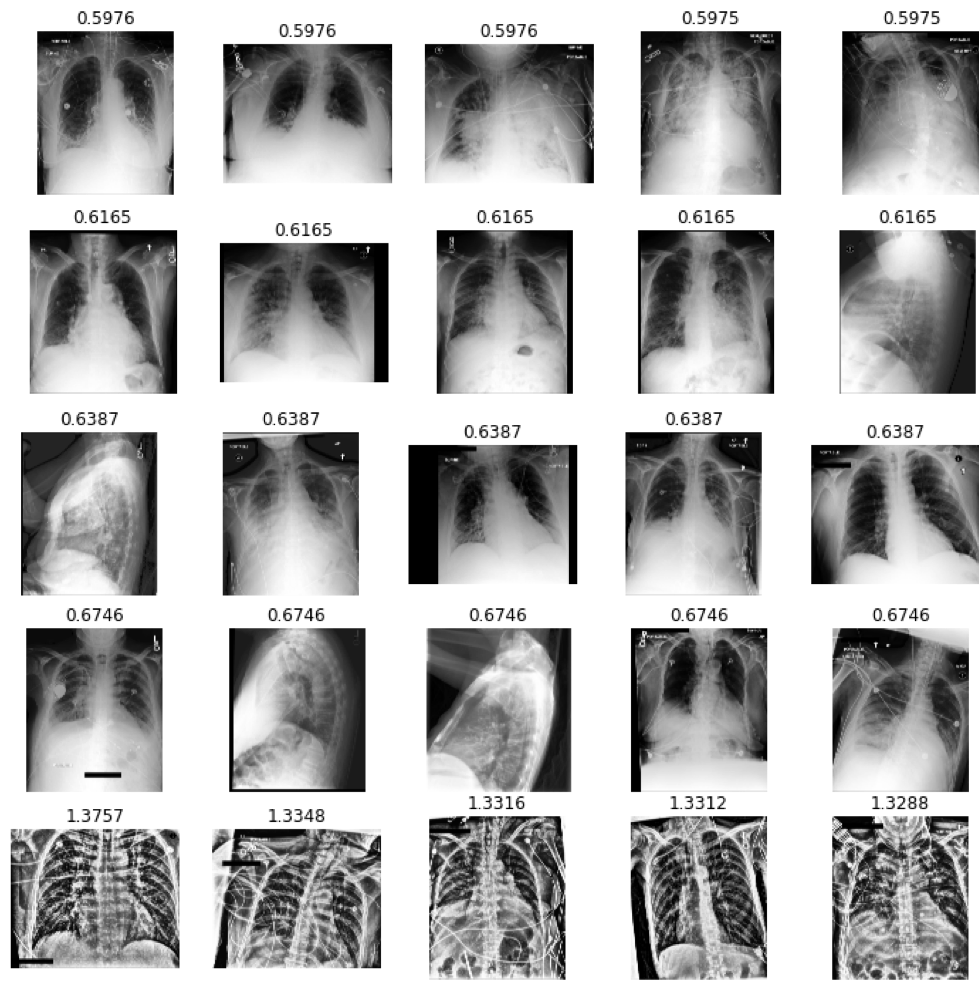}
  \caption{Edema}
\end{subfigure}
\begin{subfigure}{0.5\textwidth}
  \includegraphics[width=0.95\linewidth, height=\linewidth, frame]{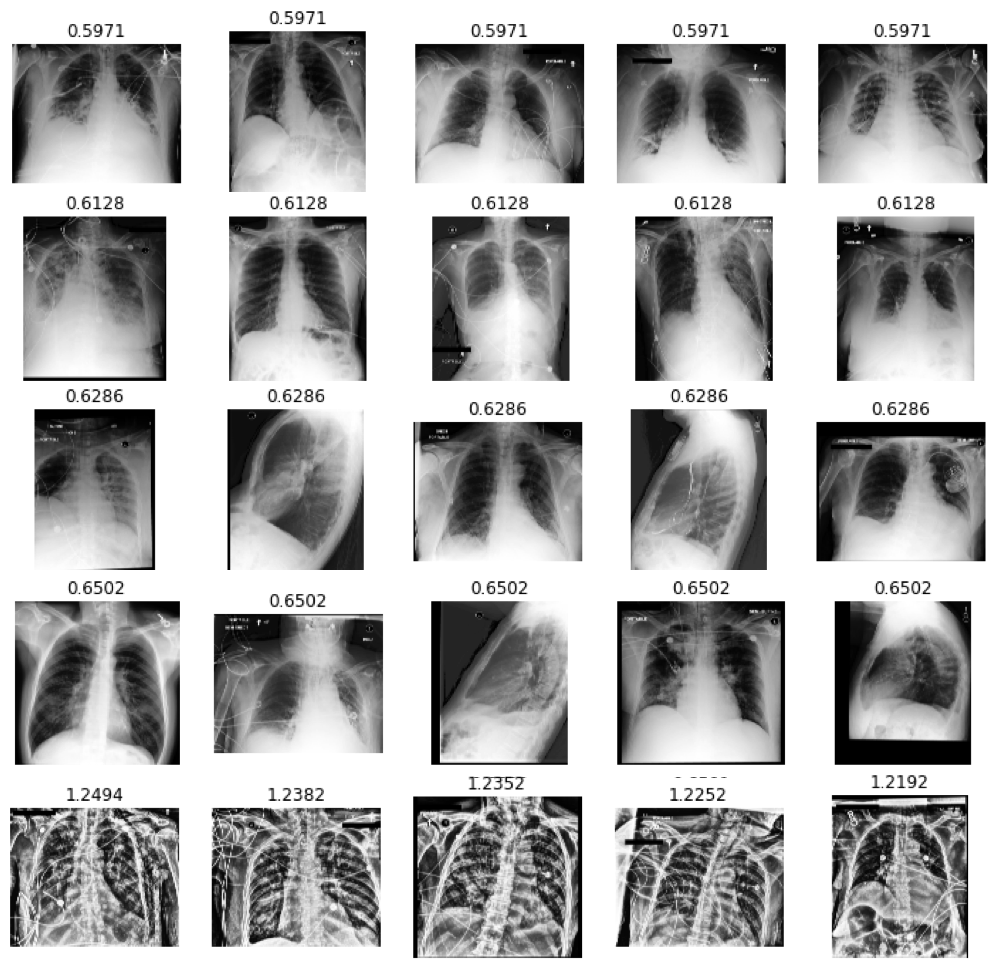}
  \caption{Consolidation}
\end{subfigure}%
\hspace*{\fill}
\begin{subfigure}{0.5\textwidth}
  \includegraphics[width=0.95\linewidth, height=\linewidth, frame]{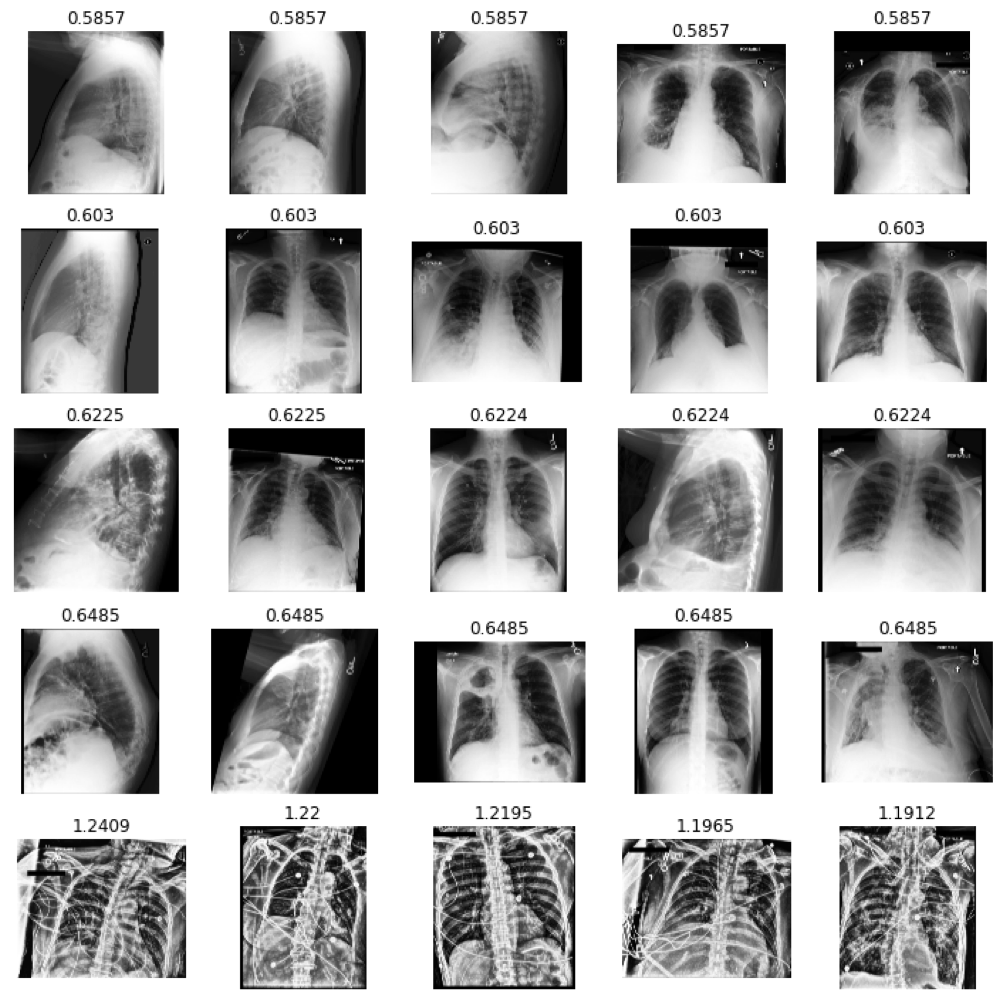}
  \caption{Pneumonia}
\end{subfigure}
\caption{\textit{MedShift\_w\_CVAD} MIMIC clustering results for classes (a) \textit{Lung Opacity}; (b) \textit{Edema}; (c) \textit{Consolidation} and (d) \textit{Pneumonia} following the same arrangement style of Fig.~\ref{fig:mura_clustering_all_1}. }
\label{fig:mimic_clustering2}
\end{figure*}

\begin{figure*}[htp]
\begin{subfigure}{0.5\textwidth}
  \includegraphics[width=0.95\linewidth, height=\linewidth, frame]{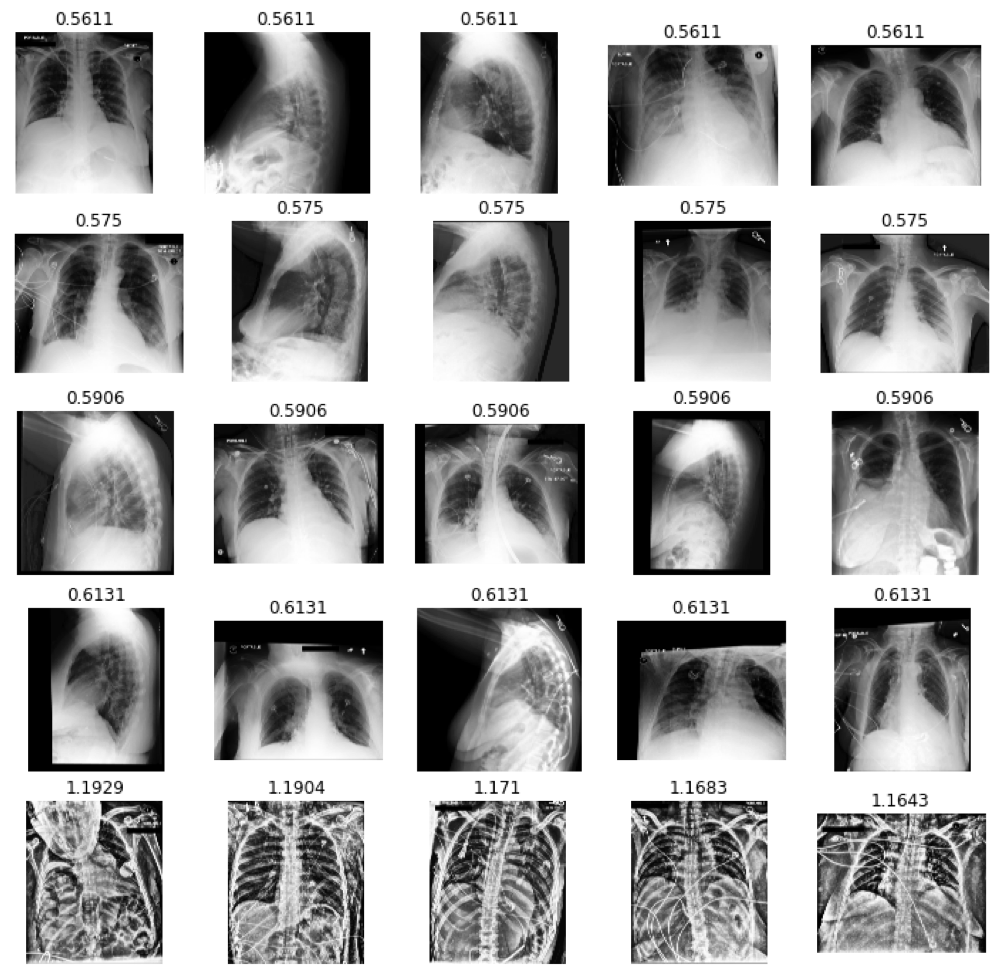}
  \caption{Atelectasis}
\end{subfigure}%
\hspace*{\fill}
\begin{subfigure}{0.5\textwidth}
  \includegraphics[width=0.95\linewidth, height=\linewidth, frame]{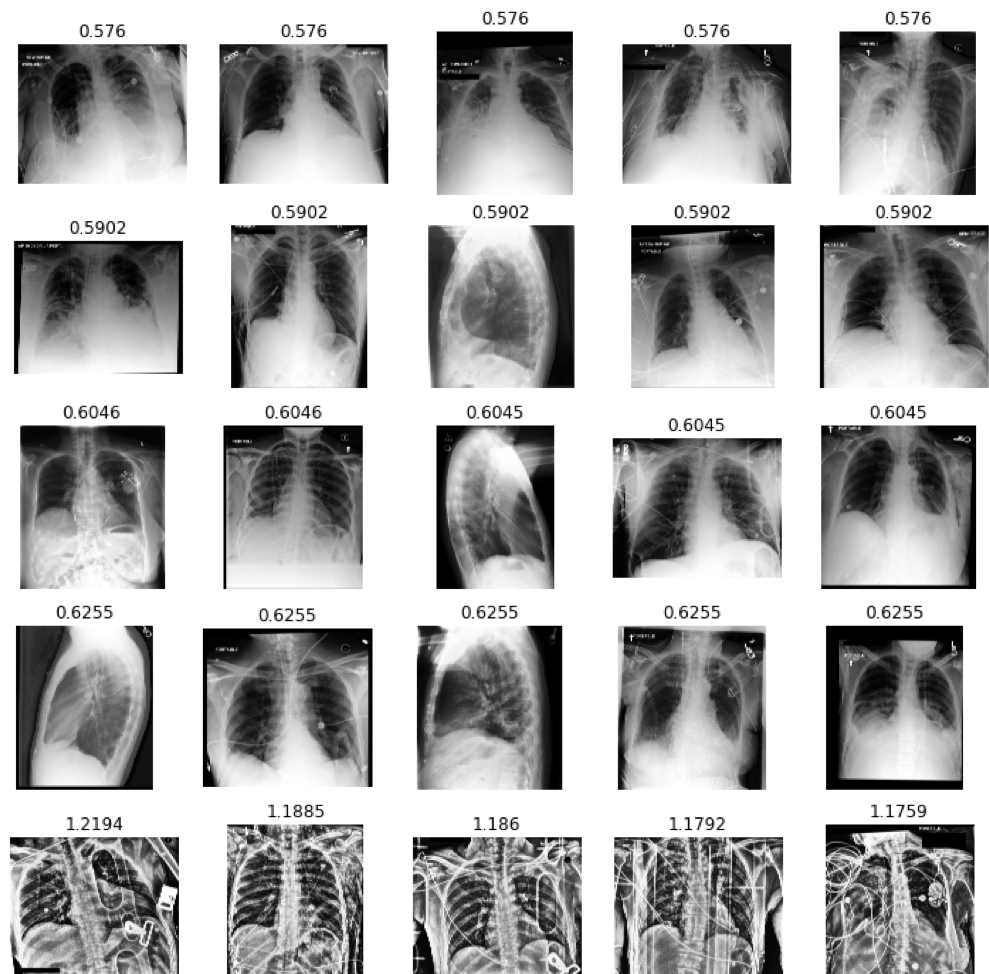}
  \caption{Pneumothorax}
\end{subfigure}
\begin{subfigure}{0.5\textwidth}
\includegraphics[width=0.95\linewidth, height=\linewidth, frame]{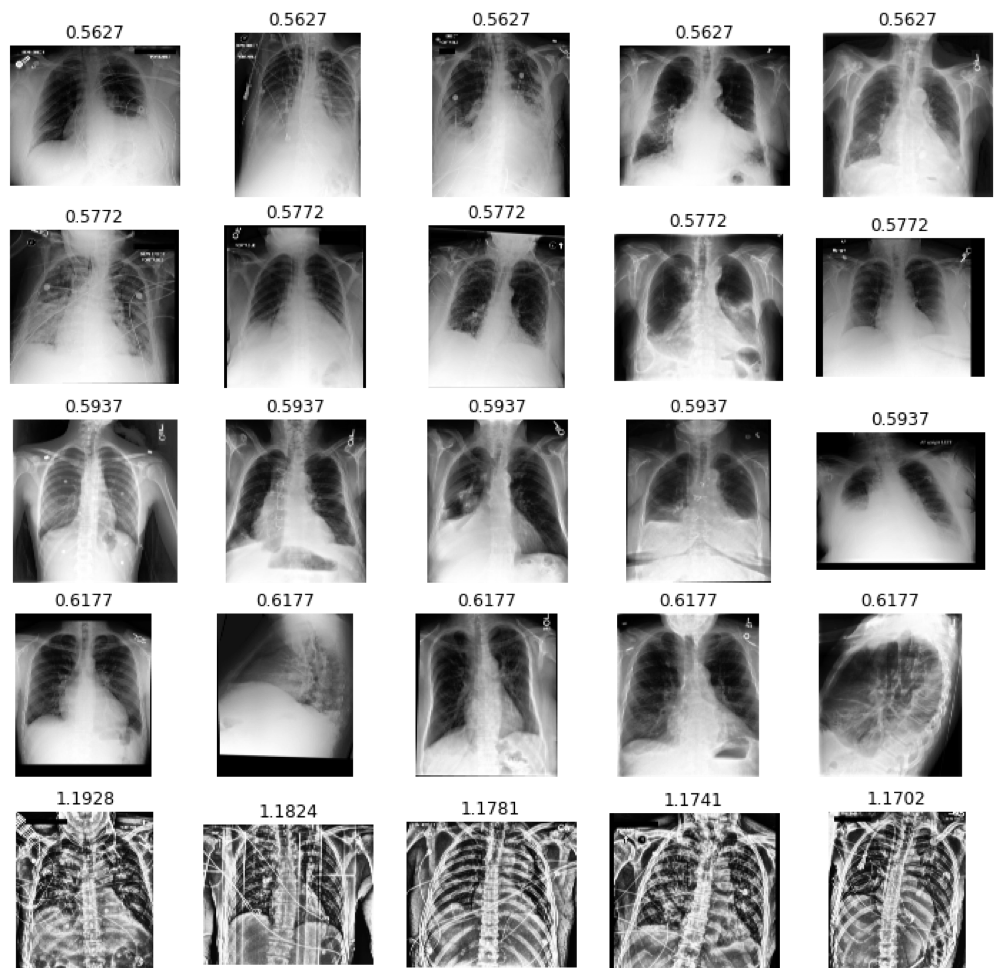}
\caption{Pleural Effusion}
\end{subfigure}
\hspace*{\fill}
\begin{subfigure}{0.5\textwidth}
  \includegraphics[width=0.95\linewidth, height=\linewidth, frame]{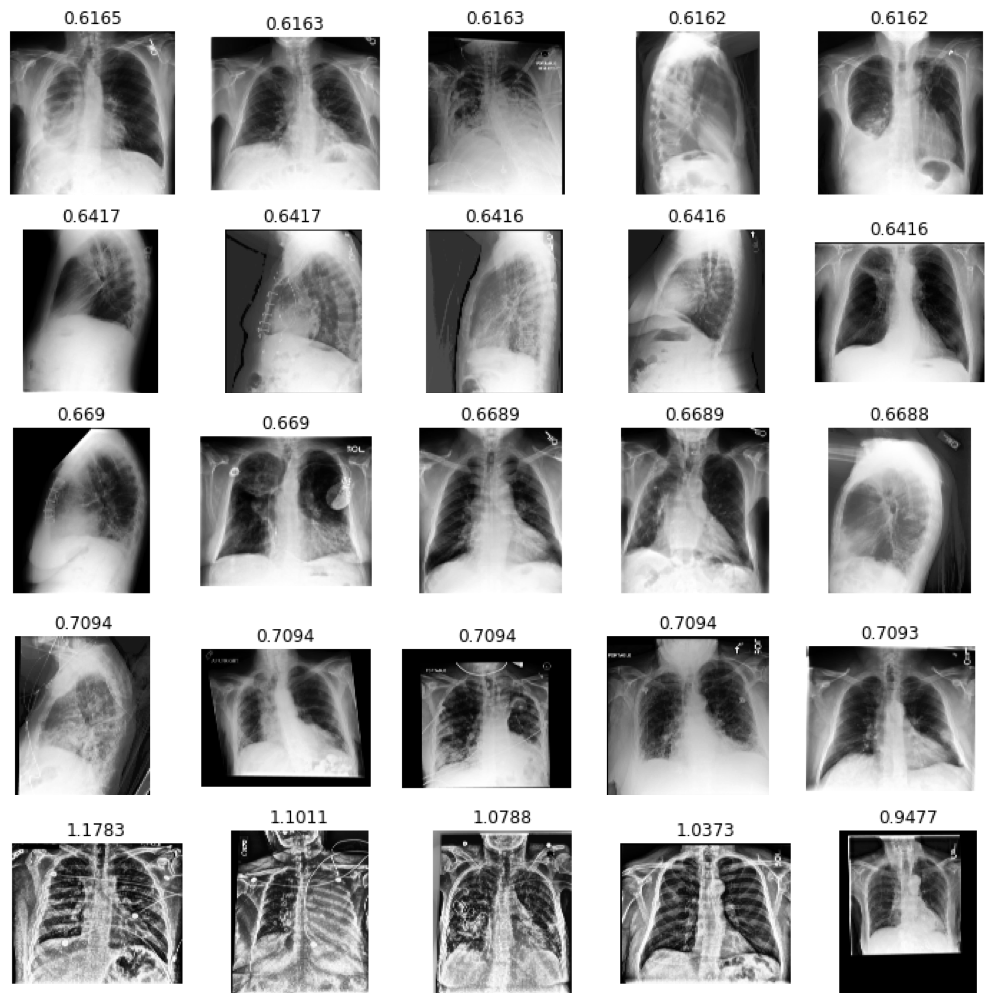}
  \caption{Pleural Other}
\end{subfigure}
\caption{\textit{MedShift\_w\_CVAD} MIMIC clustering results for classes (a) \textit{Atelectasis}; (b) \textit{Pneumothorax}; (c) \textit{Pleural Effusion} and (d) \textit{Pleural Other} following the same arrangement style of Fig.~\ref{fig:mura_clustering_all_1}. }
\label{fig:mimic_clustering3}
\end{figure*}

\begin{figure*}[tp]
\begin{subfigure}{0.5\textwidth}
\includegraphics[width=0.95\linewidth, height=\linewidth, frame]{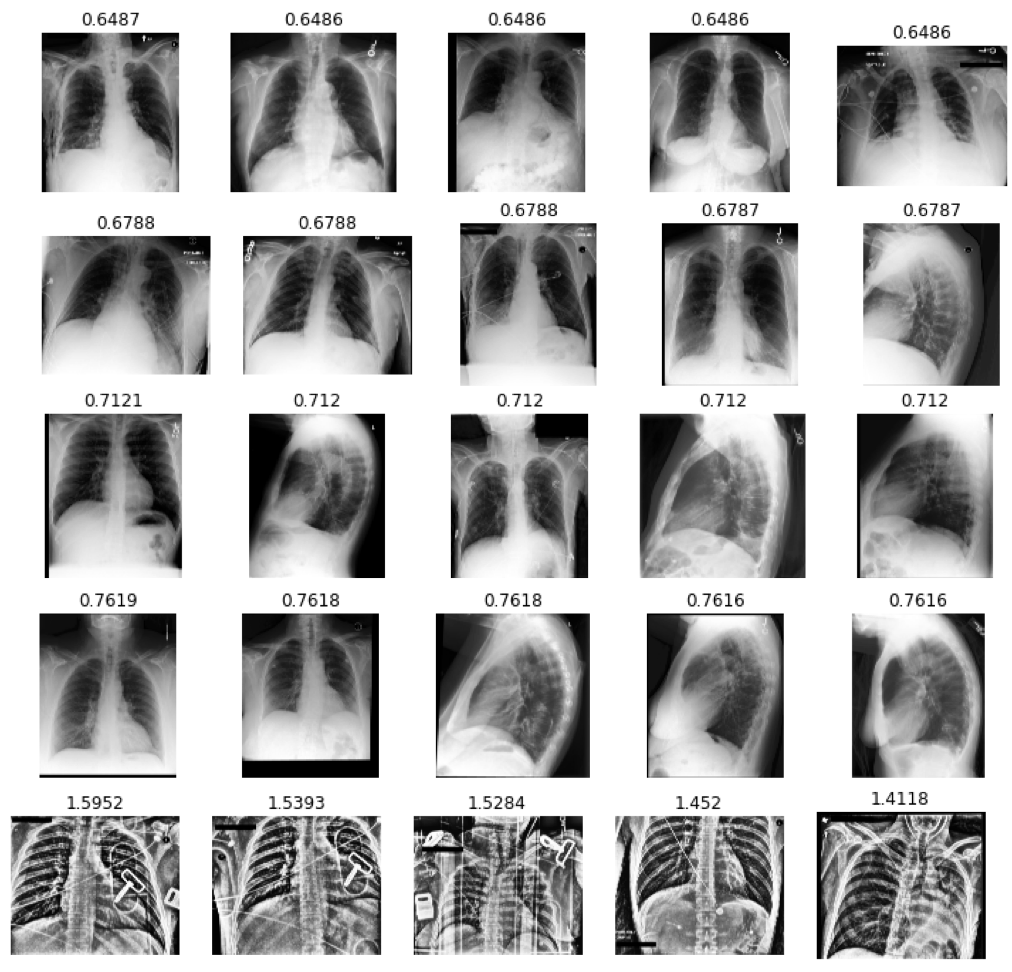}
\caption{Fracture}
\end{subfigure}
\hspace*{\fill}
\begin{subfigure}{0.5\textwidth}
  \includegraphics[width=0.95\linewidth, height=\linewidth, frame]{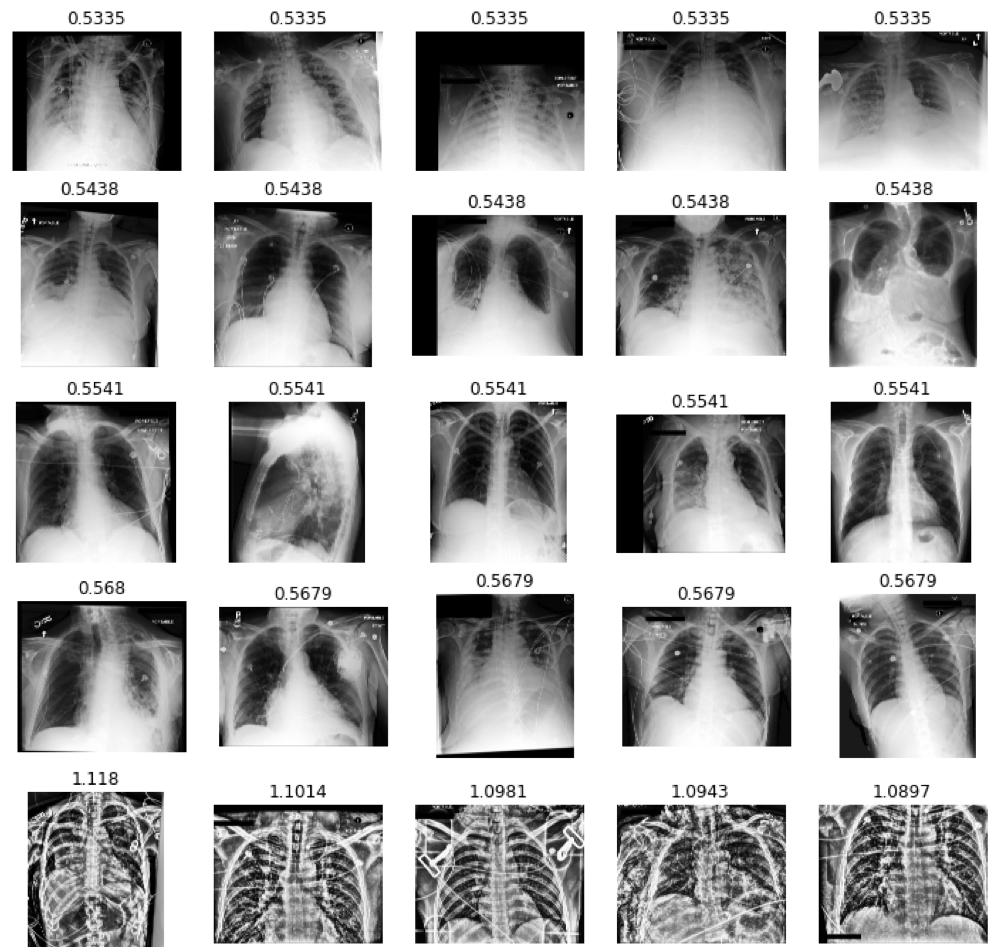}
  \caption{Support Devices}
\end{subfigure}
\caption{\textit{MedShift\_w\_CVAD} MIMIC clustering results for classes (a) \textit{Fracture}; (b) \textit{Support Devices} following the same arrangement style of Fig.~\ref{fig:mura_clustering_all_1}. }
\label{fig:mimic_clustering4}
\end{figure*}

\clearpage
\begin{landscape}
\subsection{Classification Results}~\label{chest_classification_more}

\begin{table}[htp]
\centering
\caption{Chest X-ray data classification class-wise results with CVAD (left) and f-AnoGAN (right) architectures.}
\resizebox{1.25\textwidth}{!}{%
\begin{tabular}{ll|l|cccccccccccccc|cccc}
\hline
\multicolumn{2}{l|}{\multirow{2}{*}{\textbf{Dataset}}} & \multirow{2}{*}{\textbf{Metrics}} & \multirow{2}{*}{\textbf{No Finding}} & \multirow{2}{*}{\textbf{\begin{tabular}[c]{@{}c@{}}Enlarged \\ Cardiomediastinum\end{tabular}}} & \multirow{2}{*}{\textbf{Cardiomegaly}} & \multirow{2}{*}{\textbf{\begin{tabular}[c]{@{}c@{}}Lung\\  Lesion\end{tabular}}} & \multirow{2}{*}{\textbf{\begin{tabular}[c]{@{}c@{}}Lung \\ Opacity\end{tabular}}} & \multirow{2}{*}{\textbf{Edema}} & \multirow{2}{*}{\textbf{Consolidation}} & \multirow{2}{*}{\textbf{Pneumonia}} & \multirow{2}{*}{\textbf{Atelectasis}} & \multirow{2}{*}{\textbf{Pneumothorax}} & \multirow{2}{*}{\textbf{\begin{tabular}[c]{@{}c@{}}Pleural \\ Effusion\end{tabular}}} & \multirow{2}{*}{\textbf{\begin{tabular}[c]{@{}c@{}}Pleural \\ Other\end{tabular}}} & \multirow{2}{*}{\textbf{Fracture}} & \multirow{2}{*}{\textbf{\begin{tabular}[c]{@{}c@{}}Support\\  Devices\end{tabular}}} & \multicolumn{4}{c}{\textbf{AVERAGE}} \\
\multicolumn{2}{l|}{} &  &  &  &  &  &  &  &  &  &  &  &  &  &  &  & \textbf{Micro} & \textbf{Macro} & \textbf{Weighted} & \textbf{Samples} \\ \hline

\multicolumn{1}{l|}{\multirow{5}{*}{\textbf{Emory\_CXR}}} 
& \multirow{5}{*}{\textbf{Test}} & \textit{\#images} & 7,936 & 522 & 1,256 & 397 & 2,141 & 830 & 151 & 439 & 2,315 & 150 & 711 & 98 & 177 & 9,994 &  &  &  &  \\
\multicolumn{1}{l|}{} &  & \textit{Precision} & 0.753 & 0.082 & 0.385 & 0.090 & 0.531 & 0.416 & 0.107 & 0.153 & 0.577 & 0.066 & 0.516 & 0.066 & 0.058 & 0.93 & 0.489 & 0.338 & 0.705 & 0.633 \\
\multicolumn{1}{l|}{} &  & \textit{Recall} & 0.960 & 0.232 & 0.697 & 0.446 & 0.588 & 0.627 & 0.437 & 0.490 & 0.580 & 0.433 & 0.686 & 0.520 & 0.277 & 0.109 & 0.514 & 0.506 & 0.514 & 0.537 \\
\multicolumn{1}{l|}{} &  & \textit{F1-score} & 0.844 & 0.121 & 0.496 & 0.150 & 0.558 & 0.500 & 0.171 & 0.233 & 0.579 & 0.115 & 0.589 & 0.118 & 0.096 & 0.195 & 0.502 & 0.340 & 0.477 & 0.54 \\
\multicolumn{1}{l|}{} &  & \textit{AUC} & 0.724 & 0.561 & 0.788 & 0.652 & 0.742 & 0.783 & 0.697 & 0.697 & 0.743 & 0.681 & 0.824 & 0.732 & 0.607 & 0.540 & 0.710 & 0.698 & 0.719 & 0.661 \\ \hline

\multicolumn{1}{l|}{\multirow{10}{*}{\textbf{TOP 5}}} & \multirow{5}{*}{\textbf{CheXpert}} & \textit{\#images} & 22,381 & 10,798 & 27,000 & 9,186 & 10,5581 & 52,246 & 14,783 & 6,039 & 33,376 & 19,448 & 86,187 & 3,523 & 9,040 & 116,001 & \multicolumn{4}{c}{212,273} \\
\multicolumn{1}{l|}{} &  & \textit{Precision} & 0.356 & 0.098 & 0.319 & 0.081 & 0.581 & 0.462 & 0.122 & 0.055 & 0.196 & 0.118 & 0.623 & 0.033 & 0.091 & 0.670 & 0.343 & 0.272 & 0.479 & 0.351 \\
\multicolumn{1}{l|}{} &  & \textit{Recall} & 0.684 & 0.073 & 0.610 & 0.610 & 0.832 & 0.625 & 0.556 & 0.495 & 0.818 & 0.884 & 0.835 & 0.461 & 0.173 & 0.845 & 0.752 & 0.607 & 0.752 & 0.722 \\
\multicolumn{1}{l|}{} &  & \textit{F1-score} & 0.468 & 0.084 & 0.419 & 0.143 & 0.684 & 0.531 & 0.200 & 0.098 & 0.316 & 0.208 & 0.714 & 0.061 & 0.120 & 0.747 & 0.471 & 0.342 & 0.566 & 0.446 \\
\multicolumn{1}{l|}{} &  & \textit{AUC} & 0.754 & 0.512 & 0.651 & 0.638 & 0.654 & 0.746 & 0.662 & 0.603 & 0.694 & 0.697 & 0.775 & 0.625 & 0.533 & 0.756 & 0.726 & 0.664 & 0.724 & 0.709 \\ \cline{2-21} 
\multicolumn{1}{l|}{} & \multirow{5}{*}{\textbf{MIMIC}} & \textit{\#images} & 143,352 & 10,042 & 64,346 & 10,801 & 76,423 & 36,564 & 14,675 & 26,222 & 65,047 & 14,257 & 76,957 & 3,460 & 7,605 & 84,073 &  &  &  &  \\
\multicolumn{1}{l|}{} &  & \textit{Precision} & 0.674 & 0.040 & 0.412 & 0.060 & 0.318 & 0.358 & 0.096 & 0.133 & 0.315 & 0.084 & 0.564 & 0.028 & 0.036 & 0.502 & 0.315 & 0.259 & 0.433 & 0.4 \\
\multicolumn{1}{l|}{} &  & \textit{Recall} & 0.754 & 0.078 & 0.440 & 0.539 & 0.740 & 0.619 & 0.543 & 0.428 & 0.754 & 0.719 & 0.700 & 0.379 & 0.158 & 0.739 & 0.661 & 0.542 & 0.661 & 0.661 \\
\multicolumn{1}{l|}{} &  & \textit{F1-score} & 0.712 & 0.053 & 0.425 & 0.108 & 0.445 & 0.454 & 0.164 & 0.202 & 0.445 & 0.151 & 0.624 & 0.052 & 0.058 & 0.598 & 0.427 & 0.321 & 0.504 & 0.464 \\
\multicolumn{1}{l|}{} &  & \textit{AUC} & 0.757 & 0.515 & 0.656 & 0.639 & 0.650 & 0.764 & 0.661 & 0.602 & 0.693 & 0.693 & 0.786 & 0.631 & 0.531 & 0.759 & 0.731 & 0.667 & 0.727 & 0.712 \\ \hline

\multicolumn{1}{l|}{\multirow{10}{*}{\textbf{TOP 4}}} & \multirow{5}{*}{\textbf{CheXpert}} & \textit{\#images} & 17,834 / 4,845 & 8,899 / 4,282 & 22,846 / 11,259 & 7,518 / 2,609 & 88,278 / 39,924 & 45,239 / 21,838 & 12,217 / 5,845 & 4,934 / 2,106 & 28,414 / 13,719 & 15,609 / 6,520 & 73,595 / 35,301 & 3,141 / 1,079 & 7,464 / 3,325 & 97,280 / 45,373 & \multicolumn{4}{c}{173,664 / 68,757} \\
\multicolumn{1}{l|}{} &  & \textit{Precision} & 0.382 / 0.327 & 0.099 / 0.112 & 0.315 / 0.369 & 0.085 / 0.078 & 0.583 / 0.645 & 0.463 / 0.528 & 0.119 / 0.138 & 0.054 / 0.059 & 0.199 / 0.234 & 0.115 / 0.120 & 0.626 / 0.693 & 0.037 / 0.031 & 0.098 / 0.120 & 0.670 / 0.759 & 0.351 / 0.395 & 0.275 / 0.301 & 0.483 / 0.545 & 0.358 / 0.399 \\
\multicolumn{1}{l|}{} &  & \textit{Recall} & 0.640 / 0.638 & 0.071 / 0.064 & 0.624 / 0.631 & 0.602 / 0.594 & 0.848 / 0.853 & 0.647 / 0.660 & 0.553 / 0.562 & 0.505 / 0.519 & 0.833 / 0.848 & 0.884 / 0.902 & 0.855 / 0.858 & 0.469 / 0.450 & 0.157 / 0.146 & 0.864 / 0.865 & 0.765 / 0.772 & 0.611 / 0.614 & 0.765 / 0.772 & 0.737 / 0.750 \\
\multicolumn{1}{l|}{} &  & \textit{F1-score} & 0.478 / 0.433 & 0.083 / 0.082 & 0.419 / 0.466 & 0.150 / 0.138 & 0.691 / 0.735 & 0.540 / 0.587 & 0.196 / 0.222 & 0.098 / 0.106 & 0.321 / 0.366 & 0.204 / 0.212 & 0.723 / 0.767 & 0.069 / 0.059 & 0.121 / 0.132 & 0.755 / 0.808 & 0.481 / 0.522 & 0.346 / 0.365 & 0.573 / 0.618 & 0.457 / 0.498 \\
\multicolumn{1}{l|}{} &  & \textit{AUC} & 0.760 / 0.769 & 0.518 / 0.515 & 0.709 / 0.710 & 0.655 / 0.659 & 0.611 / 0.602 & 0.691 / 0.693 & 0.622 / 0.618 & 0.623 / 0.629 & 0.589 / 0.577 & 0.607 / 0.605 & 0.740 / 0.728 & 0.623 / 0.614 & 0.546 / 0.546 & 0.661 / 0.666 & 0.729 / 0.733 & 0.640 / 0.638 & 0.715 / 0.722 & 0.661 /0.656 \\ \cline{2-21} 
\multicolumn{1}{l|}{} & \multirow{5}{*}{\textbf{MIMIC}} & \textit{\#images} & 115,457 / 25,060 & 8,503 / 4,053 & 53,833 / 24,040 & 9,033 / 2,629 & 63,674 / 26,173 & 30,636 / 14,797 & 12,159 / 6,146 & 21,621 / 8,548 & 54,894 / 25,083 & 11,529 / 4,943 & 64,695 / 30,863 & 2,915 / 1,192 & 6,223 / 2,236 & 70,786 / 34,346 & \multicolumn{4}{c}{290,657 / 89,546} \\
\multicolumn{1}{l|}{} &  & \textit{Precision} & 0.675 / 0.634 & 0.042 / 0.057 & 0.413 / 0.503 & 0.062 / 0.062 & 0.322 / 0.386 & 0.359 / 0.422 & 0.097 / 0.135 & 0.133 / 0.163 & 0.323 / 0.407 & 0.084 / 0.107 & 0.565 / 0.653 & 0.029 / 0.039 & 0.036 / 0.048 & 0.506 / 0.643 & 0.319 / 0.374 & 0.260 / 0.304 & 0.433 / 0.476 & 0.399 / 0.421 \\
\multicolumn{1}{l|}{} &  & \textit{Recall} & 0.749 / 0.726 & 0.082 / 0.070 & 0.451 / 0.472 & 0.540 / 0.510 & 0.739 / 0.786 & 0.645 / 0.675 & 0.542 / 0.556 & 0.431 / 0.471 & 0.756 / 0.809 & 0.704 / 0.715 & 0.716 / 0.750 & 0.381 / 0.400 & 0.158 / 0.140 & 0.749 / 0.764 & 0.666 / 0.681 & 0.546 / 0.560 & 0.666 / 0.681 & 0.663 / 0.673 \\
\multicolumn{1}{l|}{} &  & \textit{F1-score} & 0.710 / 0.677 & 0.056 / 0.063 & 0.431 / 0.487 & 0.111 / 0.111 & 0.449 / 0.518 & 0.461 / 0.519 & 0.165 / 0.217 & 0.203 / 0.242 & 0.453 / 0.541 & 0.150 / 0.186 & 0.632 / 0.698 & 0.055 / 0.072 & 0.058 / 0.072 & 0.604 / 0.698 & 0.431 / 0.483 & 0.324 / 0.364 & 0.507 / 0.543 & 0.464 / 0.487 \\
\multicolumn{1}{l|}{} &  & \textit{AUC} & 0.755 / 0.782 & 0.513 / 0.508 & 0.653 / 0.650 & 0.639 / 0.638 & 0.651 / 0.635 & 0.754 / 0.746 & 0.661 / 0.646 & 0.602 / 0.608 & 0.694 / 0.675 & 0.693 / 0.683 & 0.779 / 0.770 & 0.627 / 0.634 & 0.532 / 0.535 & 0.757 / 0.750 & 0.727 / 0.726 & 0.665 / 0.661 & 0.724 / 0.720 & 0.710 / 0.702 \\ \hline

\multicolumn{1}{l|}{\multirow{10}{*}{\textbf{TOP 3}}} & \multirow{5}{*}{\textbf{CheXpert}} & \textit{\#images} & 13,296 / 5,030 & 6,986 / 4,554 & 18,241 / 11,723 & 5,773 / 2,968 & 68,914 / 41,990 & 35,897 / 22,654 & 9,479 / 6,006 & 3,781 / 2,455 & 22,396 / 14,376 & 11,697 / 7,688 & 58,410 / 36,948 & 2,498 / 1,249 & 5,796 / 3,788 & 75,410 / 48,873 & \multicolumn{4}{c}{133,375 / 73,685} \\
\multicolumn{1}{l|}{} &  & \textit{Precision} & 0.394 / 0.302 & 0.104 / 0.113 & 0.315 / 0.367 & 0.089 / 0.083 & 0.586 / 0.638 & 0.462 / 0.524 & 0.118 / 0.133 & 0.054 / 0.064 & 0.202 / 0.229 & 0.114 / 0.132 & 0.630 / 0.681 & 0.041 / 0.033 & 0.100 / 0.115 & 0.667 / 0.762 & 0.357 / 0.392 & 0.277 / 0.298 & 0.485 / 0.539 & 0.364 / 0.394 \\
\multicolumn{1}{l|}{} &  & \textit{Recall} & 0.607 / 0.623 & 0.075 / 0.067 & 0.638 / 0.628 & 0.582 / 0.597 & 0.856 / 0.853 & 0.669 / 0.660 & 0.548 / 0.552 & 0.515 / 0.514 & 0.842 / 0.844 & 0.876 / 0.899 & 0.866 / 0.851 & 0.467 / 0.441 & 0.130 / 0.132 & 0.875 / 0.861 & 0.772 / 0.768 & 0.610 / 0.609 & 0.772 / 0.768 & 0.745 / 0.745 \\
\multicolumn{1}{l|}{} &  & \textit{F1-score} & 0.478 / 0.407 & 0.087 / 0.084 & 0.421 / 0.463 & 0.154 / 0.146 & 0.696 / 0.730 & 0.547 / 0.584 & 0.194 / 0.214 & 0.097 / 0.114 & 0.326 / 0.360 & 0.202 / 0.231 & 0.729 / 0.756 & 0.075 / 0.061 & 0.113 / 0.123 & 0.757 / 0.809 & 0.488 / 0.519 & 0.348 / 0.363 & 0.577 / 0.612 & 0.464 / 0.493 \\
\multicolumn{1}{l|}{} &  & \textit{AUC} & 0.752 / 0.759  & 0.520 / 0.516 & 0.709 / 0.711 & 0.656 / 0.660 & 0.604 / 0.606 & 0.691 / 0.697 & 0.617 / 0.616 & 0.625 / 0.628 & 0.585 / 0.577 & 0.610 / 0.607 & 0.735 / 0.725 & 0.629 / 0.609 & 0.538 / 0.538 & 0.653 / 0.666 & 0.732 / 0.732 & 0.637 / 0.637 & 0.719 / 0.720 & 0.656 / 0.656 \\ \cline{2-21} 
\multicolumn{1}{l|}{} & \multirow{5}{*}{\textbf{MIMIC}} & \textit{\#images} & 86,898 / 25,331 & 6,641 / 4,228 & 41,690 / 24,530 & 6,906 / 3,022 & 48,975 / 27,226 & 24,134 / 15,108 & 9,268 / 6,248 & 16,549 / 9,331 & 43,116 / 25,693 & 8,690 / 5,759 & 50,432 / 31,506 & 2,282 / 1,199 & 4,758 / 2,884 & 55,161 / 35,891 & \multicolumn{4}{c}{220,463 / 92,741} \\
\multicolumn{1}{l|}{} &  & \textit{Precision} & 0.681 / 0.618 & 0.045 / 0.059 & 0.408 / 0.494 & 0.064 / 0.070 & 0.325 / 0.385 & 0.359 / 0.414 & 0.098 / 0.131 & 0.133 / 0.167 & 0.329 / 0.401 & 0.086 / 0.120 & 0.567 / 0.639 & 0.033 / 0.039 & 0.037 / 0.054 & 0.509 / 0.641 & 0.326 / 0.371 & 0.262 / 0.302 & 0.435 / 0.466 & 0.404 / 0.413 \\
\multicolumn{1}{l|}{} &  & \textit{Recall} & 0.741 / 0.712 & 0.090 / 0.070 & 0.472 / 0.465 & 0.528 / 0.522 & 0.739 / 0.785 & 0.677 / 0.681 & 0.541 / 0.556 & 0.432 / 0.461 & 0.766 / 0.813 & 0.686 / 0.719 & 0.740 / 0.750 & 0.380 / 0.405 & 0.149 / 0.128 & 0.764 / 0.767 & 0.673 / 0.678 & 0.550 / 0.560 & 0.673 / 0.678 & 0.668 / 0.668 \\
\multicolumn{1}{l|}{} &  & \textit{F1-score} & 0.710 / 0.662 & 0.060 / 0.064 & 0.437 / 0.479 & 0.114 / 0.123 & 0.452 / 0.516 & 0.469 / 0.515 & 0.166 / 0.212 & 0.204 / 0.245 & 0.461 / 0.537 & 0.153 / 0.206 & 0.642 / 0.691 & 0.060 / 0.070 & 0.059 / 0.076 & 0.611 / 0.698 & 0.439 / 0.480 & 0.328 / 0.364 & 0.511 / 0.535 & 0.469 / 0.480 \\
\multicolumn{1}{l|}{} &  & \textit{AUC} & 0.757 / 0.773 & 0.515 / 0.508 & 0.656 / 0.647 & 0.639 / 0.643 & 0.650 / 0.632 & 0.764 / 0.747 & 0.661 / 0.645 & 0.602 / 0.602 & 0.693 / 0.674 & 0.694 / 0.685 & 0.786 / 0.766 & 0.631 / 0.637 & 0.531 / 0.528 & 0.759 / 0.748 & 0.731 / 0.723 & 0.667 / 0.660 & 0.727 / 0.716 & 0.712 / 0.698 \\ \hline

\multicolumn{1}{l|}{\multirow{10}{*}{\textbf{TOP 2}}} & \multirow{5}{*}{\textbf{CheXpert}} & \textit{\#images} & 8,808 / 5,029 & 4,911 / 4,668 & 13,019 / 11,951 & 3,927 / 4,193 & 47,978 / 43,264 & 25,529 / 22,804 & 6,498 / 6,155 & 2,585 / 2,639 & 15,707 / 14,707 & 7,841 / 8,602 & 41,341 / 37,688 & 1,754 / 1,700 & 4,022 / 3,803 & 51,977 / 50,012 & \multicolumn{4}{c}{91,428 / 76,520} \\
\multicolumn{1}{l|}{} &  & \textit{Precision} & 0.406 / 0.270 & 0.109 / 0.098 & 0.313 / 0.360 & 0.092 / 0.111 & 0.590 / 0.632 & 0.459 / 0.517 & 0.115 / 0.130 & 0.053 / 0.067 & 0.204 / 0.227 & 0.115 / 0.142 & 0.634 / 0.676  & 0.044 / 0.044 & 0.106 / 0.104 & 0.665 / 0.755 & 0.364 / 0.389 & 0.279 / 0.295 & 0.487 / 0.529 & 0.371 / 0.390 \\
\multicolumn{1}{l|}{} &  & \textit{Recall} & 0.579 / 0.603 & 0.079 / 0.057 & 0.659 / 0.611 & 0.554 / 0.603 & 0.865 / 0.848 & 0.697 / 0.659 & 0.544 / 0.548 & 0.524 / 0.519 & 0.845 / 0.839 & 0.963 / 0.892 & 0.876 / 0.843 & 0.452 / 0.459 & 0.105 / 0.133 & 0.884 / 0.855 & 0.779 / 0.762 & 0.609 / 0.605 & 0.779 / 0.762 & 0.754 / 0.738 \\
\multicolumn{1}{l|}{} &  & \textit{F1-score} & 0.477 / 0.373 & 0.092 / 0.072 & 0.425 / 0.453 & 0.158 / 0.187 & 0.701 / 0.725 & 0.554 / 0.579 & 0.190 / 0.210 & 0.096 / 0.118 & 0.329 / 0.357 & 0.202 / 0.245 & 0.736 / 0.750 & 0.080 / 0.080 & 0.106 / 0.116 & 0.759 / 0.802 & 0.496 / 0.515 & 0.350 / 0.362 & 0.582 / 0.603 & 0.473 / 0.487 \\
\multicolumn{1}{l|}{} &  & \textit{AUC} & 0.744 / 0.744 & 0.521 / 0.512 & 0.710 / 0.705 & 0.655 / 0.661 & 0.600 / 0.603 & 0.690 / 0.699 & 0.612 / 0.613 & 0.626 / 0.630 & 0.581 / 0.579 & 0.619 / 0.605 & 0.729 / 0.725 & 0.630 / 0.616 & 0.532 / 0.536 & 0.648 / 0.666 & 0.736 / 0.729 & 0.636 / 0.635 & 0.723 / 0.717 & 0.653 / 0.655 \\ \cline{2-21} 
\multicolumn{1}{l|}{} & \multirow{5}{*}{\textbf{MIMIC}} & \textit{\#images} & 57,859 / 25,256 & 4,581 / 4,386 & 28,853 / 24,588 & 4,648 / 3,590 & 33,249 / 27,724 & 17,193 / 15,187 & 6,303 / 6,311 & 11,188 / 10,120 & 30,295 / 25,939 & 5,799 / 6,271 & 34,920 / 31,486 & 1,563 / 1,502 & 3,256 / 3,008 & 38,400 / 36,113 & \multicolumn{4}{c}{147,980 / 94,532} \\
\multicolumn{1}{l|}{} &  & \textit{Precision} & 0.692 / 0.593 & 0.046 / 0.049 & 0.399 / 0.491 & 0.070 / 0.079 & 0.327 / 0.385 & 0.359 / 0.413 & 0.101 / 0.130 & 0.134 / 0.174 & 0.336 / 0.396 & 0.093 / 0.129 & 0.568 / 0.631 & 0.037 / 0.045 & 0.038 / 0.050 & 0.513 / 0.634 & 0.338 / 0.366 & 0.265 / 0.300 & 0.438 / 0.457 & 0.414 / 0.405 \\
\multicolumn{1}{l|}{} &  & \textit{Recall} & 0.731 / 0.698 & 0.100 / 0.058 & 0.505 / 0.464 & 0.492 / 0.523 & 0.740 / 0.785 & 0.721 / 0.676 & 0.545 / 0.553 & 0.434 / 0.456 & 0.778 / 0.805 & 0.658 / 0.720 & 0.766 / 0.738 & 0.367 / 0.396 & 0.120 / 0.120 & 0.780 / 0.767 & 0.683 / 0.671 & 0.553 / 0.554 & 0.683 / 0.671 & 0.676 / 0.658 \\
\multicolumn{1}{l|}{} &  & \textit{F1-score} & 0.711 / 0.641 & 0.063 / 0.053 & 0.446 / 0.477 & 0.123 / 0.138 & 0.454 / 0.516 & 0.479 / 0.512 & 0.170 / 0.210 & 0.205 / 0.252 & 0.470 / 0.531 & 0.163 / 0.218 & 0.653 / 0.680 & 0.068 / 0.081 & 0.058 / 0.071 & 0.619 / 0.694 & 0.452 / 0.474 & 0.334 / 0.363 & 0.516 / 0.526 & 0.480 / 0.471 \\
\multicolumn{1}{l|}{} &  & \textit{AUC} & 0.761 / 0.762 & 0.517 / 0.502 & 0.660 / 0.647 & 0.640 / 0.642 & 0.650 / 0.632 & 0.776 / 0.746 & 0.664 / 0.644 & 0.603 / 0.598 & 0.691 / 0.670 & 0.699 / 0.687 & 0.793 / 0.761 & 0.633 / 0.630 & 0.526 / 0.523 & 0.761 / 0.747 & 0.738 / 0.719 & 0.670 / 0.656 & 0.732 / 0.711 & 0.715 / 0.694  \\ \hline

\multicolumn{1}{l|}{\multirow{10}{*}{\textbf{TOP 1}}} & \multirow{5}{*}{\textbf{CheXpert}} & \textit{\#images} & 4,390 / 4,940 & 2,582 / 4,805 & 7,357 / 11,902 & 2,000 / 5,579 & 25,328 / 44,160 & 13,947 / 22,810 & 3,390 / 6,248 & 1,313 / 2,545 & 8,368 / 14,778 & 3,930 / 9,632 & 22,005 / 37,642 & 899 / 1,787 & 2,049 / 3,268 & 26,981 / 50,665 & \multicolumn{4}{c}{47,411 / 78,209} \\
\multicolumn{1}{l|}{} &  & \textit{Precision} & 0.422 / 0.249 & 0.116 / 0.095 & 0.319 / 0.358 & 0.100 / 0.138 & 0.596 / 0.631 & 0.453 / 0.515 & 0.113 / 0.128 & 0.052 / 0.060 & 0.206 / 0.222 & 0.117 / 0.157 & 0.638 / 0.667 & 0.050 / 0.044 & 0.108 / 0.076 & 0.661 / 0.750 & \textbf{0.374} / 0.386 & 0.282 / 0.292 & 0.489 / 0.523 & 0.381 / 0.386 \\
\multicolumn{1}{l|}{} &  & \textit{Recall} & 0.549 / 0.591 & 0.098 / 0.055 & 0.689 / 0.606 & 0.511 / 0.600 & 0.871 / 0.845 & 0.731 / 0.663 & 0.533 / 0.551 & 0.548 / 0.511 & 0.839 / 0.824 & 0.832 / 0.890 & 0.886 / 0.833 & 0.445 / 0.465 & 0.068 / 0.119 & 0.890 / 0.855 & \textbf{0.786} / 0.759 & 0.606 / 0.600 & 0.786 / 0.759 & 0.761 / 0.735 \\
\multicolumn{1}{l|}{} &  & \textit{F1-score} & 0.477 / 0.351 & 0.106 / 0.069 & 0.436 / 0.450 & 0.168 / 0.224 & 0.707 / 0.722 & 0.559 / 0.580 & 0.187 / 0.207 & 0.095 / 0.108 & 0.330 / 0.350 & 0.205 / 0.266 & 0.741 / 0.741 & 0.089 / 0.080 & 0.084 / 0.092 & 0.759 / 0.799 & \textbf{0.507} / 0.511 & 0.353 / 0.360 & 0.586 / 0.597 & 0.483 / 0.483 \\
\multicolumn{1}{l|}{} &  & \textit{AUC} & 0.736 / 0.735 & 0.527 / 0.510 & 0.709 / 0.705 & 0.655 / 0.656 & 0.596 / 0.602 & 0.681 / 0.703 & 0.606 / 0.612 & 0.632 / 0.621 & 0.572 / 0.576 & 0.632 / 0.608 & 0.725 / 0.723 & 0.640 / 0.613 & 0.521 / 0.528 & 0.644 / 0.665 & \textbf{0.741} / 0.727 & 0.634 / 0.633 & 0.729 / 0.715 & 0.649 / 0.654 \\ \cline{2-21} 
\multicolumn{1}{l|}{} & \multirow{5}{*}{\textbf{MIMIC}} & \textit{\#images} & 28,790 / 25,017 & 2,477 / 4,634 & 15,163 / 24,714 & 2,329 / 6,930 & 17,072 / 29,313 & 9,478 / 15,219 & 3,241 / 6,644 & 5,655 / 11,005 & 16,002 / 25,970 & 2,901 / 6,464 & 18,178 / 31,759 & 780 / 1,550 & 1,636 / 2,380 & 20,628 / 35,968 & \multicolumn{4}{c}{74,374 / 97,347} \\
\multicolumn{1}{l|}{} &  & \textit{Precision} & 0.706 / 0.555 & 0.048 / 0.051 & 0.390 / 0.483 & 0.083 / 0.145 & 0.334 / 0.388 & 0.351 / 0.412 & 0.102 / 0.129 & 0.136 / 0.176 & 0.341 / 0.384 & 0.109 / 0.128 & 0.567 / 0.623 & 0.046 / 0.042 & 0.044 / 0.035 & 0.532 / 0.620 & \textbf{0.357} / 0.359 & 0.271 / 0.298 & 0.442 / 0.442 & 0.434 / 0.393 \\
\multicolumn{1}{l|}{} &  & \textit{Recall} & 0.718 / 0.686 & 0.121 / 0.057 & 0.564 / 0.456 & 0.421 / 0.542 & 0.744 / 0.780 & 0.774 / 0.679 & 0.530 / 0.554 & 0.443 / 0.450 & 0.785 / 0.784 & 0.577 / 0.720 & 0.789 / 0.722 & 0.335 / 0.398 & 0.075 / 0.118 & 0.800 / 0.760 & \textbf{0.694} / 0.661 & 0.548 / 0.550 & 0.694 / 0.661 & 0.686  / 0.648\\
\multicolumn{1}{l|}{} &  & F1-score & 0.712 / 0.614 & 0.068 / 0.054 & 0.461 / 0.470 & 0.139 / 0.229 & 0.461 / 0.518 & 0.483 / 0.513 & 0.171 / 0.209 & 0.208 / 0.253 & 0.476 / 0.516 & 0.183 / 0.217 & 0.660 / 0.669 & 0.080 / 0.077 & 0.055 / 0.054 & 0.639 / 0.683 & \textbf{0.472} / 0.466 & 0.343 / 0.362 & 0.524 / 0.513 & 0.498 / 0.459\\
\multicolumn{1}{l|}{} &  & \textit{AUC} & 0.765 / 0.748 & 0.519 / 0.502 & 0.669 / 0.645 & 0.636 / 0.649 & 0.651 / 0.625 & 0.783 / 0.750 & 0.659 / 0.640 & 0.606 / 0.591 & 0.685 / 0.663 & 0.693 / 0.685 & 0.797 / 0.755 & 0.630 / 0.626 & 0.519 / 0.519 & 0.765 / 0.747 & \textbf{0.747} / 0.712 & 0.670 / 0.653 & 0.740 / 0.704 & 0.718 / 0.688 \\ \hline
\end{tabular}%
}
\label{chest_classification}
\end{table}
\end{landscape}

\clearpage

\subsection{Chest X-ray Quality Results}~\label{chest_qual_more}
\begin{figure*}[htp]
\vspace{-6mm}
\begin{subfigure}{.48\textwidth}
    \includegraphics[width=0.98\linewidth, height=0.5\linewidth]{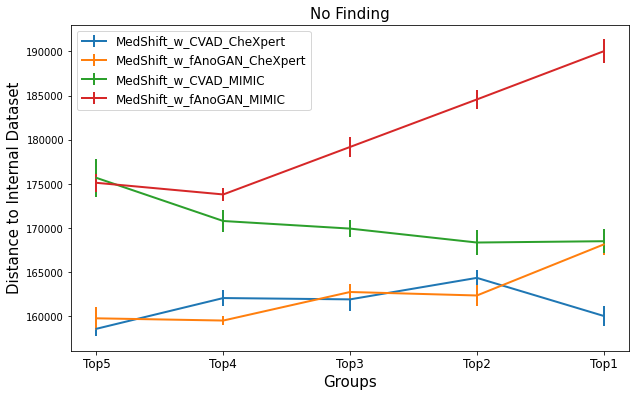}
\end{subfigure}
\begin{subfigure}{.48\textwidth}
  \includegraphics[width=0.98\linewidth, height=0.5\linewidth]{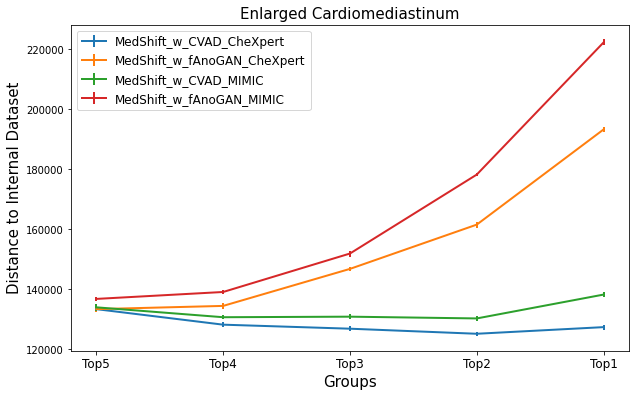}
\end{subfigure}
\begin{subfigure}{.48\textwidth}
  \includegraphics[width=0.98\linewidth, height=0.5\linewidth]{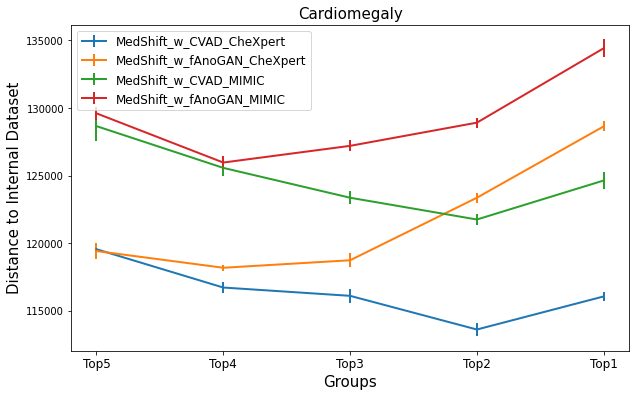}
\end{subfigure} 
\begin{subfigure}{.48\textwidth}
  \includegraphics[width=0.98\linewidth, height=0.5\linewidth]{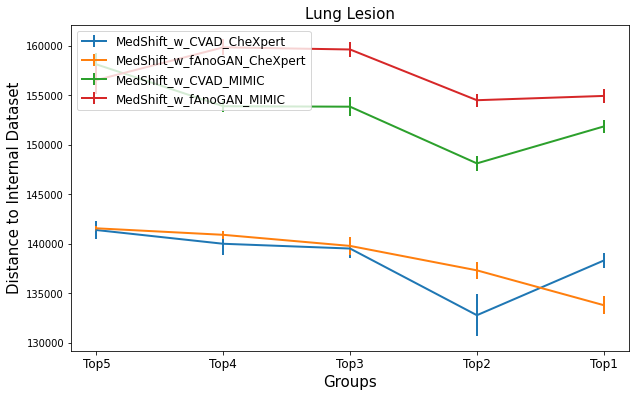}
\end{subfigure}
\begin{subfigure}{.48\textwidth}
  \includegraphics[width=0.98\linewidth, height=0.5\linewidth]{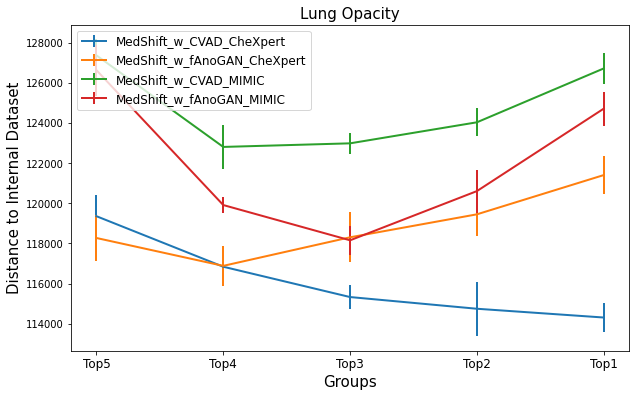}
\end{subfigure}
\begin{subfigure}{.48\textwidth}
  \includegraphics[width=0.98\linewidth, height=0.5\linewidth]{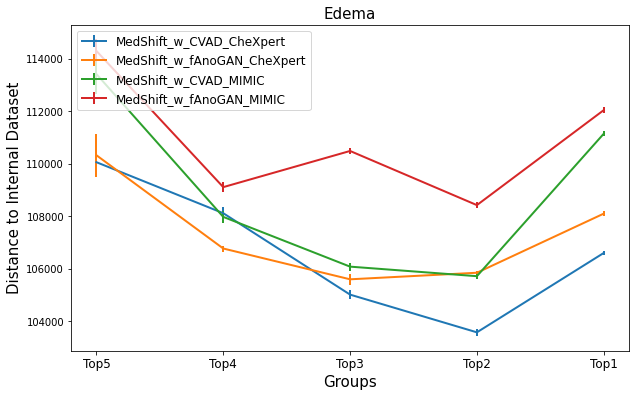}
\end{subfigure}
\begin{subfigure}{.48\textwidth}
  \includegraphics[width=0.98\linewidth, height=0.5\linewidth]{consolidation_qual.png}
\end{subfigure}
\begin{subfigure}{.48\textwidth}
  \includegraphics[width=0.98\linewidth, height=0.5\linewidth]{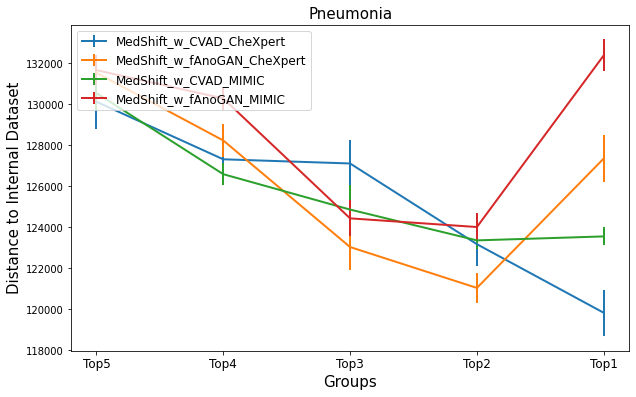}
\end{subfigure}
\caption{Chest X-ray dataset quality measurement results. From left to right, top to bottom, there are plots for classes \textit{No Finding}, \textit{Enlarged Cardiomediastinum}, \textit{Cardiomegaly},  \textit{Lung Lesion}, \textit{Lung Opacity}, \textit{Edema}, \textit{Consolidation} and \textit{Pneumonia} respectively. }
\label{fig:chest_qual1}
\vspace{-15mm}
\end{figure*}

\begin{figure*}[htp]
\begin{subfigure}{.48\textwidth}
  \includegraphics[width=0.98\linewidth]{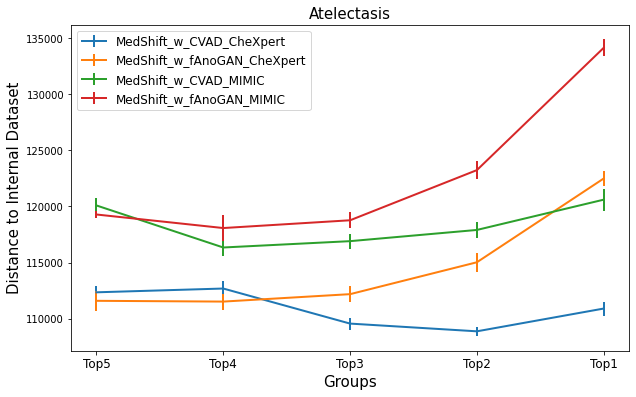}
\end{subfigure}
\begin{subfigure}{.5\textwidth}
  \includegraphics[width=0.95\linewidth]{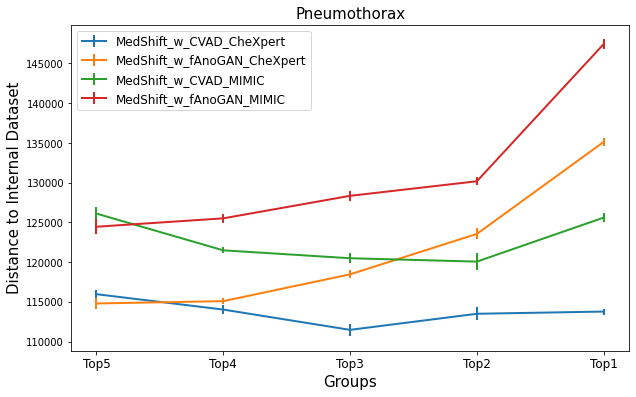}
\end{subfigure}
\begin{subfigure}{.5\textwidth}
\includegraphics[width=0.95\linewidth]{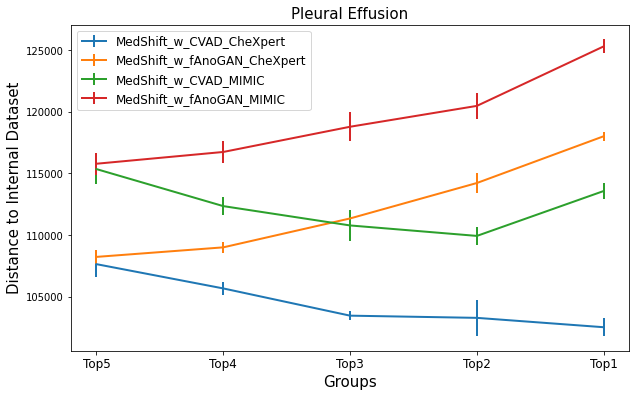}
\end{subfigure}
\begin{subfigure}{.5\textwidth}
  \includegraphics[width=0.95\linewidth]{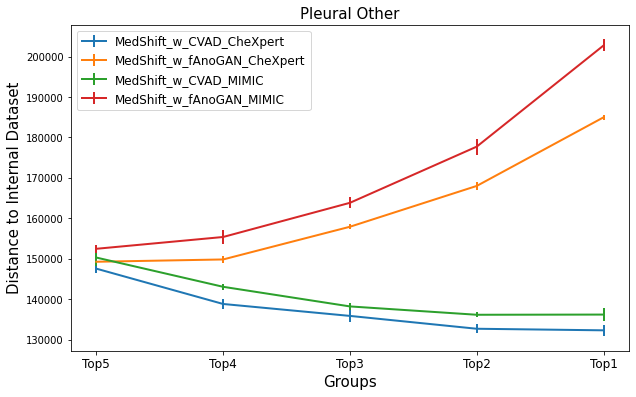}
\end{subfigure}
\begin{subfigure}{.5\textwidth}
\includegraphics[width=0.95\linewidth]{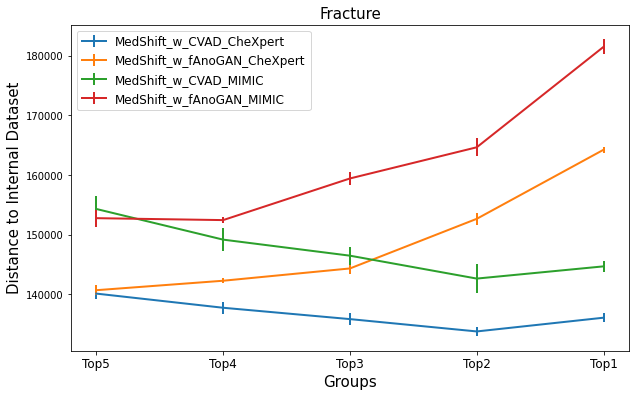}
\end{subfigure}
\begin{subfigure}{.5\textwidth}
\includegraphics[width=0.95\linewidth]{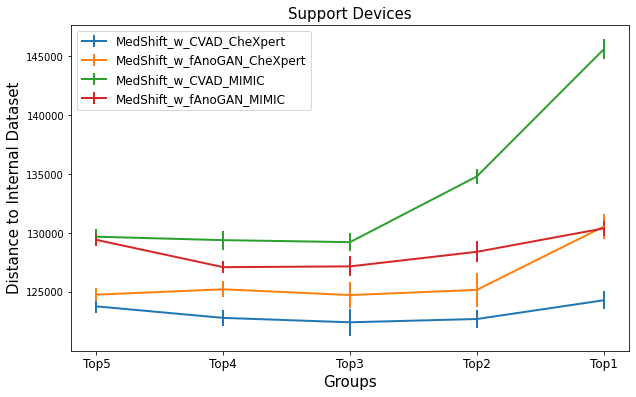}
\end{subfigure}
\caption{More Chest X-ray dataset quality measurement results. From left to right, top to bottom, there are plots for classes \textit{Atelectasis}, \textit{Pneumothorax}, \textit{Pleural Effusion},  \textit{Pleural Other}, \textit{Fracture} and \textit{Support Devices} respectively. }
\label{fig:chest_qual2}
\end{figure*}
\vspace{-15mm}


\clearpage
\twocolumn




\bibliography{mybibfile}

\begin{thebibliography}{19}
\expandafter\ifx\csname natexlab\endcsname\relax\def\natexlab#1{#1}\fi
\providecommand{\url}[1]{\texttt{#1}}
\providecommand{\href}[2]{#2}
\providecommand{\path}[1]{#1}
\providecommand{\DOIprefix}{doi:}
\providecommand{\ArXivprefix}{arXiv:}
\providecommand{\URLprefix}{URL: }
\providecommand{\Pubmedprefix}{pmid:}
\providecommand{\doi}[1]{\href{http://dx.doi.org/#1}{\path{#1}}}
\providecommand{\Pubmed}[1]{\href{pmid:#1}{\path{#1}}}
\providecommand{\bibinfo}[2]{#2}
\ifx\xfnm\relax \def\xfnm[#1]{\unskip,\space#1}\fi
\bibitem[{Alvarez~Melis and Fusi(2020)}]{alvarez2020geometric}
\bibinfo{author}{Alvarez~Melis, D.}, \bibinfo{author}{Fusi, N.},
  \bibinfo{year}{2020}.
\newblock \bibinfo{title}{Geometric dataset distances via optimal transport}.
\newblock \bibinfo{journal}{Advances in Neural Information Processing Systems}
  \bibinfo{volume}{33}.
\bibitem[{Clark et~al.(2013)Clark, Vendt, Smith, Freymann, Kirby, Koppel,
  Moore, Phillips, Maffitt, Pringle et~al.}]{clark2013cancer}
\bibinfo{author}{Clark, K.}, \bibinfo{author}{Vendt, B.},
  \bibinfo{author}{Smith, K.}, \bibinfo{author}{Freymann, J.},
  \bibinfo{author}{Kirby, J.}, \bibinfo{author}{Koppel, P.},
  \bibinfo{author}{Moore, S.}, \bibinfo{author}{Phillips, S.},
  \bibinfo{author}{Maffitt, D.}, \bibinfo{author}{Pringle, M.}, et~al.,
  \bibinfo{year}{2013}.
\newblock \bibinfo{title}{The cancer imaging archive (tcia): maintaining and
  operating a public information repository}.
\newblock \bibinfo{journal}{Journal of digital imaging} \bibinfo{volume}{26},
  \bibinfo{pages}{1045--1057}.
\bibitem[{Guo et~al.(2021)Guo, Gichoya, Purkayastha and Banerjee}]{guo2021cvad}
\bibinfo{author}{Guo, X.}, \bibinfo{author}{Gichoya, J.W.},
  \bibinfo{author}{Purkayastha, S.}, \bibinfo{author}{Banerjee, I.},
  \bibinfo{year}{2021}.
\newblock \bibinfo{title}{Cvad: A generic medical anomaly detector based on
  cascade vae}.
\newblock \bibinfo{journal}{arXiv preprint arXiv:2110.15811} .
\bibitem[{He et~al.(2016)He, Zhang, Ren and Sun}]{he2016deep}
\bibinfo{author}{He, K.}, \bibinfo{author}{Zhang, X.}, \bibinfo{author}{Ren,
  S.}, \bibinfo{author}{Sun, J.}, \bibinfo{year}{2016}.
\newblock \bibinfo{title}{Deep residual learning for image recognition}, in:
  \bibinfo{booktitle}{Proceedings of the IEEE conference on computer vision and
  pattern recognition}, pp. \bibinfo{pages}{770--778}.
\bibitem[{Huang et~al.(2017)Huang, Liu, Van Der~Maaten and
  Weinberger}]{huang2017densely}
\bibinfo{author}{Huang, G.}, \bibinfo{author}{Liu, Z.}, \bibinfo{author}{Van
  Der~Maaten, L.}, \bibinfo{author}{Weinberger, K.Q.}, \bibinfo{year}{2017}.
\newblock \bibinfo{title}{Densely connected convolutional networks}, in:
  \bibinfo{booktitle}{Proceedings of the IEEE conference on computer vision and
  pattern recognition}, pp. \bibinfo{pages}{4700--4708}.
\bibitem[{Irvin et~al.(2019)Irvin, Rajpurkar, Ko, Yu, Ciurea-Ilcus, Chute,
  Marklund, Haghgoo, Ball, Shpanskaya et~al.}]{irvin2019chexpert}
\bibinfo{author}{Irvin, J.}, \bibinfo{author}{Rajpurkar, P.},
  \bibinfo{author}{Ko, M.}, \bibinfo{author}{Yu, Y.},
  \bibinfo{author}{Ciurea-Ilcus, S.}, \bibinfo{author}{Chute, C.},
  \bibinfo{author}{Marklund, H.}, \bibinfo{author}{Haghgoo, B.},
  \bibinfo{author}{Ball, R.}, \bibinfo{author}{Shpanskaya, K.}, et~al.,
  \bibinfo{year}{2019}.
\newblock \bibinfo{title}{Chexpert: A large chest radiograph dataset with
  uncertainty labels and expert comparison}, in:
  \bibinfo{booktitle}{Proceedings of the AAAI conference on artificial
  intelligence}, pp. \bibinfo{pages}{590--597}.
\bibitem[{Johnson et~al.(2019)Johnson, Pollard, Berkowitz, Greenbaum, Lungren,
  Deng, Mark and Horng}]{johnson2019mimic}
\bibinfo{author}{Johnson, A.E.}, \bibinfo{author}{Pollard, T.J.},
  \bibinfo{author}{Berkowitz, S.J.}, \bibinfo{author}{Greenbaum, N.R.},
  \bibinfo{author}{Lungren, M.P.}, \bibinfo{author}{Deng, C.y.},
  \bibinfo{author}{Mark, R.G.}, \bibinfo{author}{Horng, S.},
  \bibinfo{year}{2019}.
\newblock \bibinfo{title}{Mimic-cxr, a de-identified publicly available
  database of chest radiographs with free-text reports}.
\newblock \bibinfo{journal}{Scientific data} \bibinfo{volume}{6},
  \bibinfo{pages}{1--8}.
\bibitem[{Qui{\~n}onero-Candela et~al.(2009)Qui{\~n}onero-Candela, Sugiyama,
  Lawrence and Schwaighofer}]{quinonero2009dataset}
\bibinfo{author}{Qui{\~n}onero-Candela, J.}, \bibinfo{author}{Sugiyama, M.},
  \bibinfo{author}{Lawrence, N.D.}, \bibinfo{author}{Schwaighofer, A.},
  \bibinfo{year}{2009}.
\newblock \bibinfo{title}{Dataset shift in machine learning}.
\newblock \bibinfo{publisher}{Mit Press}.
\bibitem[{Rabanser et~al.(2018)Rabanser, G{\"u}nnemann and
  Lipton}]{rabanser2018failing}
\bibinfo{author}{Rabanser, S.}, \bibinfo{author}{G{\"u}nnemann, S.},
  \bibinfo{author}{Lipton, Z.C.}, \bibinfo{year}{2018}.
\newblock \bibinfo{title}{Failing loudly: An empirical study of methods for
  detecting dataset shift}.
\newblock \bibinfo{journal}{arXiv preprint arXiv:1810.11953} .
\bibitem[{Rajpurkar et~al.(2017)Rajpurkar, Irvin, Bagul, Ding, Duan, Mehta,
  Yang, Zhu, Laird, Ball et~al.}]{rajpurkar2017mura}
\bibinfo{author}{Rajpurkar, P.}, \bibinfo{author}{Irvin, J.},
  \bibinfo{author}{Bagul, A.}, \bibinfo{author}{Ding, D.},
  \bibinfo{author}{Duan, T.}, \bibinfo{author}{Mehta, H.},
  \bibinfo{author}{Yang, B.}, \bibinfo{author}{Zhu, K.},
  \bibinfo{author}{Laird, D.}, \bibinfo{author}{Ball, R.L.}, et~al.,
  \bibinfo{year}{2017}.
\newblock \bibinfo{title}{Mura: Large dataset for abnormality detection in
  musculoskeletal radiographs}.
\newblock \bibinfo{journal}{arXiv preprint arXiv:1712.06957} .
\bibitem[{Schlegl et~al.(2019)Schlegl, Seeb{\"o}ck, Waldstein, Langs and
  Schmidt-Erfurth}]{schlegl2019f}
\bibinfo{author}{Schlegl, T.}, \bibinfo{author}{Seeb{\"o}ck, P.},
  \bibinfo{author}{Waldstein, S.M.}, \bibinfo{author}{Langs, G.},
  \bibinfo{author}{Schmidt-Erfurth, U.}, \bibinfo{year}{2019}.
\newblock \bibinfo{title}{f-anogan: Fast unsupervised anomaly detection with
  generative adversarial networks}.
\newblock \bibinfo{journal}{Medical image analysis} \bibinfo{volume}{54},
  \bibinfo{pages}{30--44}.
\bibitem[{Sch{\"u}tte et~al.(2021)Sch{\"u}tte, Hetzel, Gatidis, Hepp, Dietz,
  Bauer and Schwab}]{schutte2021overcoming}
\bibinfo{author}{Sch{\"u}tte, A.D.}, \bibinfo{author}{Hetzel, J.},
  \bibinfo{author}{Gatidis, S.}, \bibinfo{author}{Hepp, T.},
  \bibinfo{author}{Dietz, B.}, \bibinfo{author}{Bauer, S.},
  \bibinfo{author}{Schwab, P.}, \bibinfo{year}{2021}.
\newblock \bibinfo{title}{Overcoming barriers to data sharing with medical
  image generation: a comprehensive evaluation}.
\newblock \bibinfo{journal}{NPJ digital medicine} \bibinfo{volume}{4}.
\bibitem[{Storkey(2009)}]{storkey2009training}
\bibinfo{author}{Storkey, A.}, \bibinfo{year}{2009}.
\newblock \bibinfo{title}{When training and test sets are different:
  characterizing learning transfer}.
\newblock \bibinfo{journal}{Dataset shift in machine learning}
  \bibinfo{volume}{30}, \bibinfo{pages}{3--28}.
\bibitem[{Van~Ooijen(2019)}]{van2019quality}
\bibinfo{author}{Van~Ooijen, P.M.}, \bibinfo{year}{2019}.
\newblock \bibinfo{title}{Quality and curation of medical images and data}, in:
  \bibinfo{booktitle}{Artificial intelligence in medical imaging}.
  \bibinfo{publisher}{Springer}, pp. \bibinfo{pages}{247--255}.
\bibitem[{Van~Panhuis et~al.(2014)Van~Panhuis, Paul, Emerson, Grefenstette,
  Wilder, Herbst, Heymann and Burke}]{van2014systematic}
\bibinfo{author}{Van~Panhuis, W.G.}, \bibinfo{author}{Paul, P.},
  \bibinfo{author}{Emerson, C.}, \bibinfo{author}{Grefenstette, J.},
  \bibinfo{author}{Wilder, R.}, \bibinfo{author}{Herbst, A.J.},
  \bibinfo{author}{Heymann, D.}, \bibinfo{author}{Burke, D.S.},
  \bibinfo{year}{2014}.
\newblock \bibinfo{title}{A systematic review of barriers to data sharing in
  public health}.
\newblock \bibinfo{journal}{BMC public health} \bibinfo{volume}{14},
  \bibinfo{pages}{1--9}.
\bibitem[{Villani(2009)}]{villani2009optimal}
\bibinfo{author}{Villani, C.}, \bibinfo{year}{2009}.
\newblock \bibinfo{title}{Optimal transport: old and new}. volume
  \bibinfo{volume}{338}.
\newblock \bibinfo{publisher}{Springer}.
\bibitem[{Wang et~al.(2021)Wang, Chaudhari and Davatzikos}]{wang2021embracing}
\bibinfo{author}{Wang, R.}, \bibinfo{author}{Chaudhari, P.},
  \bibinfo{author}{Davatzikos, C.}, \bibinfo{year}{2021}.
\newblock \bibinfo{title}{Embracing the disharmony in medical imaging: A simple
  and effective framework for domain adaptation}.
\newblock \bibinfo{journal}{arXiv preprint arXiv:2103.12857} .
\bibitem[{Yamoah et~al.(2019)Yamoah, Cao, Wu, Beekman, Vandeghinste, Mannheim,
  Rosenhain, Leonardic, Kiessling and Gremse}]{yamoah2019data}
\bibinfo{author}{Yamoah, G.G.}, \bibinfo{author}{Cao, L.}, \bibinfo{author}{Wu,
  C.W.}, \bibinfo{author}{Beekman, F.J.}, \bibinfo{author}{Vandeghinste, B.},
  \bibinfo{author}{Mannheim, J.G.}, \bibinfo{author}{Rosenhain, S.},
  \bibinfo{author}{Leonardic, K.}, \bibinfo{author}{Kiessling, F.},
  \bibinfo{author}{Gremse, F.}, \bibinfo{year}{2019}.
\newblock \bibinfo{title}{Data curation for preclinical and clinical multimodal
  imaging studies}.
\newblock \bibinfo{journal}{Molecular imaging and biology}
  \bibinfo{volume}{21}, \bibinfo{pages}{1034--1043}.
\bibitem[{Yuan et~al.(2021)Yuan, Yan, Sonka and Yang}]{yuan2021large}
\bibinfo{author}{Yuan, Z.}, \bibinfo{author}{Yan, Y.}, \bibinfo{author}{Sonka,
  M.}, \bibinfo{author}{Yang, T.}, \bibinfo{year}{2021}.
\newblock \bibinfo{title}{Large-scale robust deep auc maximization: A new
  surrogate loss and empirical studies on medical image classification}, in:
  \bibinfo{booktitle}{Proceedings of the IEEE/CVF International Conference on
  Computer Vision}, pp. \bibinfo{pages}{3040--3049}.

\end{thebibliography}

\end{document}